\documentclass[pra,aps,letterpaper,longbibliography,
twocolumn,12point,superscriptaddress,floatfix]{revtex4-1}

\usepackage[utf8]{inputenc}
\usepackage{graphicx}
\usepackage{xcolor}      
\usepackage{amsmath}
\usepackage{amssymb}
\usepackage{comment}

\usepackage{esdiff}

\newcommand{\bra}[1]{\langle #1|}
\newcommand{\ket}[1]{|#1\rangle}
\newcommand{\braket}[2]{ \langle #1 | #2 \rangle }

\newcommand \be{\begin{equation}}
\newcommand \ee{\end{equation}}
\newcommand \bea{\begin{eqnarray}}
\newcommand \eea{\end{eqnarray}}
\newcommand \nn{\nonumber}

\newcommand{\red}[1]{\textcolor{black}{#1}}

\definecolor{mscolor}{rgb}{0,0.5,0.5}

\begin{document}

\title{Sensitivity of quantum gate fidelity to laser phase and intensity noise}

\author{X. Jiang}
\affiliation{
Department of Physics, University of Wisconsin-Madison, 1150 University Avenue, Madison, WI 53706}

\author{J. Scott}
\affiliation{
Department of Physics, University of Wisconsin-Madison, 1150 University Avenue, Madison, WI 53706}

\author{Mark Friesen}
\affiliation{
Department of Physics, University of Wisconsin-Madison, 1150 University Avenue, Madison, WI 53706}

\author{M. Saffman}
\affiliation{
Department of Physics, University of Wisconsin-Madison, 1150 University Avenue, Madison, WI 53706}
\affiliation{ Infleqtion, Inc.,    Madison, WI 53703}
 
\begin{abstract}
    The fidelity of gate operations on neutral atom qubits is often limited by fluctuations of the laser drive. 
    Here, we quantify the sensitivity of quantum gate fidelities to laser phase and intensity noise. 
    We first develop models to identify features observed in laser self-heterodyne noise spectra, focusing on the effects of white noise and servo bumps.
    In the weak-noise regime, characteristic of well-stabilized lasers, we show that an analytical theory based on a perturbative solution of a master equation agrees very well with numerical simulations that incorporate phase noise.
    We compute quantum gate fidelities for one- and two-photon Rabi oscillations and show that they can be enhanced by an appropriate choice of Rabi frequency relative to spectral noise peaks. We also analyze the influence of intensity noise with spectral support smaller than the Rabi frequency. 
    Our results establish requirements on laser noise levels needed to achieve desired gate fidelities.
    \end{abstract}

\date{\today}

\maketitle


\section{Introduction}

Logical gate operations on matter qubits rely on coherent driving with electromagnetic fields. For solid state qubits these are generally at microwave frequencies of 1-10 GHz.  For atomic qubits microwave as well as optical fields with carrier frequencies of several hundred THz are used for gates. High fidelity gate operations require well controlled fields with very low phase and amplitude noise. In this paper we quantify the influence of control field noise on the fidelity of gate operations on qubits. While we mainly focus on the case of optical control with lasers, our results are also applicable to high fidelity control of solid state qubits with microwave frequency fields. 

Since the limits imposed on qubit coherence and gate fidelity by control field noise  are of central importance in the quest for improving performance, the topic has been treated in a number of earlier works. Relaxation of qubits in the presence of  noise, with and without a driving field was analyzed in~\cite{Geva1995,Makhlin2003,Ithier2005,ZChen2012,FYan2013,Paladino2014,Yoshihara2014,JJing2014}. 
Using a filter function methodology the influence of control field noise on gate fidelity was analyzed in a series of papers by Biercuk and collaborators~\cite{Green2012,Green2013,Soare2014,Ball2016}. In Ref.~\cite{Soare2014} experimental measurements based on adding noise to microwave control signals were compared with  theoretical results. Subsequent work~\cite{deLeseleuc2018,MZhang2021,Day2022} has concentrated on qubit control with optical frequency fields, including the application to Rydberg gates for neutral atom qubits~\cite{deLeseleuc2018}. It was shown convincingly in~\cite{Levine2018} that filtering of the laser phase noise spectrum improves the fidelity of coherent Rydberg atom excitation and in the work of Day et. al. \cite{Day2022}, an average gate fidelity based on the filter function formalism was calculated numerically which provided a prediction of achievable performance  based on measured laser noise power spectra. Here we take a complementary approach to \cite{Day2022} and use  models for servo bump noise with Gaussian distributed amplitude as well as underlying white noise to provide compact analytical expressions that can be used to predict gate fidelity based on fits to measured laser noise spectra.         

In this paper we develop  a detailed theory of the dependence of gate fidelity on the noise spectrum of the driving field based on a perturbative solution of the master equation. Results for the cases of one- and two-photon driving are presented as well as average control fidelities together with the fidelity achieved when the qubit starts in a computational basis state, which is of particular relevance to Rydberg excitation experiments.  We show analytically using a Gaussian model for the spectral shape of servo bump noise that the spectral distribution of phase and amplitude noise relative to the Rabi frequency of the qubit drive is an important parameter that determines the extent to which noise impacts gate fidelity. Related numerical results for the impact of servo bumps on gate fidelity were presented in Ref. \cite{Day2022}.  When the noise spectrum is peaked near the Rabi frequency the deleterious effects are most prominent. Our one-photon, state-averaged results for the influence of the noise spectrum on gate error are similar to, yet quantitatively different from the predictions of filter function theory~\cite{Green2013}. As is shown in Appendix~\ref{appendix.fidelity} the differences can be traced to the use of different gate fidelity measures here, and in~\cite{Green2013}.

We proceed in Sec.~\ref{sec.lasernoise} with a summary of the theory of the laser lineshape and its relation to self-heterodyne spectral measurements. In Sec.~\ref{sec:laserparameters} we show how the theory can be used to extract parameters describing the  laser phase noise spectrum  from experimental self-heterodyne measurements.  In Sec.~\ref{sec.dm} we present a master equation description for the coherence of Rabi oscillations with a noisy drive field. A Schr\"odinger equation-based numerical simulation is given in Sec.~\ref{sec.SE}, followed by a quasi-static approximation in Sec.~\ref{sec:quasistatic}. 
The effect of intensity noise on gate fidelity is presented in Sec.~\ref{sec.intensity}.
The results obtained, as well as a comparison with filter function theory, are summarized in Sec.~\ref{sec.discussion} and the appendices.


\section{Laser Noise Analysis}
\label{sec.lasernoise}

The self-heterodyne interferometer is a powerful tool for characterizing laser noise~\cite{Okoshi1980}.
In a typical arrangement, the heterodyne circuit outputs a current $I(t)$ (or normalized current $i(t)$, as defined below) containing the noise signal.
The resulting power spectral density, $S_i(f)$, provides a convenient proxy for laser field and frequency fluctuations, $S_E(f)$ and $S_{\delta\nu}(f)$, although the correspondence is not one-to-one.
In this section, we derive these three functions and show how they are related, focusing on the regime of weak frequency noise.
While many of the results in this section have been obtained previously, we re-derive them here to establish a common framework and notation.
We then apply our results to two types of noise affecting atomic qubit experiments: white noise and servo bumps.
This analysis forms the starting point for the master equation calculations that follow.

The Rabi oscillations of a qubit are driven by a classical laser field, which we define as
\begin{equation}
\mathbf{E}(t)=\frac{\hat{\mathbf{e}}E_0(t)}{2}e^{-\imath[2\pi\nu t+\phi(t)]} +\text{c.c.}, \label{eq:E}
\end{equation} 
where $\text{c.c.}$ stands for complex conjugate.
Here we assume the polarization vector $\hat{\mathbf{e}}$ and $E_0$ may be complex.
Fluctuations of the laser field are a significant source of decoherence in current atomic qubit experiments, and are the focus of the present work. The fluctuations may occur in any of the field parameters, $\hat{\mathbf{e}}$, $E_0$, or $\phi$, where the latter is the phase of the drive.
For lasers of interest, the fluctuations predominantly occur in the phase and amplitude variables.
In this work, we therefore ignore noise in the polarization vector and focus on
 the fluctuations of $\phi(t)$. 
 The effect of relative intensity noise (RIN), where the intensity is proportional to $|E_0(t)|^2$ is considered briefly in Sec.~\ref{sec.intensity}. 
 
Phase fluctuations may alternatively be analyzed in terms of fluctuations of the frequency, $\nu(t)=\nu_0+\delta\nu(t)$, which are related to phase fluctuations through the relation
\begin{equation}
\phi(t)=\int_{t_0}^{t_0+t}2\pi\delta\nu(t')dt' , \label{eq:phasedef}
\end{equation}
where $t_0$ is a reference time.
The fluctuations of $\delta\nu(t)$ [or $\phi(t)$] have a direct influence on the Rabi oscillations, and must therefore be carefully characterized.

A compact description of a general fluctuating variable $X(t)$ is given by its autocorrelation function.
Making use of the ergodic theorem, we can equate ensemble and time averages to obtain the following definition for the autocorrelation function:
\begin{eqnarray}
R_X(\tau) &=& \langle X(t)X^*(t+\tau)\rangle 
\nonumber \\ &\equiv & \lim_{T\rightarrow\infty}\frac{1}{2T}\int_{-T}^T  X(t)X^*(t+\tau) dt . \label{eq:Rauto}
\end{eqnarray}
Throughout this work, we will only consider random variables, $X(t)$, that are wide-sense stationary.

According to the Wiener-Khintchine theorem, the autocorrelation function of $X(t)$ is related to its noise power spectrum by the Fourier transform,
\begin{equation}
S_X(f)=\int_{-\infty}^\infty R_X(\tau)e^{-i2\pi f\tau}d\tau , \label{eq:SX}
\end{equation}
and its inverse transform,
\begin{equation}
R_X(\tau)=\int_{-\infty}^\infty S_X(f)e^{i2\pi f\tau}df , \label{eq:RX}
\end{equation}
where in this work, we only consider two-sided power spectra.

The main goal of this work is to characterize the noise spectrum of $E(t)$.
However, $S_E(f)$ (also called the laser lineshape) cannot be measured directly, due to the high frequency of the carrier.
We must therefore transduce the power spectrum to lower frequencies.
Here, we consider the self-heterodyne transduction technique, in which the laser field is split, delayed, and recombined to perform interferometric measurements.
The resulting signal is read out as a photocurrent containing a direct imprint of the underlying noise spectrum.
For the dimensionless photocurrent $i(t)$, which we define below, the self-heterodyne power spectrum is denoted $S_i(f)$.

In this section, we derive the interrelated power spectra of $S_E(f)$ and $S_i(f)$, which are in turn functions of the underlying noise spectrum $S_{\delta\nu}(f)$ [or $S_\phi(f)$].
To perform noisy gate simulations, as discussed in later sections, one would like to use actual self-heterodyne experimental data to characterize the underlying noise spectra.
In principle, such a deconvolution cannot be implemented exactly~\cite{DiDomenico2010}.
However, we will show that reliable results for the noise power may indeed be obtained, particularly for lasers with very low noise levels, such as the locked and filtered lasers used in recent qubit experiments.

\subsection{Laser Lineshape}
The autocorrelation function for the laser field is defined as~\cite{Elliott1982,DiDomenico2010}
\begin{equation}
R_E(\tau) = \langle E(t)E(t+\tau)\rangle , \label{eq:RE}
\end{equation}
where we note that $E(t)$ is the real, scalar amplitude of $\mathbf{E}(t)$.
This function contains information about both the carrier signal, centered at frequency $\nu_0$, and the fluctuations, which are typically observed as a fundamental broadening of the carrier peak.
Additional features of importance for qubit experiments include structures away from the peak that may be caused by the laser locking and filtering circuitry, such as the ``servo bump", discussed in detail below.

The time average in Eq.~(\ref{eq:RE}) has been evaluated by a number of authors.
For completeness, we summarize these derivations here, following the approach of Ref.~\cite{Elliott1982}.
Let us begin by assuming the noise process is strongly stationary, so that Eq.~(\ref{eq:RE}) does not depend on $t$; for simplicity, we set $t=0$.
Using Eqs.~(\ref{eq:E}) and (\ref{eq:RE}) and trigonometric identities, defining $E_0=|E_0|e^{-i\alpha}$, and neglecting fluctuations of $E_0$, we then have
\begin{multline}
R_E(\tau)= 
\frac{|E_0|^2}{2}\Big\{\cos(2\pi\nu_0\tau)\langle\cos[\phi(\tau)-\phi(0)]\rangle \\
+\cos(2\pi\nu_0\tau)\langle\cos[\phi(\tau)+\phi(0)+2\alpha]\rangle \\
- \sin(2\pi\nu_0\tau)\langle\sin[\phi(\tau)-\phi(0)]\rangle \\
- \sin(2\pi\nu_0\tau)\langle\sin[\phi(\tau)+\phi(0)+2\alpha]\rangle\Big\} . \label{eq:RE2}
\end{multline}
We then assume the phase difference $\Phi(\tau) \equiv \phi(\tau)-\phi(0)$ to be a Gaussian random variable centered at $\Phi(\tau)=0$, with probability distribution 
\begin{equation*}
p(\Phi)=\frac{1}{\sigma_\Phi\sqrt{2\pi}}e^{-\Phi^2/2\sigma_\Phi^2} ,
\end{equation*}
and variance $\overline{\Phi^2}=\sigma_\Phi^2$. Here, the bar denotes an ensemble average.
According to the ergodic theorem, ensemble and time averages should give the same result, so that 
\begin{equation}
\sigma_\Phi^2(\tau)=\langle [\phi(\tau)-\phi(0)]^2\rangle = 2R_\phi(0) - 2R_\phi(\tau) ,
\end{equation}
where we note that
\begin{equation*}
\langle \phi^2(0)\rangle = \langle \phi^2(\tau)\rangle = R_\phi(0) .
\end{equation*}
Again, making use of the ergodic theorem, we have 
\begin{equation}
\langle \cos(\Phi)\rangle = e^{-\sigma_\Phi^2/2} 
\quad \text{and} \quad
\langle \sin(\Phi)\rangle = 0, \label{eq:moment}
\end{equation}
which is also known as the moment theorem for Gaussian random variables.
Finally we note that only biased variables like $\phi(\tau)-\phi(0)$ are Gaussian. An unbiased variable like $\phi(\tau)+\phi(0)$ is simply a random phase, for which $\langle\cos[\phi(\tau)+\phi(0)]\rangle = \langle\sin[\phi(\tau)+\phi(0)]\rangle =0$.
Combining these facts, we obtain the important relation~\cite{MZhu1993} 
\begin{equation}
R_E(\tau) = \frac{|E_0|^2}{2} \cos(2\pi\nu_0\tau) e^{\left[R_\phi(\tau)-R_\phi(0)\right]} . \label{eq:Hall}
\end{equation}
Note that since $\phi$ can take any value, $R_\phi(0)$ does not have physical significance on its own; only the difference $R_\phi(\tau)-R_\phi(0)$ is meaningful.

Another useful form for Eq.~(\ref{eq:Hall}) can be obtained from the relation $2\pi\delta\nu(t)=\partial\phi/\partial t$, together with Eq.~(\ref{eq:SX}) and the stationarity of $R_\phi(t)$, yielding
\begin{equation}
S_{\delta\nu}(f)=f^2S_\phi(f) . \label{eq:SnuSphi}
\end{equation}
Applying trigonometric identities, we then obtain the following, well-known results for the laser lineshape~\cite{DiDomenico2010}:
\begin{widetext}
\begin{equation}
R_E(\tau) = \frac{|E_0|^2}{2} \cos(2\pi\nu_0\tau) \exp\left[ -2\int_{-\infty}^{\infty}S_{\delta\nu}(f)\frac{\sin^2(\pi f\tau)}{f^2}\, df \right] ,  \label{eq:Thomann}
\end{equation}
and
\begin{equation}
S_E(f) = \frac{|E_0|^2}{2} \int_{-\infty}^\infty \cos(2\pi f\tau) \cos(2\pi \nu_0\tau) \exp\left[ -2\int_{-\infty}^{\infty}S_{\delta\nu}(f')\frac{\sin^2(\pi f'\tau)}{(f')^2}\, df' \right]
d\tau . \label{eq:ThomannSE}
\end{equation}
\end{widetext}
It is common to adopt a lineshape that is centered at zero frequency; henceforth, we therefore set $\nu_0=0$.
We note that $S_E(f)$ is properly normalized here, with $\int_{-\infty}^\infty S_E(f)df=|E_0|^2/2$.
Thus, fluctuations that broaden the lineshape also reduce the peak height.

Some additional interesting results follow from Eq.~(\ref{eq:ThomannSE}).
First, in the absence of noise [$S_{\delta\nu}=0$], we see that the laser lineshape immediately reduces to an unbroadened carrier signal: $S_E(f)=(|E_0|^2/2)\delta(f)$.
Second, when $S_{\delta\nu}$ is nonzero but small, as is typical for a locked and filtered laser, the exponential term in Eq.~(\ref{eq:ThomannSE}) may be expanded to first order, yielding~\cite{Riehle2004}
\begin{equation}
2S_E(f)/|E_0|^2 \approx [1-R_\phi(0)]\delta(f)+ S_\phi(f) . \label{eq:SEexpand}
\end{equation}
This approximation is generally very good, but breaks down in the asymptotic limit $\tau\rightarrow\infty$ of the $\tau$ integral, and therefore in the limit $f\rightarrow 0$.
To see this, we note that $\sin^2(\pi f\tau)$ may be replaced by its average value of $1/2$ in the integral; for nonvanishing values of $S_{\delta\nu}(0)$, the argument of the exponential then diverges.
To estimate the frequency $f_x$, below which Eq.~(\ref{eq:SEexpand}) breaks down, we set the argument of the exponential in Eq.~(\ref{eq:SEexpand}) to 1/2:
\begin{equation}
2\int_{f_x}^\infty\frac{S_{\delta\nu}(f)}{f^2}df\approx \frac{1}{2}. \label{eq:fx}
\end{equation}
This criterion clearly depends on the noise spectrum.

To conclude, we note that for some analytical calculations (such as the servo-bump analysis, described below), it may be convenient or pedagogical to separate the noise spectrum into distinct components: $S_{\delta\nu}(f)=S_{\delta\nu,1}(f)+S_{\delta\nu,2}(f)$, corresponding to different physical noise mechanisms.
From Eq.~(\ref{eq:Thomann}), the resulting lineshapes can then be written as
\begin{equation}
R_E(\tau)=\frac{2}{|E_0|^2} R_{E,1}(\tau)R_{E,2}(\tau), \label{eq:REcomp}
\end{equation}
where $R_{E,1}(\tau)$ and $R_{E,2}(\tau)$ are the autocorrelation functions corresponding to $S_{\delta\nu,1}(f)$ and $S_{\delta\nu,2}(f)$.
Applying the Fourier convolution theorem, we obtain
\begin{equation}
S_E(f)=2|E_0|^{-2}\int_{-\infty}^\infty S_{E,1}(f-f')S_{E,2}(f')\, df' . \label{eq:SEconv}
\end{equation}

\begin{figure}[!t]
\includegraphics[width=3.2in]{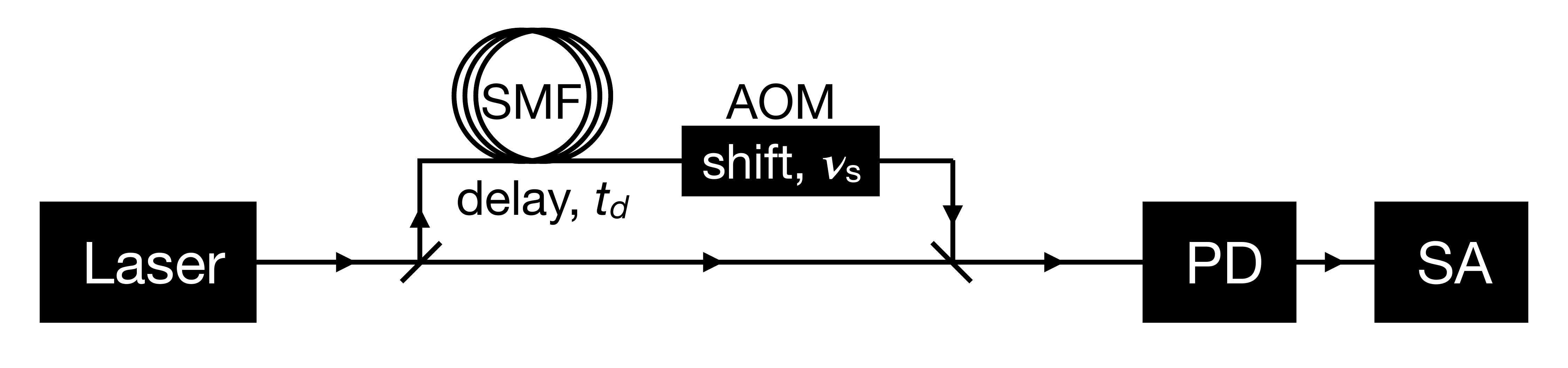}
\caption{
\label{fig:heterodyne}
Self-heterodyne setup. 
The laser signal is split equally between two paths.
One path passes through a single-mode fiber (SMF), where it is delayed by time $t_d$.
It then passes through an acousto-optic modulator (AOM), where its frequency is shifted by $\nu_s$.
The interfering signals are combined and measured by a photodiode (PD), and finally processed through a spectrum analyzer (SA).
}
\end{figure}

\subsection{Self-Heterodyne Spectrum} \label{sec:heterodyne}
We consider the self-heterodyne optical circuit shown in Fig.~\ref{fig:heterodyne}.
As depicted in the diagram, one of the paths is delayed by time $t_d$, through a long optical fiber, and then shifted in frequency by $\nu_s$, by means of an acousto-optic modulator.
Here, the delay loop allows us to interfere phases at different times, while the frequency shift provides a beat tone that is readily accessible to electronic measurements, since it occurs at submicrowave frequencies, $\nu_s\approx 100$~MHz.
The two beams are then recombined and the total intensity is measured by a photodiode, using conventional measurement techniques.

For simplicity, we assume the laser signal is split equally between the two paths, although unequal splittings may also be of interest~\cite{Gallion1984}.
The recombined field amplitude is defined as
\begin{multline}
E(t)=\frac{|E_0|}{2}\Big\{\exp [-i2\pi\nu_0 t-i\phi(t)-i\alpha] \\
+ \exp [-i2\pi(\nu_0+\nu_s) (t-t_d) -i\phi(t-t_d)-i\alpha]\Big\}+\text{c.c.} 
\end{multline}
The output current of the photodiode is then proportional to $|E(t)|^2$.
For convenience, we consider instead a dimensionless photocurrent $i(t)$, defined as
\begin{multline}
i(t) = \frac{1}{2}\Big\{ \cos\left[2\pi\nu_0 t+\phi(t)+\alpha\right]  \\
+\cos\left[2\pi(\nu_0+\nu_s) (t-t_d) +\phi(t-t_d)+\alpha\right]\Big\}^2. \label{eq:idef}
\end{multline}
The corresponding autocorrelation function is defined as
\begin{equation}
R_i(\tau)=\langle i(t)i(t+\tau) \rangle .
\end{equation}
The evaluation of $R_i(\tau)$ is greatly simplified by noting that cosine terms with $\nu_0$ in their argument average to zero in a physically realistic measurement.
Again making use of the fact that unbiased variables like $\phi(\tau)+\phi(0)$ are random phases (i.e., nongaussian), we find that 
\begin{multline}
R_i(\tau)=4 \\
+2\langle\cos[2\pi\nu_s\tau+\phi(t)-\phi(t-t_d)-\phi(t+\tau)+\phi(t+\tau-t_d)]\rangle .
\end{multline}
Taking the same approach as in the derivation of $R_E(\tau)$, we take 
$\Phi'=\phi(t)-\phi(t-t_d)-\phi(t+\tau)+\phi(t+\tau-t_d)$ to be a Gaussian random variable centered at zero, and apply the Gaussian moment relations,
\begin{equation}
\langle \cos(\Phi')\rangle=e^{-\sigma_{\Phi'}^2/2} \quad\text{and}\quad \langle \sin(\Phi')\rangle=0 ,
\end{equation}
where
\begin{equation}
\sigma_{\Phi'}^2=\langle [\phi(t)-\phi(t-t_d)-\phi(t+\tau)+\phi(t+\tau-t_d)]^2\rangle .
\end{equation}
In this way we obtain
\begin{multline} \label{eq:Ri}
R_i(\tau) = 4+2\cos(2\pi\nu_s\tau) \\
 \times \exp [2R_\phi(\tau)+2R_\phi(t_d)-2R_\phi(0) \\
 -R_\phi(\tau-t_d)-R_\phi(\tau+t_d)] .
\end{multline}
Here, the cosine function represents the beat tone, and the noise information is reflected in its amplitude.
It can be seen that the corresponding power spectrum, $S_i(f)$, includes a central peak, $\delta(f)$, which contains no information about the laser noise, and two broadened but identical satellite peaks, centered at $f=\pm\nu_s$.
We now recenter $R_i(\tau)$ at one of the satellite peaks, as consistent with typical self-heterodyne measurements, such that
\begin{multline}
R_i(\tau) \rightarrow R_i(\tau)= \exp [2R_\phi(\tau)+2R_\phi(t_d)-2R_\phi(0) \\
 -R_\phi(\tau-t_d)-R_\phi(\tau+t_d)] .  \label{eq:Rtilde} 
\end{multline}
Applying trigonometric identities, we then obtain
\begin{equation}
R_i(\tau)= \exp\left[-8\int_{-\infty}^\infty S_{\delta\nu}(f)\frac{\sin^2(\pi f\tau)\sin^2(\pi ft_d)}{f^2}df\right] . \label{eq:Ri2}
\end{equation}
Taking $S_{\delta\nu}(f)$ to be an even function, we can write
\begin{equation}
S_i(f)=\int_{-\infty}^\infty \cos(2\pi f\tau) R_i(\tau) d\tau. \label{eq:Si}
\end{equation}
We note that the self-heterodyne peak defined in this way is normalized such that $\int_{-\infty}^\infty S_i(f)df=R_i(0)=1$.

In the absence of noise [$S_{\delta\nu}=0$], we see from Eqs.~(\ref{eq:Ri2}) and (\ref{eq:Si}) that the self-heterodyne power spectrum reduces to the bare carrier:
$S_i(f)=\delta(f)$.
For nonzero but small $S_{\delta\nu}$, we can expand the exponential in Eq.~(\ref{eq:Ri2}), as was done in Eq.~(\ref{eq:SEexpand}), to obtain
\begin{multline}
S_i(f) \approx \big[1+2R_\phi(t_d)-2R_\phi(0)\big]\delta(f) \\
+4\sin^2 (\pi f t_d)S_\phi(f)  .  \label{eq:Siexpand}
\end{multline}
The second term in this expression is closely related to the envelope-ratio power spectral density described in Ref.~\cite{YLi2019}, following on the earlier work of Ref.~\cite{Tsuchida2011}, and provides a theoretical basis for the former.

As in the derivation of Eq.~(\ref{eq:SEexpand}), the expansion leading to Eq.~(\ref{eq:Siexpand}) breaks down at low frequencies.
However, the well-known ``scallop'' features in the power spectrum are seen to arise from the factor $\sin^2(\pi ft_d)$.
This result clarifies the relation between the self-heterodyne signal, the underlying laser noise, and the laser lineshape.
The latter relation is given by
\begin{equation}
|E_0|^2 S_i(f) \approx 2\sin^2 (\pi f t_d)S_E(f) ,
\label{eq:Siapprox1}
\end{equation}
where we have omitted the central carrier peak.

To conclude, we again consider the possibility that the noise spectrum may be separated into distinct components, $S_{\delta\nu}(f)=S_{\delta\nu,1}(f)+S_{\delta\nu,2}(f)$.
As for the laser lineshape, the self-heterdyne autocorrelation function may then be written as
\begin{equation}
R_i(\tau)=R_{i,1}(\tau)R_{i,2}(\tau), \label{eq:Ricomp}
\end{equation}
yielding the combined power spectrum
\begin{equation}
S_i(f)=\int_{-\infty}^\infty S_{i,1}(f-f')S_{i,2}(f')df' . \label{eq:Siconv}
\end{equation}

\subsection{White Noise}\label{sec.wn}
The self-heterodyne laser noise measurements, reported below, are well described by a combination of white noise and a Gaussian servo bump.
We now obtain analytical results for the $S_E(f)$ and $S_i(f)$ power spectra, for these two noise models.
The results for white noise are well-known~\cite{Richter1986}. 
However we reproduce them here for completeness.

\begin{figure}[t]
\includegraphics[width=2.5in]{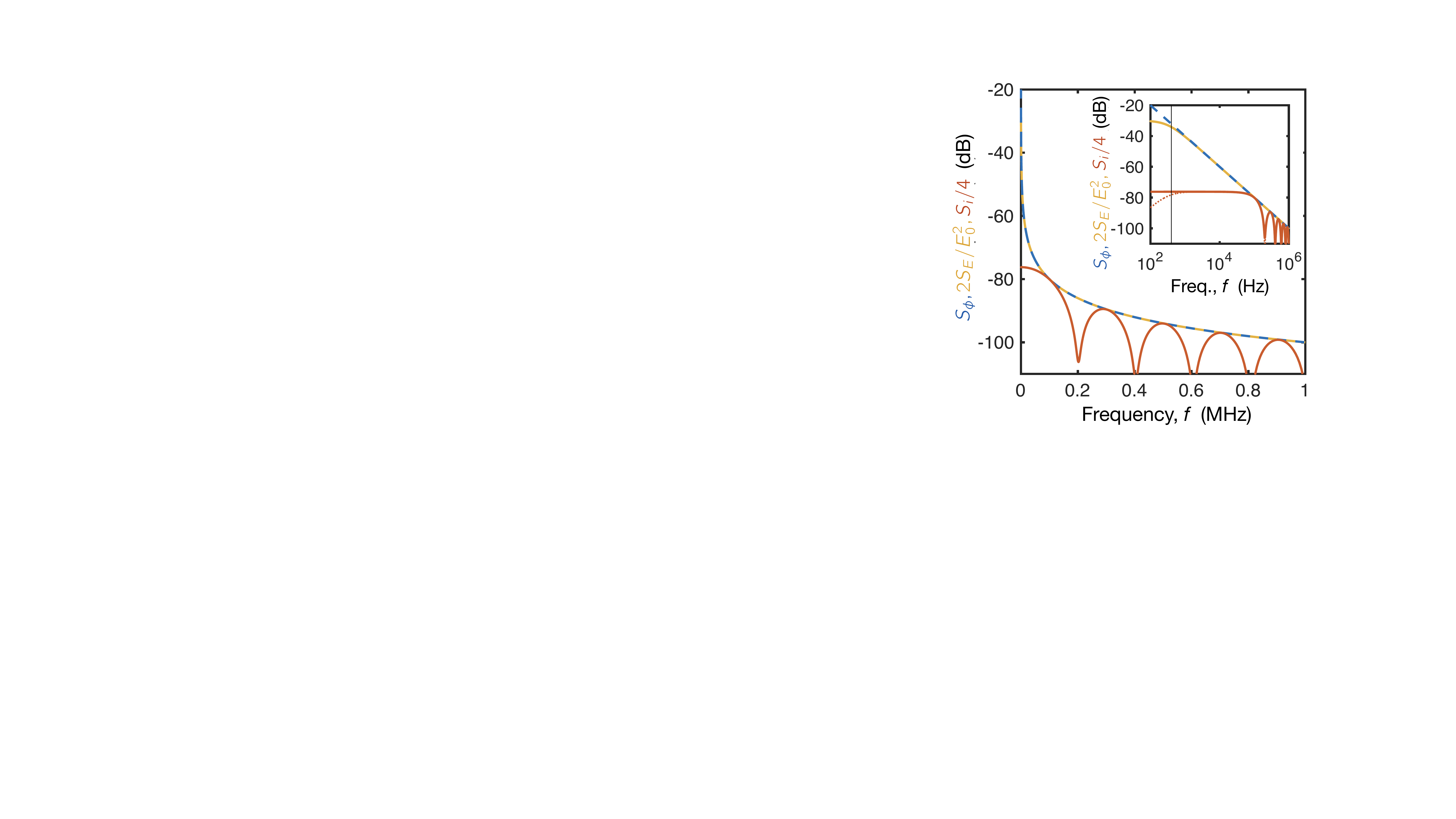}
\caption{
\label{fig:White}
White-noise power spectral densities, 
$S_\phi(f)$ (blue), $2S_E(f)/E_0^2$ (gold), and $S_i(f)/4$ (red), defined in Eqs.~(\ref{eq:Sphiwhite}), (\ref{eq:Swhite}), and (\ref{eq:Siwhite}), respectively, for noise strength $h_0=100~\rm Hz^2/Hz$.
Here we have omitted the $\delta$-function peak in $S_i(f)$.
The inset shows the same quantities plotted on a logarithmic frequency scale.
An approximate form for $S_i/4$ (red, dotted) is obtained from Eq.~(\ref{eq:Siapprox1}). 
The cutoff $f_x$ (vertical black), defined in Eq.~(\ref{eq:fx}), indicates the frequency where Eq.~(\ref{eq:SEexpand}) begins to fail.
}
\end{figure}

The underlying noise spectrum for white noise is given by
\begin{equation}
S_{\delta\nu} = h_0 \quad\quad\text{or}\quad\quad S_\phi(f) = \frac{h_0}{f^2} \label{eq:Sphiwhite}
\end{equation}
where $h_0$ is the amplitude of the power spectral density of the frequency noise and has units of $\rm Hz^2/Hz$.

Note that it is common to use a one-sided noise spectrum for such calculations; however we use a two-sided spectrum here. Our results may therefore differ by a factor of 2 from others reported in the literature.
The most straightforward calculation of $R_\phi(\tau)$, from Eq.~(\ref{eq:RX}), immediately encounters a singularity. 
We therefore proceed by calculating $R_E(\tau)$ from Eq.~(\ref{eq:Thomann}).
Setting $\nu_0=0$, to center the power spectrum, then gives
\begin{equation}
R_E(\tau) = \frac{|E_0|^2}{2} e^{-2\pi^2h_0|\tau |} .
\end{equation}
From Eq.~(\ref{eq:Hall}), we can identify
\begin{equation}
R_\phi(\tau) -R_\phi(0) = -2\pi^2h_0|\tau | , 
\end{equation}
where the singularity has now been absorbed into $R_\phi(0)$.
Solving for the laser lineshape yields
\begin{equation}
S_E(f) = \frac{|E_0|^2}{2}\frac{h_0}{f^2+(\pi h_0)^2} , \label{eq:Swhite}
\end{equation}
for which the full-width-at-half-maximum (FWHM) linewidth is $2\pi h_0$.
Away from the carrier peak, which is very narrow for a locked and well-filtered laser, we find
\begin{equation}
2S_E(f)/E_0^2 \approx \frac{h_0}{f^2} = S_\phi(f) ,  \label{eq:SEwhiteasymp}
\end{equation}
which is consistent with Eq.~(\ref{eq:SEexpand}).

We can also evaluate the self-heterodyne autocorrelation function, Eq.~(\ref{eq:Rtilde}), obtaining
\begin{equation}
R_i(\tau) = \exp\left[-2\pi^2h_0(2t_d+2|\tau|-|\tau-t_d|-|\tau+t_d|\right] , 
\end{equation}
and the corresponding power spectrum, 
\begin{eqnarray}
S_i(f) &=& \frac{2h_0}{f^2+(2\pi h_0)^2}   
+e^{-4\pi^2h_0t_d} \bigg\{ \delta(f) 
\label{eq:Siwhite} \\ && \hspace{-.3in}
-\frac{2h_0}{f^2+(2\pi h_0)^2} 
\bigg[ \cos(2\pi ft_d)+\frac{2\pi h_0}{f}\sin(2\pi ft_d) \bigg] \bigg\} . \nonumber
\end{eqnarray}

It is interesting to visualize the results and approximations employed above.
In Fig.~\ref{fig:White} we plot the white-noise power spectral densities $S_\phi(f)$, $S_E(f)$, and $S_i(f)$ corresponding to Eqs.~(\ref{eq:Sphiwhite}), (\ref{eq:Swhite}), and (\ref{eq:Siwhite}), on both linear and logarithmic scales, for the noise amplitude $h_0=100~\rm Hz^2/Hz$.
In the inset, we also plot the approximate relation between $S_i(f)$ and $S_E(f)$ given by Eq.~(\ref{eq:Siapprox1}), which clarifies how self-heterodyne measurements may be used to characterize the laser noise.
Here, we also plot the crossover frequency $f_x$ from Eq.~(\ref{eq:fx}), below which the approximations in Eq.~(\ref{eq:Siapprox1}) begin to fail. 
For the case of white noise, this expression can be evaluated analytically, giving $f_x=4h_0$.

The scallop features in Fig.~\ref{fig:White} are caused by beating between the interfering fields in the self-heterodyne circuit.
We note that, in principle, the fine-scale noise features present in $S_\phi(f)$ or $S_E(f)$ are inherited by $S_i(f)$.
However some features are obscured by the scallops, which suppress the measured signal at frequency intervals of $\Delta f=1/t_d$.

\subsection{Servo Bump}\label{sec.sb}
Lasers are commonly stabilized by locking to narrow-linewidth reference cavities~\cite{Drever1983}.
The error signal derived from the reference cavity is fed into a feedback system or servo loop~\cite{Riehle2004}, and the finite bandwidth of the servo loop induces peaks in $S_{\delta\nu}(f)$ called  servo bumps, which are typically shifted above and below  the central peak by frequencies on the order of 1~MHz.
We find that experimental servo bumps have approximately Gaussian shapes.
In fact, we find that the full noise model is well described by a Gaussian servo bump combined with white noise, as defined by
\begin{widetext}
\begin{eqnarray}
S_{\delta\nu}(f) &=& h_0 +h_g \exp \left[-\frac{(f-f_g)^2}{2\sigma_g^2}\right] 
+h_g \exp \left[-\frac{(f+f_g)^2}{2\sigma_g^2}\right] \nonumber \\
&=& h_0 +\frac{s_gf_g^2}{\sqrt{8\pi}\sigma_g} \exp \left[-\frac{(f-f_g)^2}{2\sigma_g^2}\right]
+\frac{s_gf_g^2}{\sqrt{8\pi}\sigma_g} \exp \left[-\frac{(f+f_g)^2}{2\sigma_g^2}\right]  . \label{eq:Svbump}
\end{eqnarray}
\end{widetext}
Here, $h_g$ is the bump's height, $\sigma_g$ is its width, with a FWHM given by $\sqrt{\ln 4}\,\sigma_g$, and $f_g$ is the center frequency of the bump.

In the second line of Eq.~(\ref{eq:Svbump}), we use an alternative expression for the bump height, in terms of its total, dimensionless phase-noise power, $s_g=\int_{-\infty}^\infty S_{\phi,g}(f)df$, where the subscript $g$ refers to the Gaussian noise components.
We use this expression in the simulations described below, to explore the effects of different bump shapes.
To perform the $s_g$ conversion here, we note that $S_{\phi,g}(f)$ is actually singular at $f=0$, causing its integral to diverge.
We can regularize this divergence by assuming that the servo bump is narrow (which appears to be true in many experiments), and by making the substitution
\begin{equation}
S_\phi(f)=S_{\delta\nu}(f)/f^2\approx S_{\delta\nu}(f)/f_g^2 , \label{eq:Gaussapprox}
\end{equation}
yielding 
\begin{equation}
s_g\approx \sqrt{8\pi}\sigma_gh_g/f_g^2, 
\label{eq:sg}
\end{equation}
which is the form used in Eq.~(\ref{eq:Svbump}).
We can think of this expression as describing the noise power in just the servo bump, and not the low-frequency portion of the spectrum. 
We emphasize that the latter is not ignored but it is subsumed into the white noise, which we treat separately.

\begin{figure*}[t]
\includegraphics[width=6.5in]{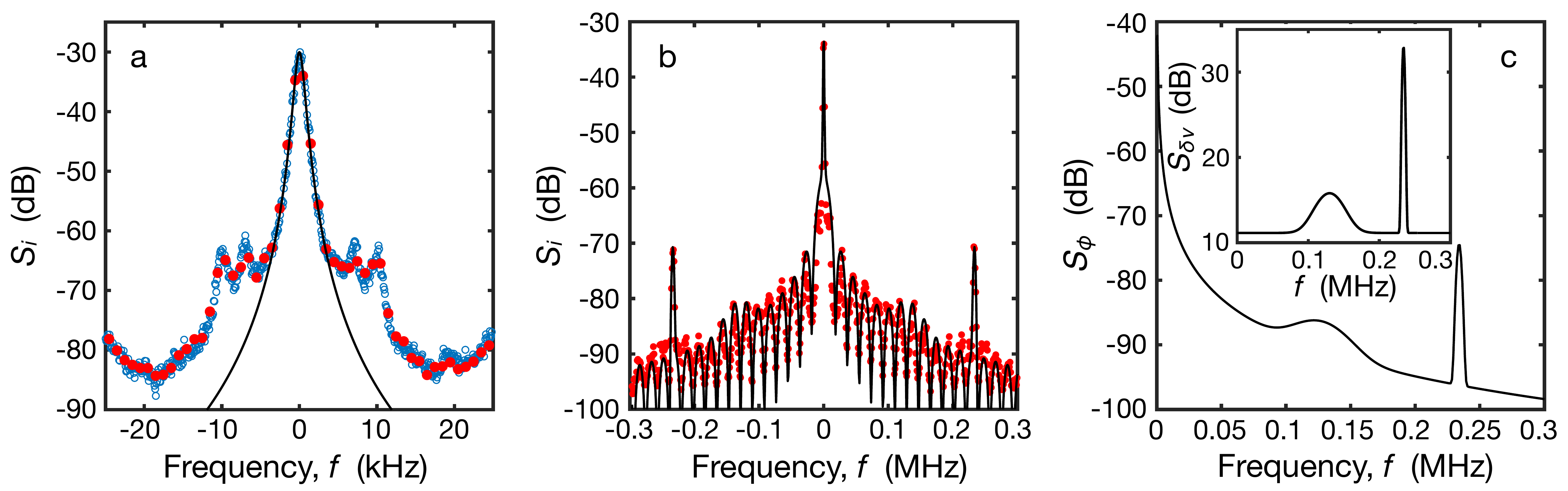}
\caption{
\label{fig:servobump}
Self-heterodyne data from a 1040~nm solid-state Ti:Sa laser (MSquared SolsTiS) with fits to white and Gaussian-bump noise models. The laser was locked to a reference cavity with linewidth of approximately 5~kHz using the Pound-Drever-Hall method~\cite{Drever1983}.
(a) Self-heterodyne power spectral density, obtained in a 50~kHz frequency window with $\text{RBW}=100$~Hz (open blue circles) and in a 600~kHz window with $\text{RBW}=300$~Hz (closed red markers).
An optical delay fiber of 11~km was used for both data sets, corresponding to a delay time of $t_d=5.445\times 10^{-5}$~s.
The data were shifted horizontally to center their peaks at zero frequency, and the peak data were then fit to Eq.~(\ref{eq:Sipeak}) (black line), obtaining $\alpha=5/2$ and $\sigma=240$~Hz.
The data were finally renormalized (i.e., shifted vertically) to ensure the correct total power, as described in the main text.
(b) Same red data set as (a), plotted over a wider frequency window.
The data were fit to Eq.~(\ref{eq:Sifull}), including white noise and two Gaussian servo bumps, obtaining $h_0=13$~Hz$^2$/Hz, 
$h_{g1}=25$~Hz$^2$/Hz, $\sigma_{g1}=18$~kHz, $f_{g1}=130$~kHz,
and
$h_{g2}=2.0\times 10^3$~Hz$^2$/Hz, $\sigma_{g12}=1.5$~kHz, $f_{g2}=234$~kHz.
(c) Phase and frequency power spectral densities, $S_\phi(f)$ and $S_{\delta\nu}(f)$ (inset), resulting from the fitting.
}
\end{figure*}

We first consider just the Gaussian term in Eq.~(\ref{eq:Svbump}), setting $h_0=0$.
Fourier transforming Eq.~(\ref{eq:Gaussapprox}), we obtain
\begin{equation}
R_\phi(\tau)\approx s_g\cos(2\pi f_g\tau)e^{-2\pi^2\sigma_g^2\tau^2} .
\end{equation}
Thus for $s_g\ll 1$, Eq.~(\ref{eq:SEexpand}) gives
\begin{multline}
2S_E(f)/|E_0|^2\approx \delta(f)+\frac{h_g}{f_g^2} \exp \left[-\frac{(f-f_g)^2}{2\sigma_g^2}\right] \\
+\frac{h_g}{f_g^2} \exp \left[-\frac{(f+f_g)^2}{2\sigma_g^2}\right] , \label{eq:SEg}
\end{multline}
and Eq.~(\ref{eq:Siexpand}) gives
\begin{multline}
S_i(f) \approx \delta(f) 
+\frac{4h_g}{f_g^2} \sin^2 (\pi ft_d)\exp \left[-\frac{(f-f_g)^2}{2\sigma_g^2}\right]  \\
+\frac{4h_g}{f_g^2} \sin^2 (\pi ft_d)\exp \left[-\frac{(f+f_g)^2}{2\sigma_g^2}\right] .
\end{multline}

The white noise component of $S_{\delta\nu}(f)$ can now be included, and using Eqs.~(\ref{eq:SEconv}) and (\ref{eq:Siconv}), yields
\begin{gather}
S_E(f)=\frac{2}{|E_0|^{2}}\int_{-\infty}^\infty S_{E,w}(f-f')S_{E,g}(f')df'  , \\
S_i(f)=\int_{-\infty}^\infty S_{i,w}(f-f')S_{i,g}(f')df' ,
\end{gather}
where the subscripts $w$ and $g$ refer to white and Gaussian power spectra, which have already been computed.
We solve these integrals, approximately, by noting that a convolution between two peaks, with very different widths, is dominated by the wider peak.
We further note that, for the lasers of interest here, the servo bump is much wider than the white-noise Lorentzian peak, which is in turn much wider than a delta function.
Hence, we find that 
\begin{multline}
2S_E(f)/E_0^2\approx \frac{h_0}{f^2+(\pi h_0)^2}+\frac{h_g}{f_g^2} \exp \left[-\frac{(f-f_g)^2}{2\sigma_g^2}\right] \\
+\frac{h_g}{f_g^2} \exp \left[-\frac{(f+f_g)^2}{2\sigma_g^2}\right] , \label{eq:SEfull}
\end{multline}
and
\begin{eqnarray}
S_i(f) &\approx & \frac{2h_0}{f^2+(2\pi h_0)^2}   
+e^{-4\pi^2h_0t_d} \bigg\{ \delta(f) 
\nonumber \\ && \hspace{-.1in}
-\frac{2h_0}{f^2+(2\pi h_0)^2} 
\bigg[ \cos(2\pi ft_d)+\frac{2\pi h_0}{f}\sin(2\pi ft_d) \bigg] \bigg\}  \nonumber \\
&& +\frac{4h_g}{f_g^2} \sin^2 (\pi ft_d)\exp \left[-\frac{(f-f_g)^2}{2\sigma_g^2}\right]  \nonumber \\
&& +\frac{4h_g}{f_g^2} \sin^2 (\pi ft_d)\exp \left[-\frac{(f+f_g)^2}{2\sigma_g^2}\right] .  \label{eq:Sifull}
\end{eqnarray}
In the following section, we apply these equations as fitting forms for experimental self-heterodyne data.

\section{Laser characterization} \label{sec:laserparameters}

To help visualize these results, we now characterize a stabilized solid-state Ti:Sa laser used in quantum gate experiments with atomic qubits~\cite{Graham2022}.
In Fig.~\ref{fig:servobump}(a), we plot two experimental data sets from the same laser (red and blue markers).
We find that the central peak is broadened more significantly than the resolution bandwidth (RBW) settings of the spectrum analyzer.
The corresponding linewidths are approximately equal, despite their different RBW, suggesting that RBW is not the only source of broadening.  

Although the central peak does not exhibit a clear characteristic form, we find that that it is well described by
\begin{equation}
S_{i,\text{peak}}(f)=\frac{s_p\sigma^{2\alpha-1}}{(f^2+\pi^2\sigma^2)^\alpha} \label{eq:Sipeak} .
\end{equation}
Fitting the data to this form yields $\alpha=5/2$ and $\sigma=240$~Hz.
The corresponding FWHM is 850~Hz, which is indeed several times larger than the RBW of the measurements.
Integrating Eq.~(\ref{eq:Sipeak}) over frequency yields $4s_p/3\pi^4$.
The data are therefore shifted vertically in Fig.~\ref{fig:servobump}(a) to give the correct normalization, $\int_{-\infty}^\infty S_i(f)df=1$.

After this normalization step, the self-heterodyne data away from the peak are fit to Eq.~(\ref{eq:Sifull}), where we introduce two distinct servo bumps, obtaining the result shown in Fig.~\ref{fig:servobump}(b) (black line).
%
The corresponding power spectral densities for the noise are plotted in Fig.~\ref{fig:servobump}(c).
We can use these results to determine the fractional noise power in different components of the $S_\phi$ spectrum
For the first servo bump, we obtain the fractional power $s_{g1}=0.00027$, and for the second servo bump, we obtain the fractional power $s_{g2}=0.00013$.
Together, these represent a small but non-negligible fraction of the total laser power.

The spectral features of the locked laser can be related to the stabilization system. The laser is stabilized using three feedback loops: a slow piezo with bandwidth of approximately 50~Hz, a faster piezo with bandwidth of 100~kHz, and an electro-optic phase modulator with bandwidth of several MHz.  The servo bumps centered at 
$f_{g1}$ and $f_{g2}$ are attributable to the fast piezo and the electro-optic modulator. 

\section{Density Matrix Solutions for Rabi Oscillations}
\label{sec.dm}
In this section, we compute the density matrix of a qubit undergoing Rabi oscillations driven by a laser (or lasers) with frequency noise. 
The calculation involves taking an average over all possible noise realizations.
In Sec.~\ref{sec.ts}, we perform a Fourier expansion of a generic Gaussian noise process, which is incorporated into the master equation calculation of Sec.~\ref{sec:Master}, and is used again in the numerical simulations of Sec.~\ref{sec.SE}.
In Sections~\ref{sec:1p} and \ref{sec:2p}, we use our master equation results to compute one and two-photon gate fidelities for the Rabi oscillations.

\subsection{Time-Series Expansion of the Laser Noise}\label{sec.ts}
A real, fluctuating Gaussian process $X(t)$, with zero mean and variance $\sigma^2_X$, can generally be expressed as a Fourier time series:
\begin{equation}
X(t)=\sum_{j=1}^\infty x_j\cos(2\pi f_jt+\varphi_j) , \label{eq:XFourier}
\end{equation}
where $f_j=j\Delta f$.
Here, we define $X(0)=0$ for convenience.
The random variables $x_j$ can be selected as Rayleigh-distributed random values~\cite{Tucker1984}, while the random variables $\varphi_j$ are uniformly distributed over $[0,2\pi]$.

We can compute the variance of $X(t)$ as
\begin{equation}
\sigma_X^2=\left\langle \sum_{j,k} x_jx_k\cos(2\pi f_jt+\varphi_j)\cos(2\pi f_kt+\varphi_k)\right\rangle,
\end{equation}
where an average is taken over the various random variables.
Since these variables are assumed to be statistically independent, the double sum vanishes, except for the case $j=k$.
Hence,
\begin{equation}
\sigma_X^2=\sum_{j=1}^\infty \langle x_j^2\rangle/2 . \label{eq:Xvar1}
\end{equation}
The variance is also related to the same-time autocorrelation function, defined in Eq.~(\ref{eq:Rauto}), such that
\begin{equation}
\sigma_X^2=R_X(0)=2\int_0^\infty S_X(f)df  
=2\sum_{j=1}^\infty S_{X,j}\,\Delta f , \label{eq:Xvar2}
\end{equation}
where we have assumed that $S_X(f)$ is an even function, and converted the integral to a series representation, with $S_{X,j}\equiv S_X(f_j)$.
Comparing Eqs.~(\ref{eq:Xvar1}) and (\ref{eq:Xvar2}), we note the correspondence 
\begin{equation}
\langle x_j^2\rangle \leftrightarrow 4S_{X,j}\,\Delta f .
\end{equation}
To generate time traces of $X(t)$, it is then standard practice to make the following replacement for the random variable $x_j$ in Eq.~(\ref{eq:XFourier})~\cite{Tucker1984}: 
\begin{equation}
x_j\rightarrow 2\sqrt{S_{X,j}\,\Delta f}.
\end{equation}
Defined in this way, $x_j$ is deterministic rather than random.
The resulting time trace $X(t)$ inherits the correct statistical properties of $x_j$.
Although this procedure cannot account for all random behavior of $X(t)$~\cite{Saulnier2009}, it successfully describes most behavior.

The method described above is used to generate random time traces in our numerical simulations, as described below in Sec.~\ref{sec.de}.
Specifically, the simulations employ time traces of the laser phase fluctuations, defined as
\begin{equation}
\phi(t)=\sum_{j=1}^\infty 2\sqrt{S_\phi(f_j)\Delta f}\cos(2\pi f_jt+\varphi_j) . \label{eq:randomphi}
\end{equation}
Here, we note that, while time-series amplitude coefficients have been replaced by their deterministic averages, the random phases $\varphi_j$ must still be chosen from the uniform distribution $[0,2\pi]$.
In the following section, we employ time traces of the laser frequency fluctuations, defined as
\begin{equation}
\delta \nu(t)=\frac{1}{2\pi}\frac{d\phi}{dt} = \sum_{j=1}^\infty\delta\nu_j\sin(2\pi f_jt+\varphi_j) , \label{eq:dnu}
\end{equation}
where
\begin{equation}
\delta\nu_j=-2\sqrt{S_{\delta\nu}(f_j)\Delta f} , \label{eq;dnudef}
\end{equation}
and we make use of the relation $S_{\delta\nu}(f)=f^2S_\phi(f)$.

\subsection{Time-Series Master Equation} \label{sec:Master}
We consider a two-level system $\{\ket{g},\ket{e}\}$, with corresponding energy levels $E_g$ and $E_e$, and qubit energy $h\nu_0=E_e-E_g$.
The Hamiltonian for a qubit driven resonantly by a monochromatic laser with frequency $\nu_0$ and angular Rabi frequency $\Omega_0$ (assumed to be real) is then
\begin{multline}
H=\frac{h\nu_0}{2}(\ket{e}\bra{e}-\ket{g}\bra{g}) \\
+ \hbar\Omega_0\cos(2\pi\nu_0t) 
\left[e^{i\phi(t)}\ket{e}\bra{g}+e^{-i\phi(t)}\ket{g}\bra{e} \right], \label{eq:Hinit}
\end{multline}
where we have explicitly included phase fluctuations $\phi(t)$.
Moving to a rotating frame defined by $U(t)=\exp[-i\pi\nu_0t(\ket{e}\bra{e}-\ket{g}\bra{g})]$ and applying a rotating wave approximation (RWA), we obtain the transformed Hamiltonian
\begin{eqnarray}
H' &\approx& \frac{\hbar\Omega_0}{2}\left[e^{i\phi(t)}\ket{e}\bra{g}+e^{-i\phi(t)}\ket{g}\bra{e} \right] ,\nonumber \\
&=& \frac{\hbar\Omega_0}{2}[\cos\!\phi(t)\,\sigma_x-\sin\!\phi(t)\,\sigma_y]
, \label{eq:Hp}
\end{eqnarray}
where the Pauli matrices are defined as $\sigma_x=\ket{g}\bra{e}+\ket{e}\bra{g}$, $\sigma_y=i(\ket{g}\bra{e}-\ket{e}\bra{g})$, and $\sigma_z=\ket{g}\bra{g}-\ket{e}\bra{e}$.

In Eqs.~(\ref{eq:Hinit}) and (\ref{eq:Hp}), the axis of Rabi rotations shifts with the fluctuating phase $\phi(t)$.
Although this description captures the physics of the problem, it is inconvenient for our calculations.
We therefore consider a frame that follows the fluctuating phase, in which the rotation axis is fixed~\cite{Haslwanter1988}.
This fluctuating frame is defined by the transformation $U_\varphi(t)=\text{diag}[e^{i\varphi/2},e^{-i\varphi/2}]$, yielding the Hamiltonian
\begin{eqnarray}
H''&\approx& \frac{\hbar\Omega_0}{2}\sigma_x-\frac{\hbar (d\phi/dt)}{2}\sigma_z \nonumber \\
&&=\frac{\hbar\Omega_0}{2}\sigma_x-\frac{h\delta\nu(t)}{2}\sigma_z  \nonumber \\
&&= \frac{\hbar\Omega_0}{2}\sigma_x-\sum_{j=1}^\infty \frac{h\delta\nu_j}{2} \sin(2\pi f_jt+\varphi_j) \sigma_z ,
\label{eq:Hpp}
\end{eqnarray}
where we make use of Eq.~(\ref{eq:dnu}), and the only approximation employed is the standard RWA in Eq.~(\ref{eq:Hp}).

In Eq.~(\ref{eq:Hpp}), we have moved to a frame where $\Omega_0$ now represents the quantizing field, and where $\delta\nu_j$ represents a Rabi driving term, applied simultaneously at multiple frequencies.
To formalize this correspondence, we move to the frame where $\Omega_0$ points towards the north pole of the Bloch sphere, as defined by the transformation $U=\exp[-i(\pi/4)\sigma_y]$, obtaining 
\begin{equation}
H'''\approx \frac{\hbar\Omega_0}{2}\sigma_z+\sum_{j=1}^\infty\frac{h\delta\nu_j}{2} \sin(2\pi f_jt+\varphi_j) \sigma_x . \label{eq:Hppp}
\end{equation}
For simplicity, we drop the primed notation on $H'''$ in the following derivations.

We now solve for the time evolution of the density operator, for a two-level system governed by Eq.~(\ref{eq:Hppp}):
\begin{equation}
\hbar\frac{d\rho}{dt}=i[\rho,H] . \label{eq:rhot}
\end{equation}
Although Eq.~(\ref{eq:Hppp}) has the standard form of a Rabi rotation, we note that conventional Rabi techniques are not applicable here, because in the frame of Eq.~(\ref{eq:Hppp}), the initial state of the system is along the driving axis ($\hat{x}$), as discussed below.
As such, the time evolution arises entirely from the counterrotating terms in Eq.~(\ref{eq:Hppp}), rather than the co-rotating terms.
(Note that counterrotating and co-rotating refer, here, to the $\delta\nu_j$ fluctuations, not the original Rabi drive.)
Moreover, we will need to consider perturbative corrections to $\rho(t)$ of order ${\cal O}[\delta\nu_j^2]$ in the frequency fluctuations.

To construct a perturbation theory, we first note that $2\pi\delta\nu_j$ is typically smaller than $\Omega_0$, allowing us to define the dimensionless small parameter, $\delta_j=2\pi\delta\nu_j/\Omega_0\lesssim 1$.
Defining $\omega_j=2\pi f_j$, the Hamiltonian becomes
\begin{equation}
\frac{H}{\hbar\Omega_0}=\frac{\sigma_z}{2}+\sum_{j=1}^\infty\delta_j\frac{\sigma_x}{2}\sin(\omega_jt+\varphi_j) . \label{eq:Hdelta}
\end{equation}
We can then solve the density matrix by expanding in powers of the small parameter,
\begin{eqnarray}
\rho &=& \rho_0+\rho_1+\rho_2+\dots \nonumber \\
& =& \rho_0+\sum_{j=1}^\infty\delta_j\rho_1^{(j)}+\sum_{j,k=1}^\infty\delta_j\delta_k\rho_2^{(j,k)}+\dots, \label{eq:rhoexpand}
\end{eqnarray}
where $\rho_m^{(j,\dots)}$ are assumed to be independent of $\delta_j$.
Inserting Eqs.~(\ref{eq:Hdelta}) and (\ref{eq:rhoexpand}) into (\ref{eq:rhot}), collecting terms of equal order in $\delta_j$, and solving up to ${\cal O}[\delta_j^2]$ gives
\begin{gather}
\frac{1}{\Omega_0}\frac{d\rho_0}{dt}=i\left[\rho_0,\frac{\sigma_z}{2}\right] , \\
\hspace{-2in} \sum_{j=1}^\infty\frac{\delta_j}{\Omega_0}\frac{d\rho_1^{(j)}}{dt}=  \label{eq:rho1} \\
\hspace{.2in} i\sum_{j=1}^\infty\delta_j\left\{\left[\rho_1^{(j)},\frac{\sigma_z}{2}\right]
+ \left[\rho_0,\frac{\sigma_x}{2}\sin(\omega_jt+\varphi_j)\right]\right\}, \nonumber \\
\hspace{-2in} \sum_{j,k=1}^\infty\frac{\delta_j\delta_k}{\Omega_0}\frac{d\rho_2^{(j,k)}}{dt}= \label{eq:rho2} \\
\hspace{.2in} i\sum_{j,k=1}^\infty\delta_j\delta_k\left\{\left[\rho_2^{(j,k)},\frac{\sigma_z}{2}\right]
+ \left[\rho_1^{(j)},\frac{\sigma_x}{2}\sin(\omega_kt+\varphi_k)\right]\right\}.  \nonumber
\end{gather}

For a Rabi driving experiment, in the frame of Eq.~(\ref{eq:Hinit}), we consider a qubit initialized to the north pole of the Bloch sphere.
In the frame of Eq.~(\ref{eq:Hdelta}), the corresponding initial state on the Bloch sphere is $\rho(0)=\frac{1}{2}(1+\sigma_x)$.
Since $\rho_0$, $\rho_1^{(j)}$, and $\rho_2^{(j,k)}$ are independent of $\delta_j$, the initial conditions for the different terms in the density operator expansion are given by $\rho_0(0)=\frac{1}{2}(1+\sigma_x)$, with
$\rho_1(0)=\rho_2(0)=0$.

Taking into account these initial conditions, the $j^\text{th}$ term of the expansion in Eq.~(\ref{eq:rho1}), can be solved independently of the other terms, as follows:
\begin{equation}
\frac{1}{\Omega_0}\frac{d\rho_1^{(j)}}{dt}=  
 i\left[\rho_1^{(j)},\frac{\sigma_z}{2}\right]
+ i\left[\rho_0,\frac{\sigma_x}{2}\sin(\omega_jt+\varphi_j)\right] . \label{eq:rho1b} 
\end{equation}
Now defining $\rho_1\equiv\sum_{j=1}^\infty\delta_j\rho_1^{(j)}$, rewriting Eq.~(\ref{eq:rho1}) in the form
\begin{equation}
\frac{1}{\Omega_0}\frac{d\rho_1}{dt}=  
 i\left[\rho_1,\frac{\sigma_z}{2}\right]
+ i\left[\rho_0,\frac{\sigma_x}{2}\left\{\sum_{j=1}^\infty\delta_j\sin(\omega_jt+\varphi_j)\right\}\right] ,
\end{equation}
and making use of the uniqueness theorem for differential equations, we see that this solution for $\rho_1$ is unique.

In Eq.~(\ref{eq:rho2}), we note the presence of mixed terms, involving parameters $j$ and $k$.
This is inconvenient; however, in the following derivations, we perform an average over the independent, fluctuating variables $\{\varphi_j,\varphi_k\}\in [0,2\pi]$, which leads to a helpful simplification.
Let us define the averaging procedure as
\begin{equation}
\langle f(\varphi)\rangle_\varphi = \frac{1}{2\pi}\int_0^{2\pi}f(\varphi)d\varphi .
\end{equation}
In the derivations described below, it can be shown that
\begin{multline}
\left\langle \left[\rho_1^{(j)},\frac{\sigma_x}{2}\sin(\omega_kt+\varphi_k)\right]\right\rangle_{\varphi_j,\varphi_k} = \\
\delta_{jk}\left\langle \left[\rho_1^{(j)},\frac{\sigma_x}{2}\sin(\omega_jt+\varphi_j)\right]\right\rangle_{\varphi_j} ,
\end{multline}
where $\delta_{jk}$ is the Kronecker $\delta$-function.
As a result, we find that $\langle\rho_2^{(j,k)}\rangle_{\varphi_j,\varphi_k}=\delta_{jk}\langle\rho_2^{(j,j)}\rangle_{\varphi_j}\equiv \delta_{jk}\langle\rho_2^{(j)}\rangle_{\varphi_j}$.
Anticipating this step, we can preemptively eliminate the mixed terms in Eq.~(\ref{eq:rho2}), so that the sum runs only over the variable $j$.
As was the case for $\rho_1^{(j)}$, we can then independently solve for each $\rho_2^{(j)}$, obtaining a unique solution for $\rho_2$.
Equation~(\ref{eq:rho2}) can therefore be replaced by the decoupled equation
\begin{equation}
\frac{1}{\Omega_0}\frac{d\rho_2^{(j)}}{dt}=
i\left[\rho_2^{(j)},\frac{\sigma_z}{2}\right]
+ i\left[\rho_1^{(j)},\frac{\sigma_x}{2}\sin(\omega_jt+\varphi_j)\right].  \label{eq:rho2b} 
\end{equation}
Thus, we may solve for the density matrix terms $\rho_1^{(j)}$ and $\rho_2^{(j)}$ independently, and combine the results for different $j$ values afterwards.

Following the procedure described above, we perturbatively solve for $\rho$, apply initial conditions, and perform an average over the fluctuating variable $\varphi_j$, obtaining
\begin{widetext}
\begin{multline}
\langle\rho(t)\rangle \approx \frac{1}{2}
+\left[\frac{1}{2}\cos(\Omega_0t)
-\sum_{j=1}^\infty\delta_j^2\frac{2\cos(\Omega_0t)-2\cos(\omega_jt)+(\Omega_0^2-\omega_j^2)(t/\Omega_0)\sin(\Omega_0t)}
{8(\Omega_0^2-\omega_j^2)^2/\Omega_0^4} \right]\sigma_x \\
+\left[\frac{1}{2}\sin(\Omega_0t)
-\sum_{j=1}^\infty\delta_j^2\frac{2\sin(\Omega_0t)-2(\omega_j/\Omega_0)\sin(\omega_jt)+(\Omega_0^2-\omega_j^2)(t/\Omega_0)\cos(\Omega_0t)}
{8(\Omega_0^2-\omega_j^2)^2/\Omega_0^4}
\right]\sigma_y . \label{eq:rhonon}
\end{multline}
\end{widetext}
The perturbative expansion leading up to this result is formally related to a cumulant expansion~\cite{Kubo1963}, with an explicit, generalized averaging procedure.
 
Finally we note that certain terms in the $\sigma_y$ sum of Eq.~(\ref{eq:rhonon})   diverge when $\omega_j\rightarrow\Omega_0$.
However, we emphasize that the continuum limit is implied for both of the sums in Eq.~(\ref{eq:rhonon}), as discussed below. 
In this limit, the singularity is found to be integrable, provided that $S_{\delta\nu}(f)$ is smooth at $2\pi f=\Omega_0$, as discussed in Appendix~\ref{appendix.gates}.
The singularity therefore poses no problems.

\subsection{One-Photon Gate Fidelity}\label{sec:1p}
We now use Eq.~(\ref{eq:rhonon}) to compute quantum gate errors incurred during Rabi oscillations.
In Appendix~\ref{appendix.gates}, we provide expressions that allow gate errors to be computed numerically, for general gate periods, $t=t_g$.
However analytical results are available for special gate periods.
We specifically consider gates defined by the periods $t=2\pi N/\Omega_0$ with $N=1/2,1,3/2,\dots$, where $N=1/2$ corresponds to a $\pi$ rotation, $N=1$ corresponds to a $2\pi$ rotation, and so on.
In the absence of fluctuations ($\delta_j=0$), the ideal solution for such gates is given by
\begin{equation}
\rho_\text{ideal}(N)=\frac{1}{2}+\frac{1}{2}(-1)^{2N}\sigma_x . \label{eq:rhoideal}
\end{equation}
Defining the gate errors as ${\mathcal E}=1-{\mathcal F}$, where ${\mathcal F}=\text{Tr}[\langle\rho\rangle\rho_\text{ideal}]$ is the gate fidelity, and making the substitutions $\omega_j\rightarrow 2\pi f$, $\Delta f\rightarrow df$, and
\begin{equation}
\sum_{j=1}^\infty\delta_j^2\rightarrow\int_0^\infty\left(\frac{4\pi}{\Omega_0}\right)^2S_{\delta\nu}(f)df , \label{eq:integ}
\end{equation}
we obtain
\begin{widetext}
\begin{equation}
\mathcal{E}=4\pi^2
\int_0^\infty S_{\delta\nu}(f)\frac{\Omega_0^2[1-(-1)^{2N}\cos(4\pi^2Nf/\Omega_0)]}{(\Omega_0^2-4\pi^2f^2)^2}df .
\label{eq:error_init0}
\end{equation}
As noted above, this expression remains finite for all $f$, including $2\pi f=\Omega_0$. 
Repeating these calculations for the initial conditions $\rho(0)=\frac{1}{2}(1+\sigma_y)$ and $\rho(0)=\frac{1}{2}(1+\sigma_z)$, and performing an average over the results yields the average gate error~\cite{Bowdrey2002,Nielsen2002}
\begin{equation}
\overline{\mathcal{E}}=\frac{8\pi^2}{3}
\int_0^\infty S_{\delta\nu}(f)
\frac{(\Omega_0^2+4\pi^2f^2)\left[1-(-1)^{2N}\cos(4\pi^2Nf/\Omega_0)\right]}{(\Omega_0^2-4\pi^2f^2)^2}df .
\label{eq:error_init}
\end{equation}
In the remainder of the paper we will calculate $\mathcal{E}$ and $\overline{\mathcal{E}}$ in various scenarios. The error $\overline{\mathcal{E}}$ gives a state averaged error which is of interest for characterizing the typical performance of gate operations. Alternatively  $\mathcal{E}$
is the gate error for the particular case of the qubit starting in $\ket{0}$ in the computational basis, which is of particular relevance for optical excitation of Rydberg states. 

\end{widetext}

Equation~(\ref{eq:error_init}) is our main result, which may now be applied to cases of interest, including white noise and servo bumps.
For the $t=2\pi N/\Omega_0$ gates, defined above, with white noise defined in Eq.~(\ref{eq:Sphiwhite}),
we obtain the simple result  
\begin{equation}
\mathcal{E}=\frac{\pi^3h_0N}{\Omega_0},~~~\overline{\mathcal{E}}=\frac{4\pi^3h_0N}{3\Omega_0}
\hspace{0.3in}\text{(white noise)}. \label{eq:error_white}
\end{equation}
As a benchmark, we can determine the white noise level needed to implement a $\pi$ pulse ($N=1/2$) with errors below $10^{-4}$: starting in a computational basis state and assuming a Rabi rate of $\Omega_0/2\pi=1 ~\rm MHz$, we find this is given by $h_0\le 40~\rm Hz^2/Hz.$

For servo bumps, the frequency noise is defined in Eq.~(\ref{eq:Svbump}).
To simplify the error calculation, we make use of the fact that the peak in $S_{\delta\nu}(f)$ is typically sharper and narrower than other frequency-dependent terms in Eq.~(\ref{eq:error_init}). 
This sharp peak can be observed, for example, in Fig.~\ref{fig:servobump}(c).
We therefore make the following substitution in Eq~(\ref{eq:error_init}):
\begin{equation}
S_{\delta\nu}(f)\rightarrow\frac{s_gf_g^2}{2}\delta(f-f_g) ,
\end{equation}
which yields the following expressions for the gate error  
\begin{multline}
\mathcal{E}\approx 2s_g(\pi f_g\Omega_0)^2
\frac{1-(-1)^{2N}\cos(4\pi^2 Nf_g/\Omega_0)}{(\Omega_0^2-4\pi^2f_g^2)^2} 
\\ 
\hspace{0.5in}\text{(servo bump)}, \label{eq:servo_error0}
\end{multline}
\begin{eqnarray}
\overline{\mathcal{E}} &\approx& \frac{4\pi^2s_gf_g^2(\Omega_0^2+4\pi^2f_g^2)}{3(\Omega_0^2-4\pi^2f_g^2)^2} \label{eq:servo_error}
\\ && \nonumber \times
\left[1-(-1)^{2N}\cos(4\pi^2 Nf_g/\Omega_0)\right]
\hspace{0.1in}\text{(servo bump)}. 
\end{eqnarray}

\red{
Due to its narrow bandwidth, servo-bump noise causes the qubit to evolve coherently at a well-defined frequency, with interference occurring at its other characteristic frequency, that of the Rabi drive. In contrast, the broadband nature of white noise precludes any type of interference. }
As shown in later sections, the largest errors due to servo bumps occur when $2\pi f_g\approx\Omega_0$.
For integer or half-integer values of $N$, evaluating Eqs.~(\ref{eq:servo_error0}) and (\ref{eq:servo_error}) in the limit $2\pi f_g\rightarrow \Omega_0$ (worst-case scenario) gives 
\begin{equation}
\label{eq:errormax}
\mathcal{E}\approx\frac{s_g(\pi N)^2}{4}, ~~~ \overline{\mathcal{E}}\approx\frac{s_g(\pi N)^2}{3} . 
\end{equation}

Comparing Eqs. (\ref{eq:error_white}) and (\ref{eq:errormax}) we see that the worst-case error due to a servo bump is smaller than the error due to the background white noise when 
\begin{equation}
s_g<  \frac{4\pi h_0}{N\Omega_0}.   
\label{eq:white_vs_servo}
\end{equation} 
For the measured laser spectrum, shown in Fig.~\ref{fig:servobump}, the corresponding requirement for
a $\pi$ pulse with $\Omega_0/2\pi=1 ~\rm MHz$ is 
$s_g< 1.0\times 10^{-4}$. 
In this case, the measured  $s_g$ values for the two servo bumps ($s_{g1}=1.3\times 10^{-4}$ and $s_{g2}=2.7\times 10^{-4}$) do not satisfy this criterion. 
However, there is a known interplay between the stabilized white noise and the servo bumps~\cite{Day2022}.
From Eq.~(\ref{eq:servo_error}), we see that servo bumps are most dangerous when peaked near the Rabi frequency.
When the servo bump peak is well separated from the Rabi frequency, the gate error is dominated by the white-noise background.

\begin{figure}[t]
\includegraphics[width=1.6in]{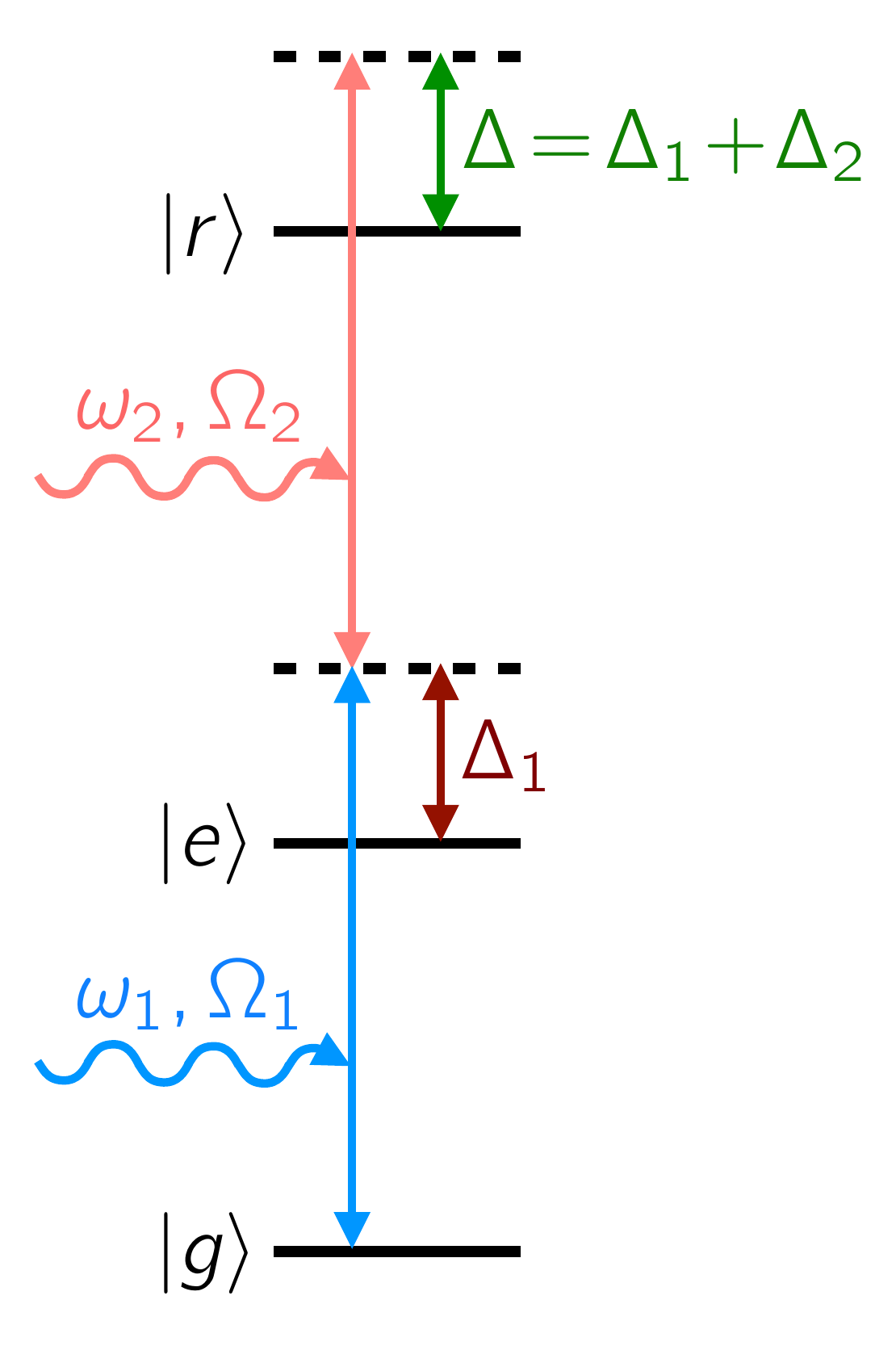}
\caption{
\label{fig:3lvlscheme}
Coupling scheme for two-photon Rabi oscillations in a ladder configuration, for states $|g\rangle$, $|e\rangle$, and $|r\rangle$.
Two lasers are employed, with photon angular frequencies $\omega_{1,2}$ and associated Rabi angular frequencies $\Omega_{1,2}$.
The lasers are detuned away from ladder excitations by angular frequencies $\Delta_1$ and $\Delta_2$.}
\end{figure}

\subsection{Two-Photon Gate Fidelity}\label{sec:2p}
We can extend the time-series master equation approach to describe two-photon Rabi oscillations in the ladder geometry  shown in Fig.~\ref{fig:3lvlscheme}. This approach is widely used for Rydberg excitation in quantum gate experiments with atomic qubits~\cite{Johnson2008}.
We consider a three-level system $\{\ket{g},\ket{e},\ket{r}\}$ with the corresponding energy levels $E_g$, $E_e$, and $E_r$.
We also consider two monochromatic lasers with angular frequencies $\omega_1$ and $\omega_2$ and Rabi angular frequencies $\Omega_1$ and $\Omega_2$.
\red{ 
As is well known there is an additional error source associated with two-photon excitation due to photon scattering from the intermediate level\cite{Saffman2010,Graham2019}. This can lead to depolarization errors of the qubit as well as leakage errors when the atomic ground state includes additional sub-levels outside of the computational basis. Our analysis is focused on the effects of laser noise, and does not account for additional scattering related errors. 
}

As before, we allow for phase fluctuations in both lasers, characterized by their individual noise spectral densities, $S_{1,\delta\nu}(f)$ and $S_{2,\delta\nu}(f)$:
\begin{equation}
\phi_i(t)=-\sum_{j=1}^N\delta\nu_{i,j}\cos(2\pi f_jt+\varphi_{i,j}) ,
\label{eq.phi_noise}
\end{equation}
where
\begin{equation}
\delta\nu_{i,j}=-2\sqrt{S_{i,\delta\nu}(f_j)\Delta f} ,
\end{equation}
and $i=1,2$.
Here, the random phases $\{\varphi_{i,j}\}$ are assumed to be independent for all $i$ and $j$.

In analogy with Eq.~(\ref{eq:Hinit}), the full system Hamiltonian in the laboratory frame is now given by
\begin{multline}
H=E_g\ket{g}\bra{g}+E_e\ket{e}\bra{e}+E_r\ket{r}\bra{r} \\
+\hbar\Omega_1\cos(\omega_1t)\left(e^{-i\phi_1}\ket{e}\bra{g}+e^{i\phi_1}\ket{g}\bra{e}\right) \\
+\hbar\Omega_2\cos(\omega_2t)\left(e^{-i\phi_2}\ket{r}\bra{e}+e^{i\phi_2}\ket{e}\bra{r}\right) .
\end{multline}
Moving to the rotating frame defined by 
\begin{multline}
U(t)=\exp\left[i\left(\frac{2\omega_1t}{3}+\frac{\omega_2t}{3}\right)(\ket{e}\bra{e}-\ket{g}\bra{g}) 
\right. \\ \left.
+i\left(\frac{\omega_1t}{3}+\frac{2\omega_2t}{3}\right)(\ket{r}\bra{r}-\ket{e}\bra{e}) \right] ,
\end{multline}
and applying a RWA, we obtain
\begin{multline}
H'\approx \frac{\hbar\Delta}{2}\ket{g}\bra{g}-\frac{\hbar\delta}{2}\ket{e}\bra{e} -\frac{\hbar\Delta}{2}\ket{r}\bra{r} \\
+\frac{\hbar\Omega_1}{2}\left(e^{-i\phi_1}\ket{e}\bra{g}+e^{i\phi_1}\ket{g}\bra{e}\right) \\
+\frac{\hbar\Omega_2}{2}\left(e^{-i\phi_2}\ket{r}\bra{e}+e^{i\phi_2}\ket{e}\bra{r}\right), \label{eq:Hp2p}
\end{multline}
where we have removed a constant energy term and defined $\hbar\Delta_1=\hbar\omega_1+E_g-E_e$, $\hbar\Delta_2=\hbar\omega_2+E_e-E_r$, and $\Delta=\Delta_1+\Delta_2$, as illustrated in Fig.~\ref{fig:3lvlscheme}, with $\delta=\Delta_1-\Delta_2$.

As in the previous section, we next move to a fluctuation frame, defined by the transformation
\begin{multline}
U_\phi(t)=e^{-i\phi_1/2-i\phi_2/2}\ket{g}\bra{g} \\
+e^{i\phi_1/2-i\phi_2/2}\ket{e}\bra{e}+e^{i\phi_1/2+i\phi_2/2}\ket{r}\bra{r},
\end{multline}
yielding the Hamiltonian
\begin{multline}
H''\approx 
\frac{\hbar}{2}\left(\Delta+\dot\phi_1+\dot\phi_2\right)\ket{g}\bra{g} \\
+\frac{\hbar}{2}\left(-\delta-\dot\phi_1+\dot\phi_2\right)\ket{e}\bra{e} 
+\frac{\hbar}{2}\left(-\Delta-\dot\phi_1-\dot\phi_2\right)\ket{r}\bra{r} \\
+\frac{\hbar\Omega_1}{2}\left(\ket{e}\bra{g}+\ket{g}\bra{e}\right) 
+\frac{\hbar\Omega_2}{2}\left(\ket{r}\bra{e}+\ket{e}\bra{r}\right).
\end{multline}

Now if we assume that $|\delta|\gg\Omega_1,\Omega_2,|\dot\phi_1|,|\dot\phi_2|$, and that the system wavefunction $\ket{\psi}$ is not initialized into state $\ket{e}$, then at later times we still have $|\braket{\psi}{e}|^2\ll 1$.
Hence, it is a good approximation to slave $\ket{e}$ to states $\ket{g}$ and $\ket{r}$, such that
\begin{equation}
\ket{e}\approx\frac{\Omega_1}{\delta}\ket{g}+\frac{\Omega_2}{\delta}\ket{r}.
\end{equation}
Eliminating $\ket{e}$ from $H''$, we arrive at an effective 2D Hamiltonian that describes the dynamical evolution of $\ket{g}$ and $\ket{r}$:
\begin{equation}
H_\text{2D}\approx \frac{\hbar\tilde\Omega_0}{2}\sigma_x
-\frac{\hbar}{2}\left(\Delta_++\dot\phi_1+\dot\phi_2\right)\sigma_z , 
\end{equation}
where we define $\Delta_+=\Delta+(\Omega_1^2-\Omega_2^2)/2\delta$, $\tilde\Omega_0=\Omega_1\Omega_2/\delta$, $\sigma_z=\ket{r}\bra{r}-\ket{g}\bra{g}$, and $\sigma_x=\ket{r}\bra{g}+\ket{g}\bra{r}$, and we have again removed a constant energy term. The shift $\Delta_+$ is due to the dynamic Stark shift of the bare atomic levels that arises from the intermediate state detuning.  

In the absence of noise ($\dot\phi_1,\dot\phi_2=0$), $H_\text{2D}$ describes rotations tilted slightly away from the $x$ axis, which is undesirable from a gating perspective.
This situation can be avoided by adopting the special detuning value defined by the relation $\Delta_+=0$, or equivalently,
\begin{equation}
\Delta=\Delta_1\left(1-\sqrt{1+\frac{\Omega_1^2-\Omega_2^2}{2\Delta_1^2}}\right) . \label{eq:Delta}
\end{equation}
For this case, we obtain 
\begin{equation}
H_\text{2D}\approx \frac{\hbar\tilde\Omega_0}{2}\sigma_x
-\frac{\hbar}{2}\left(\dot\phi_1+\dot\phi_2\right)\sigma_z , \label{eq:H2D}
\end{equation}
which maps immediately onto Eq.~(\ref{eq:Hpp}) of our previous one-photon analysis.

In the one-photon calculation, we were able to make progress by noting that the random phases $\varphi_j$, corresponding to frequency variables $\omega_j$, were independent, yielding additive contributions to the total error in the quantum gates.
Now in the two-photon case, the random variables $\varphi_{j1}$ and $\varphi_{j2}$, corresponding to lasers 1 and 2, are also independent; therefore their contributions to the total error should also be additive. 
Accounting for the separate power spectral densities in the two lasers, we obtain the following two-photon results, for gates defined by $t=2\pi N/\tilde\Omega_0$, with $N=1/2,1,3/2,\dots$.
For white noise defined by the parameters $h_1$ and $h_2$ in the two lasers, and for the initial state $\rho(0)=\frac{1}{2}(1+\sigma_x)$, we obtain
\begin{equation}
\mathcal{E}=\frac{\pi^3(h_1+h_2)N}{\tilde\Omega_0}
\hspace{0.5in}\text{(white noise)}, \label{eq:error_white2p0}
\end{equation}
which is relevant for Rydberg excitations.
Averaging over initial states, we obtain the average gate fidelity
\begin{equation}
\overline{\mathcal{E}}=\frac{4\pi^3(h_1+h_2)N}{3\tilde\Omega_0}
\hspace{0.5in}\text{(white noise)}. \label{eq:error_white2p}
\end{equation}

For the servo-bump model of laser phase noise, the differences in the lasers are characterized by the total power of the phase noise in the two servo bumps ($s_{g1}$ and $s_{g2}$) and their corresponding peak frequencies ($f_{g1}$ and $f_{g2}$).
The resulting error for two-photon gates, for the initial state $\rho(0)=\frac{1}{2}(1+\sigma_x)$, is given by
\begin{multline}
\mathcal{E}\approx 2s_{g1}(\pi f_{g1}\tilde\Omega_0)^2
\frac{1-(-1)^{2N}\cos(4\pi^2 Nf_{g1}/\tilde\Omega_0)}{(\tilde\Omega_0^2-4\pi^2f_{g1}^2)^2} \\
+2s_{g2}(\pi f_{g2}\tilde\Omega_0)^2
\frac{1-(-1)^{2N}\cos(4\pi^2 Nf_{g2}/\tilde\Omega_0)}{(\tilde\Omega_0^2-4\pi^2f_{g2}^2)^2}
\\ 
\hspace{0.5in}\text{(servo bump)}. \label{eq:servo_error2p0}
\end{multline}
Averaging over initial states gives
\begin{eqnarray}
\overline{\mathcal{E}} &\approx& \frac{4\pi^2s_{g1}f_{g1}^2(\tilde\Omega_0^2+4\pi^2f_{g1}^2)}{3(\tilde\Omega_0^2-4\pi^2f_{g1}^2)^2} \label{eq:servo_error2p}
\\ && \nonumber \times
\left[1-(-1)^{2N}\cos(4\pi^2 Nf_{g1}/\tilde\Omega_0)\right]
\hspace{0.1in}
\\ &+ & \nonumber 
\frac{4\pi^2s_{g2}f_{g2}^2(\tilde\Omega_0^2+4\pi^2f_{g2}^2)}{3(\tilde\Omega_0^2-4\pi^2f_{g2}^2)^2}
\\ && \nonumber \times
\left[1-(-1)^{2N}\cos(4\pi^2 Nf_{g2}/\tilde\Omega_0)\right]
\hspace{0.1in}\text{(servo bump)}. 
\end{eqnarray}

\begin{figure}[t]
\includegraphics[width=1.9in]{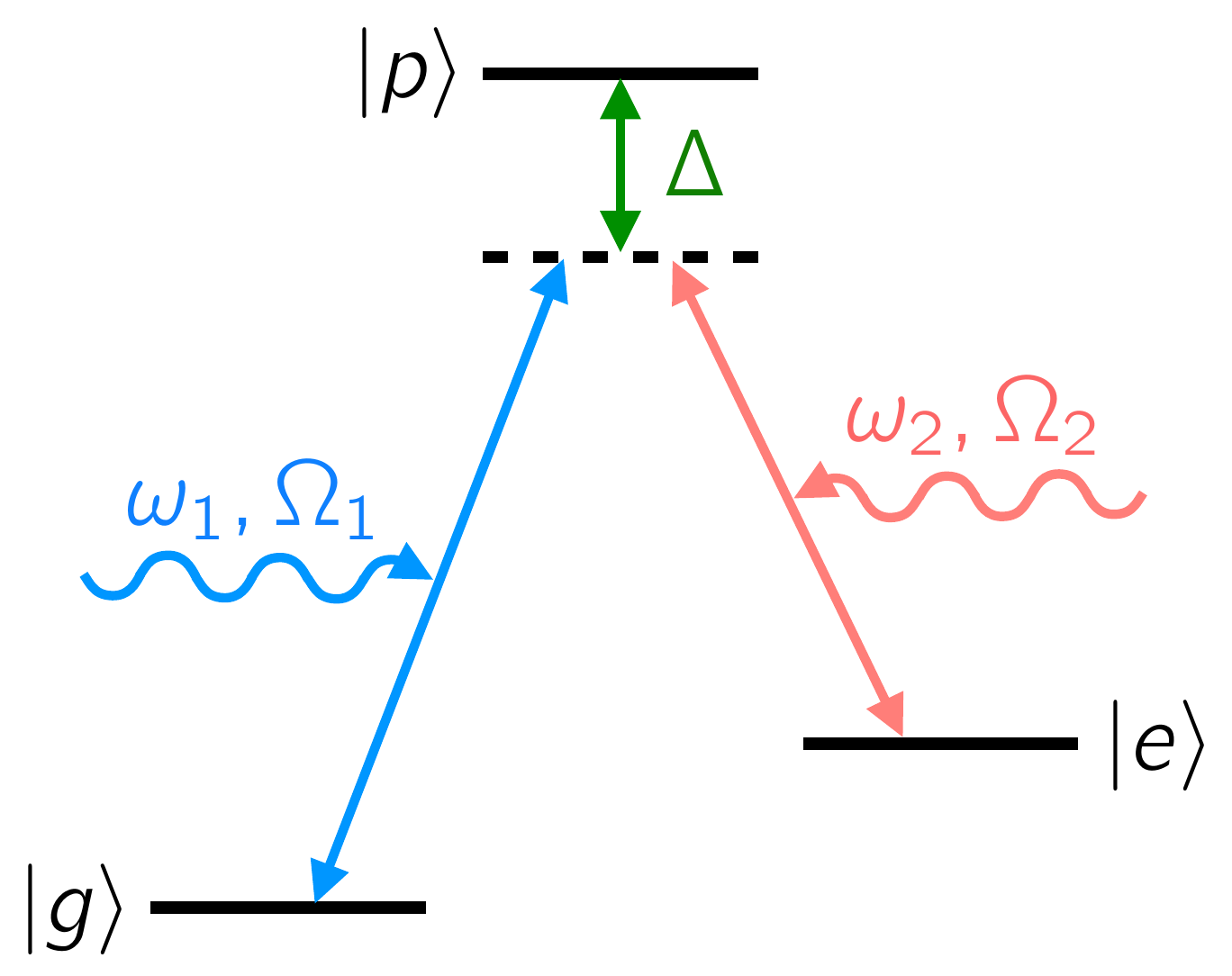}
\caption{
\label{fig:Lambda}
\red{
Coupling scheme for two-photon Rabi oscillations in a $\Lambda$ configuration, for states $|g\rangle$, $|e\rangle$, and $|p\rangle$.
In the setup considered here, the two driving frequencies are assumed to differ by the qubit frequency: $\hbar(\omega_1-\omega_2)=E_e-E_g$.
The figure also shows the associated Rabi angular frequencies $\Omega_1$ and $\Omega_2$, and the detuning $\Delta$.}}
\end{figure}

\subsection{\red{Two-Photon Raman Transitions}\label{sec:Lambda}}

\red{While the main focus of this work is on two-photon transitions in the ladder configuration (Fig.~\ref{fig:3lvlscheme}), we also briefly consider two-photon Raman transitions in the $\Lambda$ configuration (Fig.~\ref{fig:Lambda}).
An important difference in the latter case is that only one laser is used in a typical setup.
The single laser field  is modulated  such that it acquires sidebands of frequency $\omega_1$ and $\omega_2$, separated by the qubit frequency, $\hbar(\omega_1-\omega_2)=E_e-E_g$.
Both fields are then made to co-propagate in the same spatial mode when exciting the atom. In practice this is most often done by modulating the current of a laser diode\cite{Knoernschild2010}, or with  an electro-optic modulator to add sideband frequencies\cite{Akerman2015}. Either of these techniques results in correlated phase noise at both sideband frequencies.    
As is well known, such an arrangement is resilient to dephasing induced by the Doppler effect~\cite{Kasevich1992}.}

\red{
Here we show explicitly that the $\Lambda$ configuration is also resilient to laser phase noise.
While we do not show all the steps of the derivation, we follow the same procedure as the preceding sections.
In the rotating frame equivalent to Eq.~(\ref{eq:Hp2p}), the two-photon Hamiltonian becomes
\begin{multline}
H'\approx \hbar\Delta \ket{p}\bra{p} 
+\left(\frac{\hbar\Omega_1}{2}e^{i\phi}\ket{p}\bra{g} +\text{H.c.}\right) \\
+\left(\frac{\hbar\Omega_2^*}{2}e^{-i\phi}\ket{e}\bra{p} +\text{H.c.}\right), \label{eq:HpLambda}
\end{multline}
where we have subtracted a constant energy.
The phase $\phi$ depends on the noise spectrum of the laser as defined in  Eq. (\ref{eq.phi_noise}) and the parallel treatment of $|g\rangle$ and $|e\rangle$ is evident.
We note that the Rabi frequencies associated with the two drives, $\Omega_1$ and $\Omega_2$, include complex phases  $\xi_1, \xi_2$ which  determine the azimuthal angle on the Bloch sphere of the  Rabi rotation axis according to $\phi_{\rm rot}=\xi_1-\xi_2$.
Now, moving to the fluctuation frame, assuming $|\Delta|\gg|\Omega_1|,|\Omega_2|,|\dot\phi|$, and eliminating the detuned level $|p\rangle$, we obtain the effective 2D Hamiltonian describing the states $|g\rangle$ and $|e\rangle$ in the  $\Lambda$ configuration:
\begin{multline}
H_\text{2D}\approx -\frac{\hbar(|\Omega_1|^2-|\Omega_2|^2)}{8\Delta}(|g\rangle\langle g|-|e\rangle\langle e|) \\
-\left(\frac{\hbar\tilde\Omega_R}{2}|e\rangle\langle g|+\text{H.c.}\right)
-\hbar\dot\phi(|g\rangle\langle g|+|e\rangle\langle e|),
\end{multline}
where we have again subtracted a constant energy and defined the two-photon Rabi frequency $\tilde\Omega_R=\Omega_1\Omega_2^*/2\Delta$.
Notice that the phase fluctuations $\dot\phi$ are now proportional to the identity operator in the qubit subsystem, and therefore only contribute to the global phase.
The phase fluctuations are therefore harmless, and do not affect Rabi gates at linear  order, although they do affect the dynamics at higher order.
However, that is beyond the scope of the current work.}

\begin{figure}[t]
    \centering
    \includegraphics[width = 0.4\textwidth]{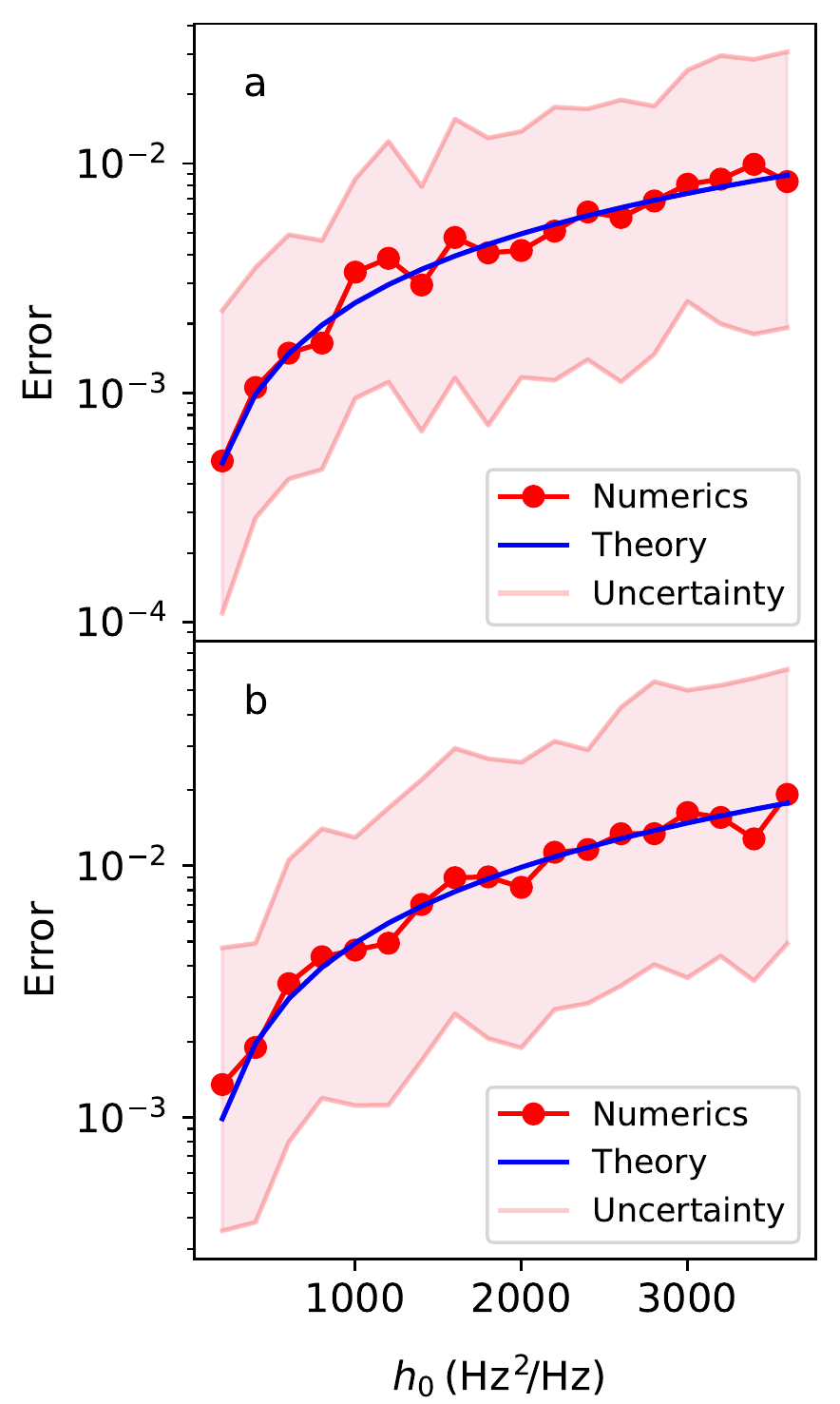}
    \caption{
    Rabi errors for one-photon Rabi oscillations due to white phase noise, plotted as a function of noise amplitude $h_0$. 
    Results are obtained following the procedure described in Sec.~\ref{sec.de} for (a) $\pi$ rotations and (b) $2\pi$ rotations.
    Red markers represent averages from numerical simulations, while pink shading shows the corresponding $1\sigma$ error bars. Blue curves show theoretical results for $\mathcal E$  from Eq.~(\ref{eq:error_white}).
    }
    \label{fig:white_1p}
\end{figure}

\section{Dynamical simulation of Rabi oscillations with phase noise}
\label{sec.SE}
In this section, we perform simulations of Rabi oscillations, including laser phase fluctuations. 
We specifically consider the effects of white noise, defined in Eq.~(\ref{eq:Sphiwhite}), for a range of noise amplitudes, $h_0\in (0,4000)$~Hz$^2$/Hz.
We also consider servo bumps, defined in Eq.~(\ref{eq:Svbump}), for fixed parameter values of $h_g=1100$~Hz$^2$/Hz and $\sigma_g=1.4$~kHz, which are similar to the experimental values obtained in Sec.~\ref{sec:laserparameters}.
Since gate errors caused by servo bumps are maximized when the bump peak occurs near the Rabi frequency, $f_g\approx(\Omega_0/2\pi)$, we focus on the parameter range $f_g\in(0,2)\times(\Omega_0/2\pi)$. 
In the latter simulations, we set the white-noise amplitude to $h_0=0$, to focus exclusively on the servo bump.
For all simulations, we adopt the typical Rabi frequency $\Omega_0/2\pi=1$~MHz.
For the two-photon gates, for simplicity, we assume that both lasers have the same noise spectra.

\subsection{One-photon gates}\label{sec.de}
We first consider gates implemented with one-photon Rabi drives  with laser phase noise.
The gates are defined in Sec.~\ref{sec:1p}, with $N=1/2$ ($\pi$ rotations) and $N=1$ ($2\pi$ rotations).
As previously, we assume that the qubit is driven resonantly.
Defining $\ket{\psi(t)}=c_g(t)\ket{g}+c_e(t)\ket{e}$, the Schr\"odinger equation associated with Eq.~(\ref{eq:Hp}) can be written as
\begin{eqnarray}
\dot{c}_g &=&-i\frac{\Omega_0}{2}e^{-i\phi(t)}c_e , \label{eq:cg1p}\\
\dot{c}_e &=&-i\frac{\Omega_0}{2}e^{i\phi(t)}c_g . \label{eq:ce1p}
\end{eqnarray}

\begin{figure}[t]
    \centering
    \includegraphics[width = 0.4\textwidth]{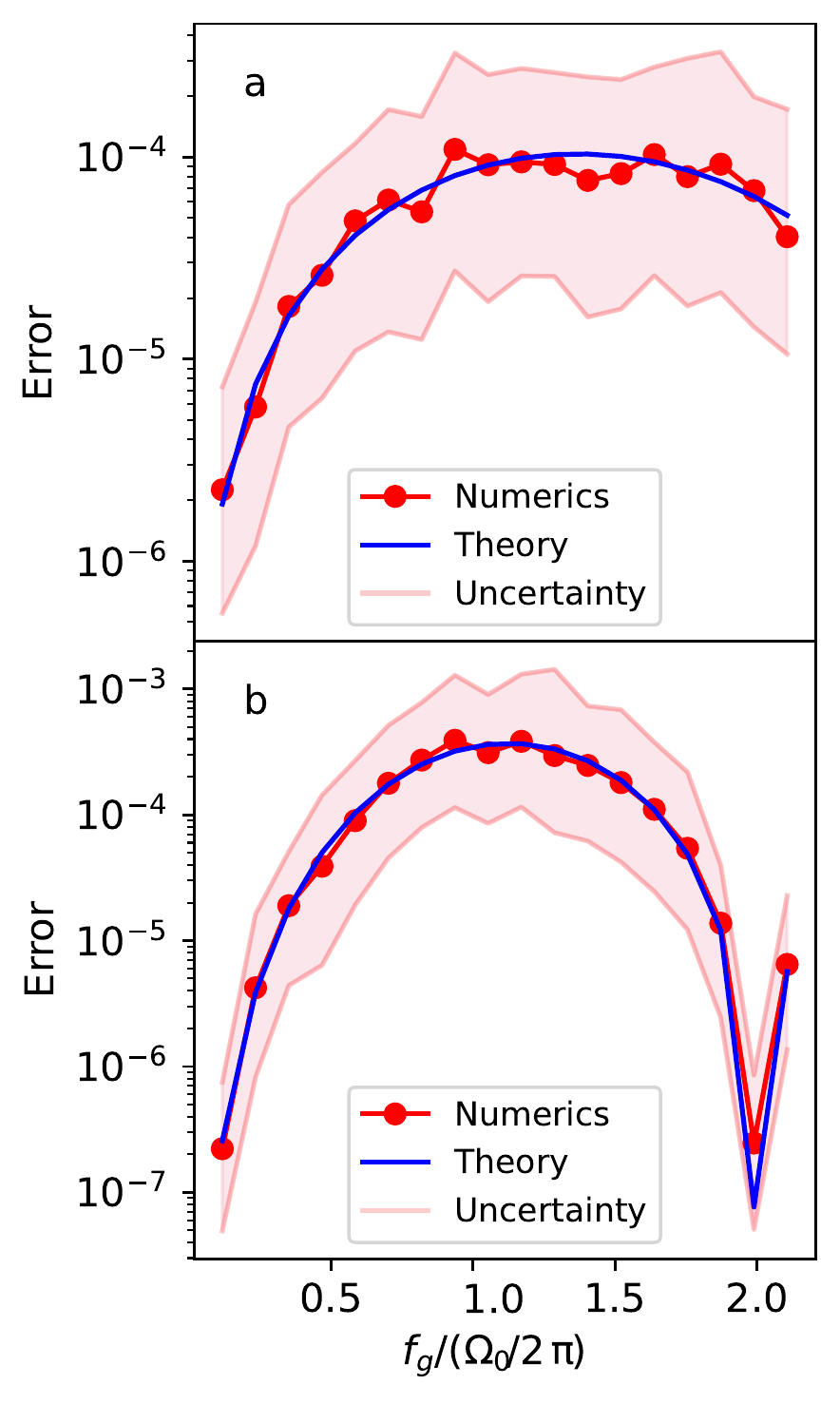}
    \caption{
    Rabi errors for one-photon Rabi oscillations due to servo-bump phase noise.  
    Results are plotted as a function of the center frequency of the servo bump $f_g$, scaled by the Rabi frequency $\Omega_0/2\pi=1$~MHz, with additional noise parameters described in Sec.~\ref{sec.de}.
    Similar to Fig.~\ref{fig:white_1p}, (a) shows results for $\pi$ rotations and (b) shows results for $2\pi$ rotations.
    Theory results are obtained from Eq.~(\ref{eq:servo_error0}).
    }
    \label{fig:bumpfreq1}
\end{figure}

To solve these equations, we first obtain a random time trace for $\phi(t)$ from Eq.~(\ref{eq:randomphi}).
The infinite series expansion must be truncated and we therefore write
\begin{equation}
\phi(t_k)=\sum_{j=1}^{M/2} 2\sqrt{S_\phi(f_j)\Delta f}\cos(2\pi f_jt_k+\varphi_j) , \label{eq:randomphifin}
\end{equation}
where $\Delta f=f_{j+1}-f_j=1/T$ for times sampled in the range $0\leq t_k\leq T$, with $t_{k+1}-t_k=T/M$, according to the Nyquist sampling theorem.
For the simulations described below, white noise poses the most serious challenge to numerical convergence. 
In this case, we find that a frequency bandwidth of $f_{M/2}=10$~MHz is sufficient for our purposes, and that convergence is achieved when $M\approx 10^3$.
Equations~(\ref{eq:cg1p}) and (\ref{eq:ce1p}) are then solved numerically, using the computed time series.
Statistical properties are obtained by performing averages over results based on many random time series.
As in previous sections, the resulting gate errors are defined as ${\mathcal E}=1-F$, where $F=\text{tr}[\langle\rho\rangle\rho_\text{ideal}]$.

In Fig.~\ref{fig:white_1p}, we plot the numerical gate errors for $\pi$ and $2\pi$ rotations, obtained by solving the the Schr\"odinger equations in the presence of white noise. In this figure, as in all figures that follow, the numerical averages are shown as red markers, while $1\sigma$ error bars are shown with pink shading. All reported errors $\mathcal E$, both theoretical and numerical, correspond to the qubit initial state $\ket{0}$.
Theoretical results are shown as blue curves.
For the case of white noise, these correspond to Eq.~(\ref{eq:error_white}).
For typical white-noise amplitudes ($h_0< 100$~Hz$^2$/Hz), the observed error levels are low.
Theoretical results are found to reproduce the numerical ones quite accurately.

\begin{figure}[t]
    \centering
    \includegraphics[width = 0.4\textwidth]{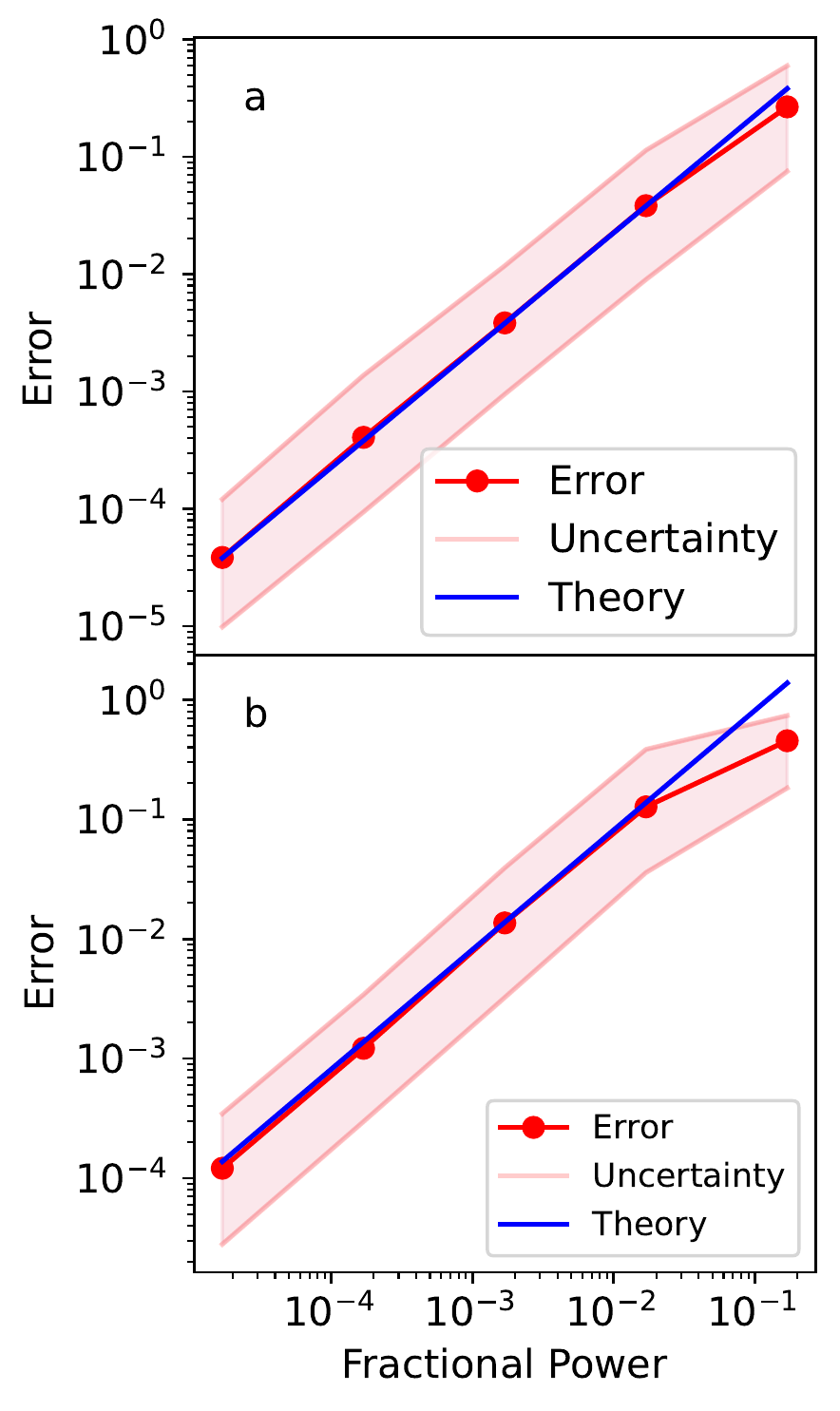}
    \caption{
    Rabi errors for one-photon Rabi oscillations due to servo-bump phase noise: (a) $\pi$ rotations, (b) $2\pi$ rotations.
    Calculations are very similar to Fig.~\ref{fig:bumpfreq1}, except that here, the central peak frequency of the servo bump is held fixed at $1.2~\textrm{MHz}$, while the total power $s_g$ is varied, where $s_g$ is defined in Eq.~(\ref{eq:sg}). 
    }
    \label{fig:bumppower}
\end{figure}

In Fig.~\ref{fig:bumpfreq1}, we plot $\pi$ and $2\pi$  gate errors for Rabi oscillations in the presence of servo-bump noise. 
The results are plotted as a function of the center frequency of the servo bump, scaled by the Rabi frequency $\Omega_0/2\pi=1$~MHz. 
The calculations are performed while holding the total power $s_g$ and peak height $h_g$ fixed at the values obtained for the larger servo bump observed in Fig.~\ref{fig:servobump}, while simultaneously varying the peak frequency $f_g$ and width $\sigma_g$ according to Eq.~(\ref{eq:sg}).
Corresponding theory results are also shown, based on Eq.~(\ref{eq:servo_error0}).
Again, the theoretical results are found to describe well the non-monotonic behavior of the numerical results.
As expected, gate errors are maximized for servo bumps centered near the Rabi frequency.
Also as expected from the theoretical calculations of Sec.~\ref{sec:Master}, the gate errors are strongly suppressed for the condition $f_g=2(\Omega_0/2\pi)$.
This is an interesting interference effect induced by the shape of Gaussian noise peaks, which isn't observed, for example, in the case of white noise.
Such error suppression could potentially be leveraged for reducing Rabi gate errors.

\begin{figure}[t]
    \centering
    \includegraphics[width = 0.4\textwidth]{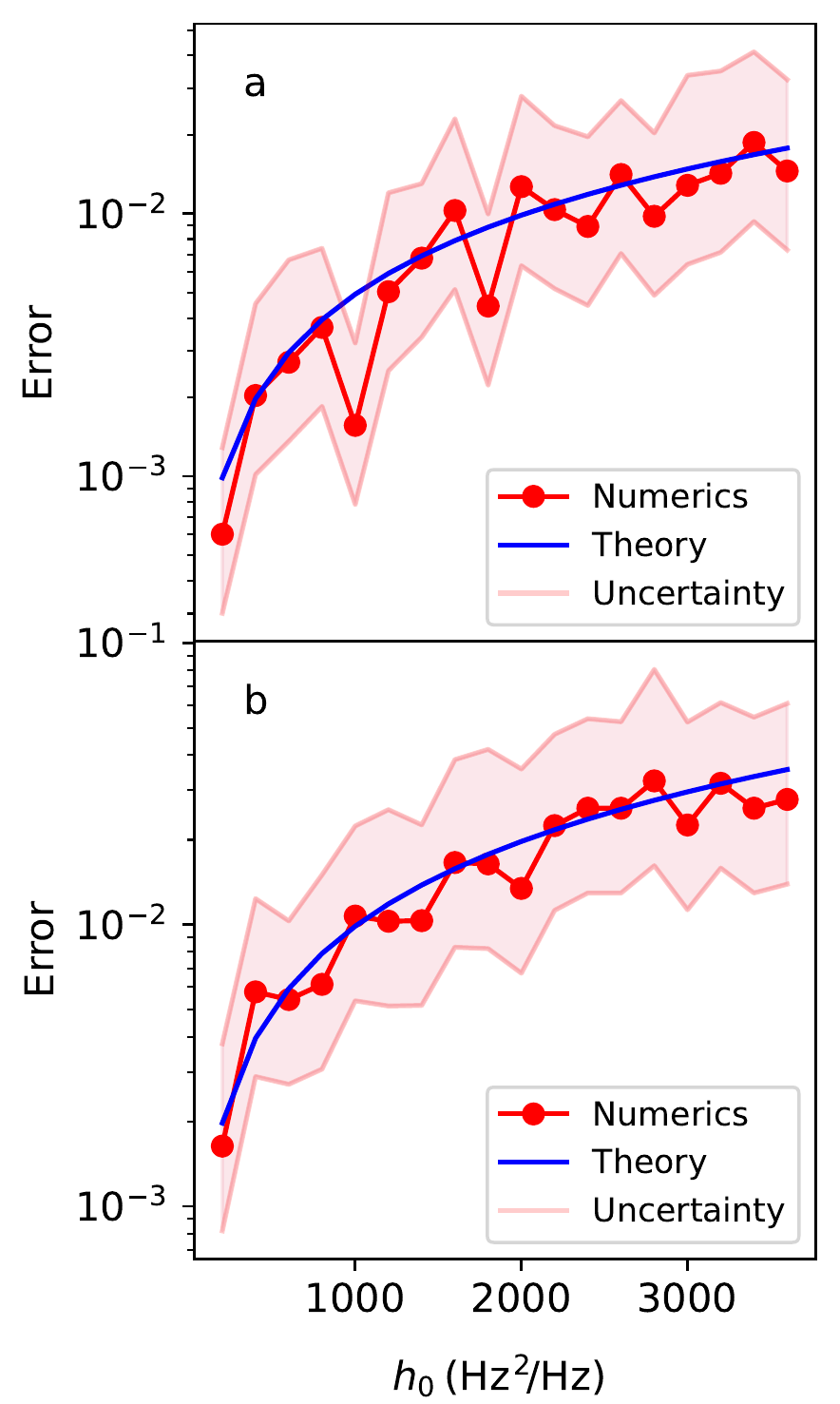}
    \caption{
    Rabi errors for two-photon Rabi oscillations due to white phase noise: (a) $\pi$ rotations, (b) $2\pi$ rotations. 
    All simulations and plots are analogous to Fig.~\ref{fig:white_1p}, while theory results are given by Eq.~(\ref{eq:error_white2p0}).}
    \label{fig:white_2p}
\end{figure}

In Fig.~\ref{fig:bumppower}, we also show results as a function of the servo-bump noise level. 
In this case, the center frequency of the servo bump is held fixed at the experimentally observed value $f_g=234$~kHz reported in Sec.~\ref{sec:laserparameters}, while the total noise power in the servo bump is varied, using the definition of $s_g$ in Eq.~(\ref{eq:sg}). Plotted on a log-log scale, we observe an initial linear dependence of the gate error on noise power.

\subsection{Two-photon gates}\label{sec.2psims}
Two-photon gates are described in Sec.~\ref{sec:2p}.
For the gates considered here, both lasers are detuned, in contrast with the one-photon gates described above.
Defining $\ket{\psi(t)}=c_g(t)e^{-i\Delta t/2}\ket{g}+c_e(t)e^{i\delta t/2}\ket{e}+c_r(t)e^{i\Delta t/2}\ket{r}$, the Schr\"odinger equation associated with Eq.~(\ref{eq:Hp2p}) can be written as
\begin{eqnarray}
\dot{c}_g &=&-i\frac{\Omega_1}{2}e^{i\phi_1(t)}e^{i\Delta_1t}c_e , \label{eq:cg2p}\\
\dot{c}_e &=&
-i\frac{\Omega_1}{2}e^{-i\phi_1(t)}e^{-i\Delta_1t}c_g 
-i\frac{\Omega_2}{2}e^{i\phi_2(t)}e^{i\Delta_2t}c_r ,\,\, \label{eq:ce2p} \\
\dot{c}_r &=&-i\frac{\Omega_2}{2}e^{-i\phi_2(t)}e^{-i\Delta_2t}c_e . \label{eq:cr2p}
\end{eqnarray}

\begin{figure}
    \centering
    \includegraphics[width = 0.4\textwidth]{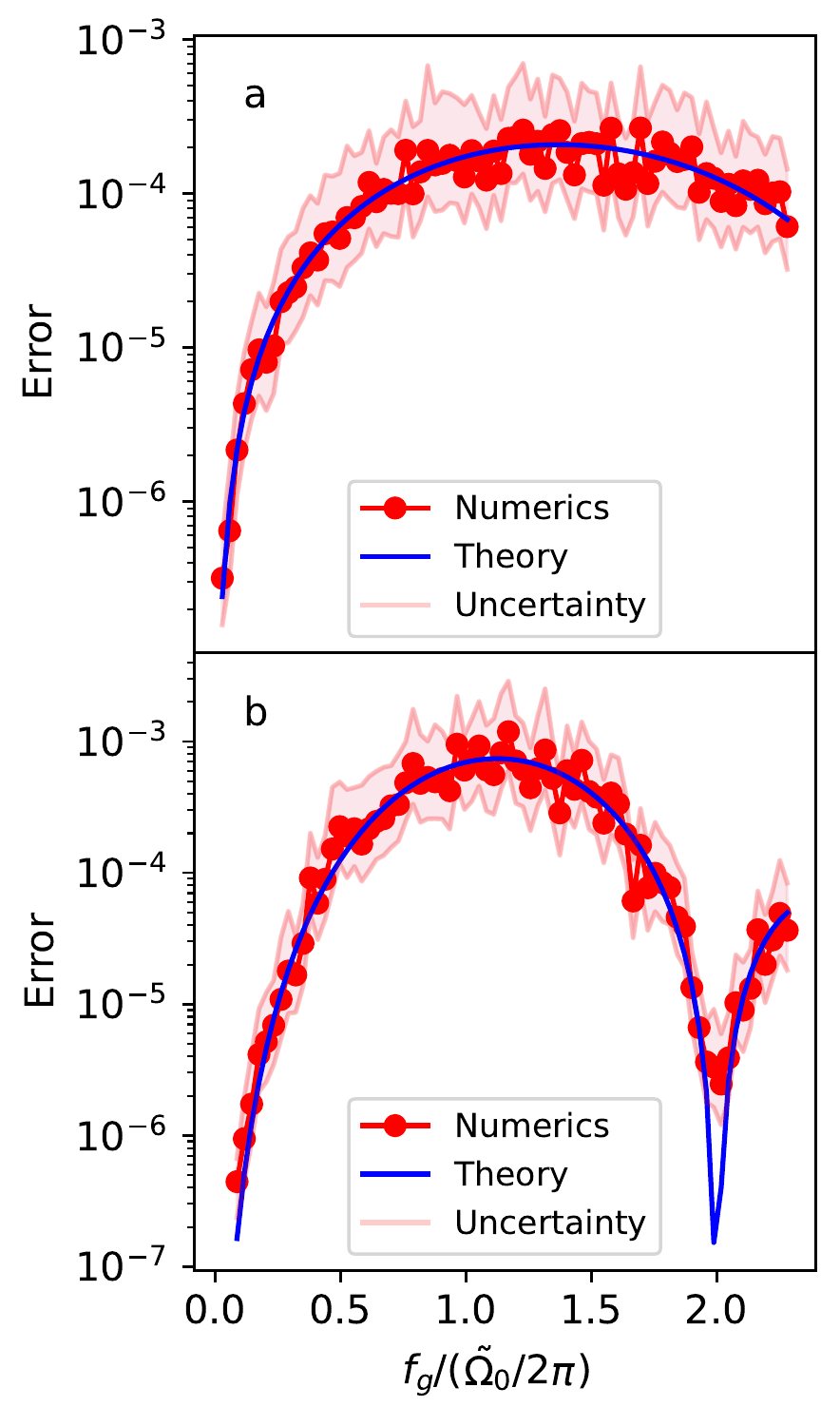}
    \caption{
    Rabi errors for two-photon Rabi oscillations due to servo-bump phase noise: (a) $\pi$ rotations, (b) $2\pi$ rotations. 
    All simulations and plots are analogous to Fig.~\ref{fig:bumpfreq1}, while theory results are given by Eq.~(\ref{eq:servo_error2p0}).
    }
    \label{fig:bumpfreq2}
\end{figure}

As in experiments, the detuning parameters in two-photon simulations should be chosen carefully.
Referring to the notation of Sec.~\ref{sec:2p}, we adopt the following criteria: (1) the effective, two-photon Rabi frequency is chosen to be $\tilde\Omega_0=\Omega_1\Omega_2/\delta=2\pi \times 1$~MHz (as in the one-photon simulations), (2) the ratio $\delta/\tilde\Omega_0$ should be large enough to avoid populating the intermediate level $\ket{e}$, but small enough for simulations to complete in a reasonable time (here we choose $\delta/\tilde\Omega_0=100$).
For convenience, we also choose $\Omega_1=\Omega_2$, and we note that the resonance condition, Eq.~(\ref{eq:Delta}), must be satisfied.
These combined criteria yield $\Delta_1=-\Delta_2=2\pi\times 5$~GHz, $\Omega_1=\Omega_2=2\pi\times 100$~MHz, and $\delta=2\pi\times 10$~GHz. 
These choices yield a resonant excitation of the state $\ket{r}$, defined as $\Delta=0$ (see Fig.~\ref{fig:3lvlscheme}), as consistent with the resonant excitation scheme for one-photon gates.
For convenience, we assume identical noise parameters for the two lasers (e.g., $h_1=h_2$, etc.), although the time series for $\phi_1(t)$ and $\phi_2(t)$ are still generated independently via Eq.~(\ref{eq:randomphifin}).
Finally, Eqs.~(\ref{eq:cg2p})-(\ref{eq:cr2p}) are solved numerically for many random time series, and averaged.

Figure~\ref{fig:white_2p} shows two-photon results for the case of white phase noise.
The corresponding theoretical results are given in Eq.~(\ref{eq:error_white2p0}).
These results may be compared directly to one-photon gates.
Indeed, the two-photon gate errors appear similar in shape, but doubled in magnitude, as compared to Fig.~\ref{fig:white_1p}.
This is consistent with the expectation that errors should be additive in the limit of weak noise, as discussed in Sec.~\ref{sec:2p}.

Figure~\ref{fig:bumpfreq2} shows two-photon results for the case of servo-bump phase noise.
Here, the theoretical results are given in Eq.~(\ref{eq:servo_error2p0}).
Again, we observe an approximate doubling of the gates error as compared to the single-photon case, shown in Fig~\ref{fig:bumpfreq1}. 
In all cases, the theoretical results of Sec.~\ref{sec.dm} appear quite accurate.

\section{Bandwidth-Limited Phase Noise and the Quasistatic Limit}
\label{sec:quasistatic}

Narrow bandwidth noise may provide a good approximation for certain highly filtered lasers.
In this section, we extend the previous theoretical approach to the case of quasistatic phase noise, where the spectral content of the noise is restricted to very low frequencies.
To begin, we consider the more general situation of bandwidth-limited white noise, defined as
\begin{equation}
S_{\delta\nu}(f) = \left\{ 
\begin{array}{ccc} \vspace{0.03in}
h_0 &\quad\text{when}\quad&  |f|\leq f_c, \\ \vspace{0.03in}
0 &\quad\text{when}\quad&  |f|>f_c.
\end{array}
\right. \label{eq:fb}
\end{equation} 
A noise spectrum of this type could describe a strongly filtered laser with no noise except a broadened carrier signal.

Di Domenico et al.~\cite{DiDomenico2010} have studied how band-limited white noise is manifested in laser field noise, $S_E(f)$. 
They observe two distinct behaviors, with an abrupt transition between them occurring at $f_c=\pi^2h_0/8\ln(2)\approx 1.78h_0$.
When $f_c\gtrsim 1.78h_0$, $S_E(f)$ takes the form appropriate for white noise, which we previously characterized in Sec.~\ref{sec.wn}.
When $f_c\lesssim 1.78h_0$, Eqs.~(\ref{eq:ThomannSE}) and (\ref{eq:Si}) are readily solved, giving
\begin{gather}
S_E(f)\approx \frac{|E_0|^2}{\sqrt{16\pi h_0f_c}}e^{-f^2/4h_0f_c} , \label{eq:S_Eqs} \\
S_i(f)\approx \sqrt{\frac{3}{16\pi^3h_0t_d^2f_c^3}}e^{-3f^2/16\pi^2h_0t_d^2f_c^3} . \label{eq:S_iqs}
\end{gather}
In the context of Rabi gate operations, when we also have $f_c\ll \Omega_0/2\pi$, we refer to this compressed-noise regime as quasistatic.

The singular nature of quasistatic noise causes the interrelations between $S_\phi(f)$, $S_E(f)$, and $S_i(f)$,  embodied in Eqs.~(\ref{eq:Siexpand}) and (\ref{eq:Siapprox1}), to collapse.
This is particularly evident in Eq.~(\ref{eq:S_iqs}) which exhibits no scallop features typical of self-heterodyne measurements.
We also note that the FWHM of the broadened carrier signals in Eqs.~(\ref{eq:S_Eqs}) and (\ref{eq:S_iqs}) are no longer related, and exhibit different scaling properties.
In this limit, $S_i(f)$ can no longer be taken as a proxy for $S_E(f)$.

\subsection{Master Equation Approach}

While $S_{\delta\nu}(f), S_E(f)$, and $S_i(f)$ depend only on $h_0$ and $f_c$, Rabi gate errors also depend on $\Omega_0$.
Single-photon gate errors caused by finite-bandwidth white noise can be computed from Eq.~(\ref{eq:error_init}), without approximation, giving  
\begin{multline}
\mathcal{E}=\frac{\pi h_0}{2\Omega_0}\bigg\{\frac{2y[1-(-1)^{2N}\cos(2\pi Ny)]}{1-y^2} \\
+2\text{Arctanh}(y)+\text{Ci}[2\pi N(1-y)]-\text{Ci}[2\pi N(1+y)] \\
-2\pi N\, \text{Si}[2\pi N(1-y)]+2\pi N\, \text{Si}[2\pi N(1+y)] \bigg\} , \label{eq:blwnerror_0}
\end{multline}
\vspace{-.3in}
\begin{multline}
\overline{\mathcal{E}}=\frac{4\pi h_0}{3\Omega_0}\bigg\{\frac{y[1-(-1)^{2N}\cos(2\pi Ny)]}{1-y^2} \\
-\pi N\, \text{Si}[2\pi N(1-y)]+\pi N\, \text{Si}[2\pi N(1+y)] \bigg\} , \label{eq:blwnerror_ave}
\end{multline}
where $\text{Ci}(x)$ and $\text{Si}(x)$ are cosine and  sine integral functions~\footnote{The sine and cosine integral functions are defined as $\text{Si}(z)=\int_0^z(\sin t)/t\,dt$ and $\text{Ci}(z)=-\int_z^\infty(\cos t)/t\,dt$.}, and $y\equiv 2\pi f_c/\Omega_0$.
 
In the limit $f_c\gg \Omega_0$, Eqs.~(\ref{eq:blwnerror_0})-(\ref{eq:blwnerror_ave}) reduce to the previously obtained results in Eq.~(\ref{eq:error_white}) for wide-bandwidth white noise.
In the opposite limit, $f_c\lesssim \Omega_0$, we obtain the following results for $2\pi N$ Rabi gates: 
\begin{gather}
\mathcal{E}\approx\left\{ 
\begin{array}{cc} \vspace{0.03in}
\frac{8\pi^2h_0f_c}{\Omega_0^2} & (N=1/2,3/2,\dots) ,
\\ \vspace{0.03in}
\frac{32\pi^6h_0f_c^3N^2}{3\Omega_0^4} 
& (N=1,2,\dots) ,
\end{array}\right. \label{eq:error1pqs_0}\\
\overline{\mathcal{E}}\approx\left\{ 
\begin{array}{cc} \vspace{0.03in}
\frac{16\pi^2h_0f_c}{3\Omega_0^2} & (N=1/2,3/2,\dots) ,
\\ \vspace{0.03in}
\frac{64\pi^6h_0f_c^3N^2}{9\Omega_0^4} 
& (N=1,2,\dots) ,
\end{array}\right. \label{eq:error1pqs_ave}
\end{gather} 
where we have also taken $h_0\lesssim\Omega_0$, as consistent with the weak-noise approximation.
In this regime, we note that the results do not depend on $f_c$ being larger or smaller than $h_0$.
As in Sec.~\ref{sec:2p}, the two-photon gate errors are additive, yielding
\begin{gather}
\mathcal{E}=\left\{ 
\begin{array}{cc} \vspace{0.03in}
\frac{8\pi^2(h_1f_{c1}+h_2f_{c2})}{\tilde\Omega_0^2} & (N=1/2,3/2,\dots) ,
\\ \vspace{0.03in}
\frac{32\pi^6(h_1f_{c1}^3+h_2f_{c2}^3)N^2}{3\tilde\Omega_0^4} 
& (N=1,2,\dots) ,
\end{array}\right. \label{eq:error2pqs_0}\\
\overline{\mathcal{E}}=\left\{ 
\begin{array}{cc} \vspace{0.03in}
\frac{16\pi^2(h_1f_{c1}+h_2f_{c2})}{3\tilde\Omega_0^2} & (N=1/2,3/2,\dots) ,
\\ \vspace{0.03in}
\frac{64\pi^6(h_1f_{c1}^3+h_2f_{c2}^3)N^2}{9\tilde\Omega_0^4} 
& (N=1,2,\dots) .
\end{array}\right. \label{eq:error2pqs_ave}
\end{gather}

Qualitatively different types of behavior are observed in Eqs.~(\ref{eq:error1pqs_0})-(\ref{eq:error2pqs_ave}) for half vs.\ full rotations, which may be understood as follows.
Frequency noise causes the Rabi rotation axis to tilt away from the equator of the Bloch sphere, resulting in gate errors.
However, for the special case of full rotations, the Bloch state returns to its initial value (to leading order in a noise expansion), regardless of the tilted rotation axis.
A secondary effect of the tilt is to increase the rotation speed, resulting in over-rotation.
This effect is higher-order, however, yielding smaller errors for full rotations.
In other words, for the special case of full rotations, in the limit of quasistatic noise, the leading-order contribution to the gate error vanishes; for all other cases, lower-order contributions are still present.

\begin{figure}
    \centering
    \includegraphics[width = 0.4\textwidth]{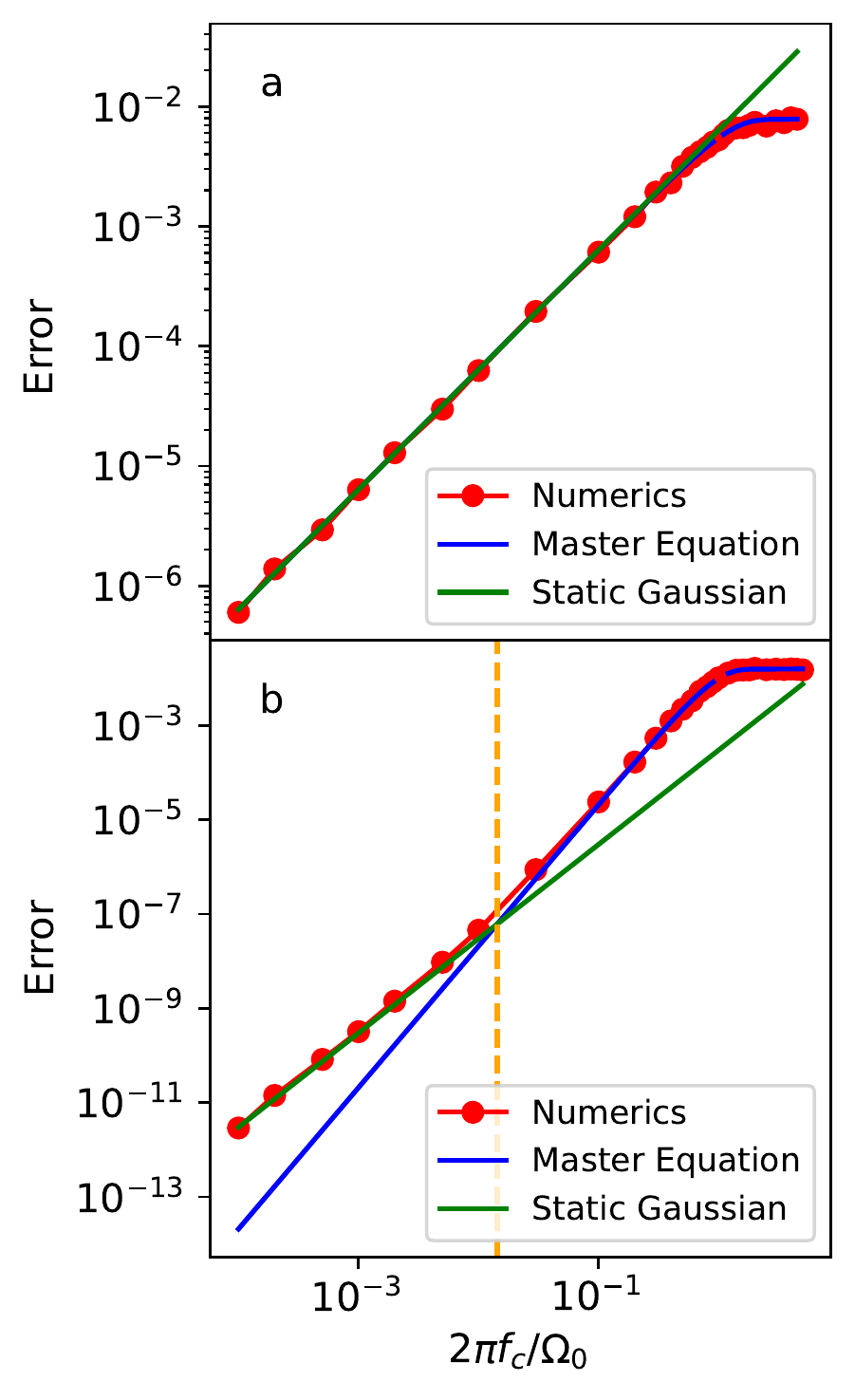}
    \caption{
    One-photon Rabi errors for white, band-limited frequency noise with fixed amplitude $h_0=3.18$~kHz$^2/$kHz, as a function of the noise bandwidth, $f_c$. 
    (a) $\pi$ rotations. 
    (b) $2\pi$ rotations.
    Here, red data correspond to numerical simulations, while
    the blue curves correspond to master-equation theory, Eq.~(\ref{eq:blwnerror_0}), and the green curves correspond to the quasistatic theory, Eq.~(\ref{eq:errorqs0}).
    The master-equation result fails only for the special case of $2\pi$ rotations, for bandwidths below the (vertical-dashed) crossover frequency, $f_c\approx 1.43h_0$.
    The quasistatic theory works well for both $\pi$ and $2\pi$ rotations, for the case of narrow bandwidths.
    }
    \label{fig:fh_bswp}
\end{figure}

In Fig.~\ref{fig:fh_bswp}, we show results of numerical simulations for bandwidth-limited white noise, for both $\pi$ and $2\pi$ rotations.
The corresponding theoretical predictions from Eq.~(\ref{eq:blwnerror_0}) are also shown as blue curves.
The theory clearly captures the majority of the errors arising from such noise spectra.
However, for the case of $2\pi$ rotations, the theory breaks down in the limit of small $f_c/h_0$. 
Specifically, we find that Eq.~(\ref{eq:blwnerror_0}) fails when $f_c\lesssim 1.43 h_0$. 
(Failure also requires that $f_c\lesssim \Omega_0/2\pi$.)
We attribute this failure to the fact that the master equation derivation in Sec.~\ref{sec.dm} employs an expansion in powers of the noise strength, keeping only the leading-order term. 
Hence, for the special case of full rotations, in the quasistatic limit (where the leading-order contribution to the gate error vanishes), our theory does not capture the central physics. 

\subsection{Quasistatic Gaussian-Distributed Noise}\label{sec:qausistaticfrequency}

To obtain an accurate solution for the singular problem of full rotations in the quasistatic limit, we modify the master equation approach of Sec.~\ref{sec:Master}.
The starting point for these calculations is the fluctuating frame Hamiltonian of Eq.~(\ref{eq:Hppp}),
\begin{equation}
H=\frac{\hbar\Omega_0}{2}\sigma_z+\frac{h\delta\nu}{2}\sigma_x , \label{eq:Hqs1p}
\end{equation}
which describes a rotation tilted slightly away from the desired Rabi rotation axis.
In the quasistatic limit, the frequency fluctuation $\delta\nu$ remains constant for the duration of the gate operation.

The dynamics of the density matrix is readily solved for a static Hamiltonian, yielding
\begin{multline}
\rho(t)=\frac{1}{2}+\left\{\frac{(2\pi\delta\nu)^2}{2{\Omega'_0}^2}+\frac{\Omega_0^2}{2{\Omega'_0}^2}\cos(\Omega'_0t)\right\}\sigma_x \\
+\left\{\frac{\Omega_0}{2{\Omega'_0}}\sin(\Omega'_0t)\right\}\sigma_y \\
+\left\{\frac{2\pi\delta\nu\,\Omega_0}{2{\Omega'_0}^2}\left[1-\cos(\Omega'_0t)\right]\right\}\sigma_z, \label{eq:qsrho}
\end{multline}
where $\Omega'_0\equiv\sqrt{\Omega_0^2+(2\pi\delta\nu)^2}$, and we have adopted the same initial conditions as in Sec.~\ref{sec:Master}, namely, $\rho(0)=\frac{1}{2}(1+\sigma_x)$.

Equation~(\ref{eq:qsrho}) describes the evolution of a pure state in a rotating frame.
We now assume the fluctuation $\delta\nu$ is drawn from a Gaussian distribution with probability
\begin{equation}
p_{\delta\nu}=\frac{1}{\sigma_{\delta\nu}\sqrt{2\pi}}e^{-(\delta\nu)^2/2\sigma_{\delta\nu}^2} .
\end{equation}
Here, the connection to the bandwidth-limited white-noise power spectrum in Eq.~(\ref{eq:fb}) is provided through the variance:
\begin{equation}
\sigma_{\delta\nu}^2=\langle(\delta\nu)^2\rangle_{\delta\nu}=\int_{-\infty}^\infty S_{\delta\nu}(f)df=2h_0f_c .
\end{equation}

As in Sec.~\ref{sec:1p}, the error in a Rabi gate defined by the gate period $t=2\pi N/\Omega_0$ is given by $\mathcal{E}=1-\text{tr}[\langle\rho\rangle_{\delta\nu}\rho_\text{ideal}]$, where $\rho_\text{ideal}$ is defined in Eq.~(\ref{eq:rhoideal}).
Expanding Eq.~(\ref{eq:qsrho}) in leading powers of $2\pi\delta\nu/\Omega_0$, we obtain the following result for one-photon quasistatic gate errors:
\begin{equation}
\mathcal{E}\approx\left\{ 
\begin{array}{cc} \vspace{0.03in}
\frac{8\pi^2h_0f_c}{\Omega_0^2} & (N=1/2,3/2,\dots) ,
\\ \vspace{0.03in}
\frac{48\pi^6h_0^2f_c^2N^2}{\Omega_0^4} 
& (N=1,2,\dots) .
\end{array}\right.\label{eq:errorqs0}
\end{equation}
Repeating these calculations for the initial states $\rho(0)=\frac{1}{2}(1+\sigma_y)$ and $\rho(0)=\frac{1}{2}(1+\sigma_z)$, and averaging the results, gives the average gate error
\begin{equation}
\overline{\mathcal{E}}\approx\left\{ 
\begin{array}{cc} \vspace{0.03in}
\frac{16\pi^2h_0f_c}{3\Omega_0^2} & (N=1/2,3/2,\dots) ,
\\ \vspace{0.03in}
\frac{32\pi^6h_0^2f_c^2N^2}{\Omega_0^4} 
& (N=1,2,\dots) .
\end{array}\right. \label{eq:errorqs}
\end{equation}
For the case of half-rotations ($N=1/2,3/2,\dots$), we note that Eq.~(\ref{eq:errorqs}) agrees with Eq.~(\ref{eq:error1pqs_ave}).
However, for full rotations ($N=1,2,\dots$), the results disagree.
We plot Eq.~(\ref{eq:errorqs}) as a green line in Fig.~\ref{fig:fh_bswp}, finding that the quasistatic theory accurately captures the physics of both $\pi$ and $2\pi$ rotations, for narrow bandwidths.

For two-photon gates, we follow a similar procedure.
The Hamiltonian in the fluctuating frame, Eq.~(\ref{eq:H2D}), can be rewritten as 
\begin{equation}
H=\frac{\hbar\tilde\Omega_0}{2}\sigma_z
+\frac{h(\delta\nu_1+\delta\nu_2)}{2}\sigma_x , \label{eq:Hqs2p}
\end{equation}
where $\delta\nu_1$ and $\delta\nu_2$ are the fluctuations of the two different lasers.
After making the replacements $\Omega_0\rightarrow\tilde\Omega_0=\Omega_1\Omega_2/\delta$ and $\delta\nu\rightarrow\delta\nu_1+\delta\nu_2$, the problem is then identical to Eq.~(\ref{eq:Hqs1p}).
We note that, although $\tilde\Omega_0$ depends on the Rabi frequencies of the two lasers, the fluctuations are contained in parameters $\delta\nu_1$ and $\delta\nu_2$, not $\tilde\Omega_0$.
Assuming independent Gaussian distributions for $\delta\nu_1$ and $\delta\nu_2$, as defined by their variances, $\sigma_{\delta\nu_1}^2=2h_1f_{c1}$ and $\sigma_{\delta\nu_2}^2=2h_2f_{c2}$, the error calculation then gives
\begin{equation}
\mathcal{E}\approx\left\{ 
\begin{array}{cc} \vspace{0.03in}
\frac{8\pi^2(h_1f_{c1}+h_2f_{c2})}{\tilde\Omega_0^2} & (N=1/2,3/2,\dots) ,
\\ \vspace{0.03in}
\frac{48\pi^6(h_1f_{c1}+h_2f_{c2})^2N^2}{\tilde\Omega_0^4} 
& (N=1,2,\dots) ,
\end{array}\right. 
\end{equation}
for the initial condition $\rho(0)=\frac{1}{2}(1+\sigma_x)$, and \begin{equation}
\overline{\mathcal{E}}\approx\left\{ 
\begin{array}{cc} \vspace{0.03in}
\frac{16\pi^2(h_1f_{c1}+h_2f_{c2})}{3\tilde\Omega_0^2} & (N=1/2,3/2,\dots) ,
\\ \vspace{0.03in}
\frac{32\pi^6(h_1f_{c1}+h_2f_{c2})^2N^2}{\tilde\Omega_0^4} 
& (N=1,2,\dots) 
\end{array}\right. \label{eq:errorqs2p}
\end{equation}
for the average gate fidelity.

\section{Intensity Noise}
\label{sec.intensity}

\begin{figure}[!t]
    \centering
    \includegraphics[width = 8.5cm]{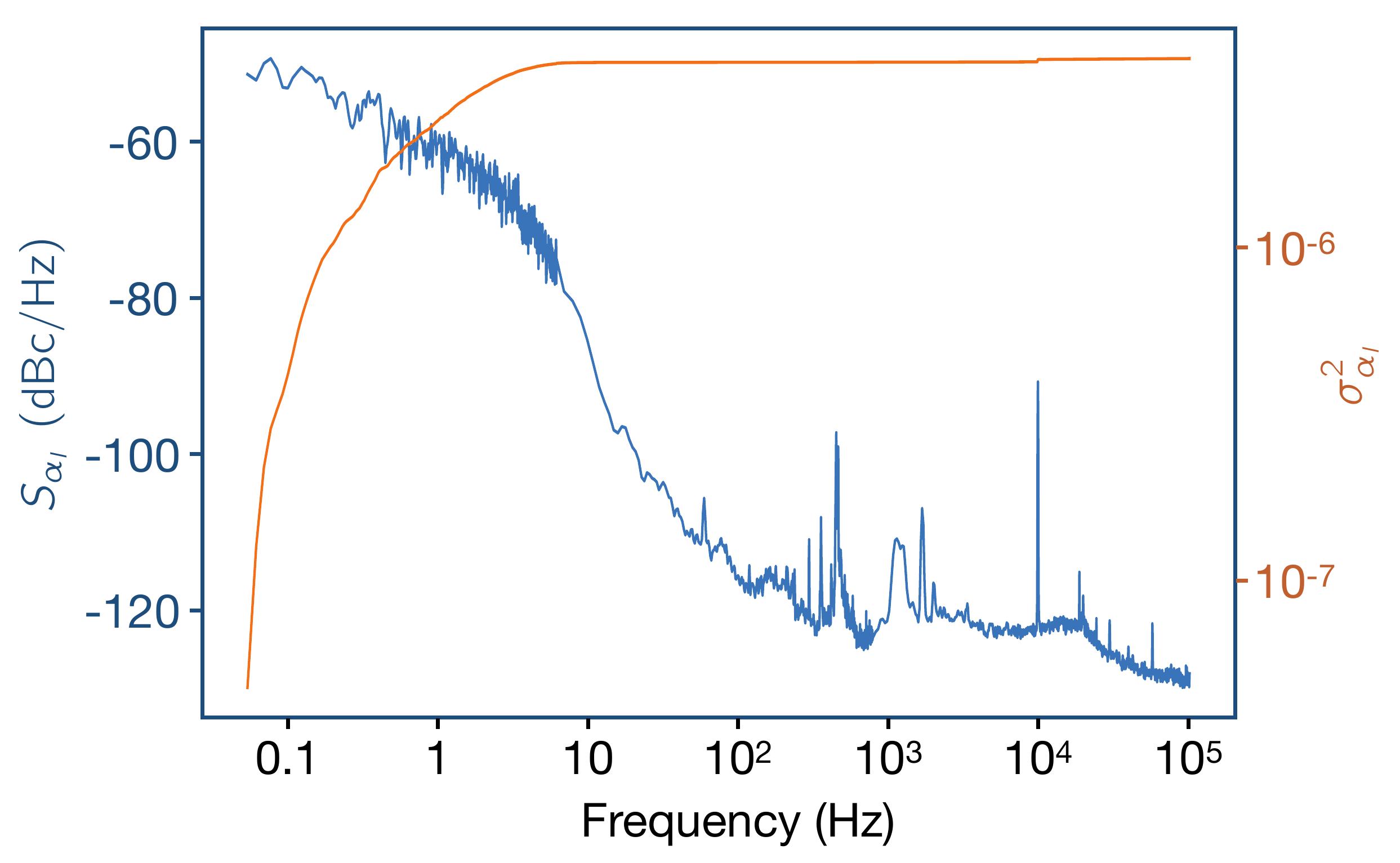}
    \caption{Measured RIN spectrum of the Ti:Sa laser used in Fig.~\ref{fig:servobump}. The orange curve shows the integrated variance at frequency $f$, and indicates that nearly all intensity noise occurs below 10~Hz, so the quasistatic approximation should be excellent for this laser.
    The variance integrated from 0.1 - $10^5$ Hz is $\sigma_{\alpha_I}^2=3.7\times 10^{-6}$.} 
    \label{fig:RINmeasurement}
\end{figure}

Up to this point, we have only considered phase fluctuations of the laser field.
We now also consider fluctuations of the field amplitude, or more specifically, the intensity $I(t)$, which is proportional to the laser power, or $|E_0|^2$.
We define the  fluctuating intensity as
\begin{equation}
    I(t)=I_0+\delta I(t)=I_0[1+\alpha_I(t)]
\end{equation}
where $I_0$ is the intensity of the noise-free laser and $\alpha_I(t)$ is  the time-varying relative intensity fluctuation. 
The relative intensity noise (RIN) is a quantity frequently used to characterize the laser quality, defined as~\cite{Riehle2004}
\begin{equation}
\text{RIN}=S_{\delta I}(f)/I_0^2=S_{\alpha_I}(f) ,
\end{equation}
where $S_{\delta I}(f)$ and $S_{\alpha_I}(f)$ are power spectral densities corresponding to the autocorrelation functions for $\delta I$ and $\alpha_I$, as defined in Eq.~(\ref{eq:SX}).

In certain types of lasers, including semiconductor diode lasers, relaxation oscillations may lead to intensity noise at frequencies of several GHz~\cite{Vahala1983}. 
In principle, the effect of such wide-band intensity noise on gate fidelities can be calculated for arbitrary $S_{\alpha_I}(f)$, using methods similar to those derived in previous sections for phase noise.
However, for optically pumped solid-state lasers, relaxation oscillations tend to be limited to much lower, sub-MHz frequencies~\cite{Koechner1972}. 
These fluctuations are  typically  well below the Rabi frequency, and can therefore be considered as quasistatic for our purposes.
A typical measured RIN spectrum for a solid-state Ti:Sa laser is shown in Fig.~\ref{fig:RINmeasurement}. Apart from narrow spikes at multiples of the 60~Hz power-line frequency, the noise is seen to decrease rapidly with frequency, such that the variance $\sigma^2_{\alpha_I}$ arises primarily from frequencies below 10~Hz. We therefore expect a quasistatic analysis, as provided below,  to be accurate. 

Similar to the approach in the previous section, we assume the fluctuations of $\alpha_I$ are drawn from a Gaussian distribution with probability 
\begin{equation}
p_{\alpha_I}=\frac{1}{\sigma_{\alpha_I}\sqrt{2\pi}}e^{-\alpha_I^2/2\sigma_{\alpha_I}^2} .
\end{equation}
The connection to conventional RIN measurements is made through the variance:
\begin{equation}
\sigma_{\alpha_I}^2=\langle\alpha_I^2\rangle_{\alpha_I}=\int_{-\infty}^\infty S_{\alpha_I}(f)df . \label{eq:sigalph}
\end{equation}

The time-varying Rabi frequency due to intensity noise, for one-photon gate operations, is given by
$\Omega(t)=\Omega_0\sqrt{1+\alpha_I(t)}$, where $\Omega_0$ is the noise-free  Rabi frequency. 
However, to avoid unphysical behavior in this expression, for the extremely unlikely event that $\alpha_I < -1$, we simply assume that $|\alpha_I|\ll 1$, to obtain the leading order expression for the fluctuating Hamiltonian in the rotating frame, given by
\begin{equation}
H\approx\frac{\hbar\Omega_0}{2}\left(1+\frac{\alpha_I}{2}\right)\sigma_z .
\end{equation}

\begin{figure}
    \centering
    \includegraphics[width = 0.8in]{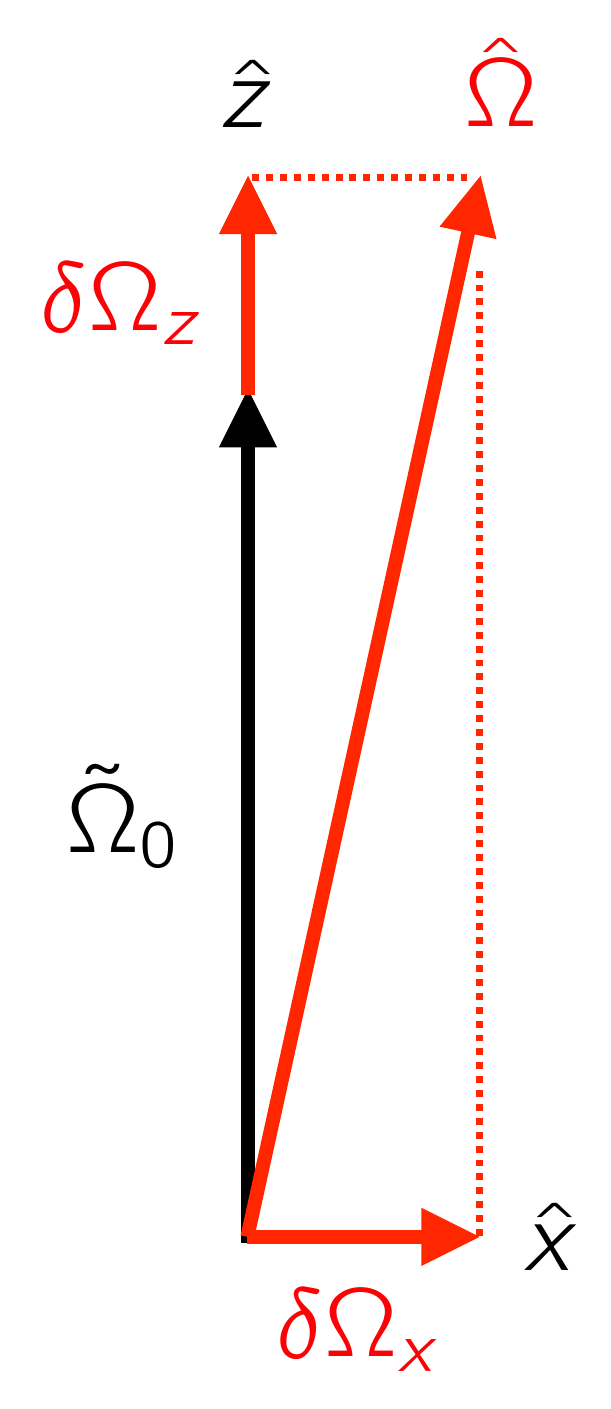}
    \caption{
    Tilted rotation axis ($\hat\Omega$) for two-photon Rabi gates, due to intensity fluctuations.
    }
    \label{fig:tilt}
\end{figure}

In this section we consider only the worst-case scenario for intensity-induced errors, which corresponds to the initial condition $\rho(0)=\frac{1}{2}(1+\sigma_x)$, on the equator of the Bloch sphere in the fluctuating frame, or the north pole of the Bloch sphere in the laboratory frame of Eq.~(\ref{eq:Hinit}).
Solving Eq.~(\ref{eq:rhot}) for a given, time-independent fluctuation $\alpha_I$, we then obtain
\begin{multline}
\rho(t)=\frac{1}{2}
+\frac{1}{2}\cos\left[\Omega_0\left(1+\frac{\alpha_I}{2}\right)t\right]\sigma_x \\
+\frac{1}{2}\sin\left[\Omega_0\left(1+\frac{\alpha_I}{2}\right)t\right]\sigma_y .
\end{multline}
We again compute the error $\mathcal E$ for a Rabi gate defined by gate period $t=2\pi N/\Omega_0$, where $\mathcal{E}=1-\text{tr}[\langle\rho\rangle_{\alpha_I}\rho_\text{ideal}]$ and $\rho_\text{ideal}$ is defined in Eq.~(\ref{eq:rhoideal}), obtaining
\begin{equation}
\mathcal{E}\approx\left\langle\frac{(\pi N\alpha_I)^2}{4}\right\rangle_{\alpha_I}
=\frac{\pi^2N^2\sigma_{\alpha_I}^2}{4} .
\label{Eq.intonephoton}
\end{equation}

\begin{figure*}[t]
    \centering
    \includegraphics[width=5in]{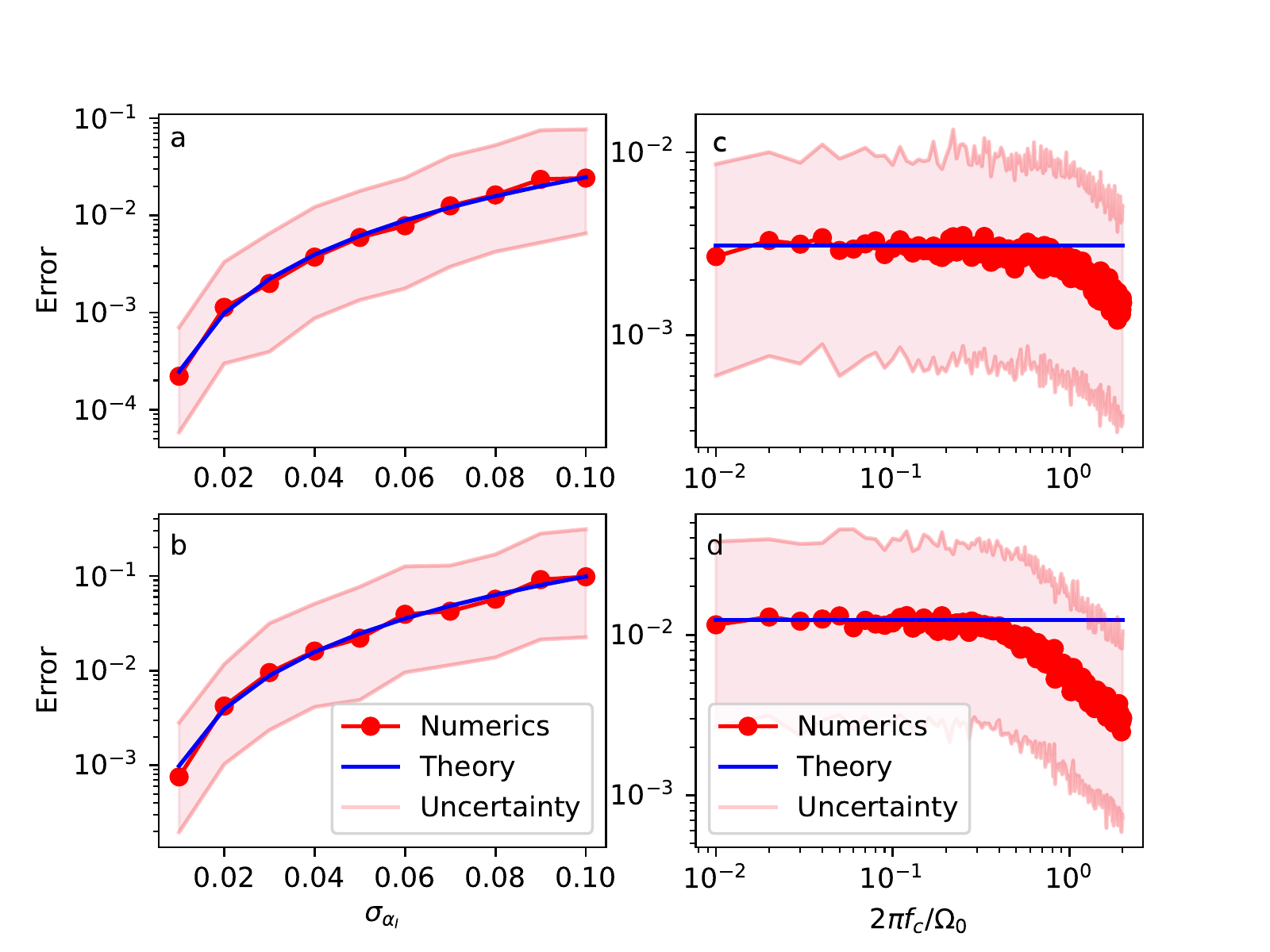}
    \caption{
    \label{fig:int1}
    Rabi errors due to RIN for one-photon gate operations. 
    The first row, (a), (c), corresponds to $\pi$ rotations, while the second row, (b), (d), corresponds to $2\pi$ rotations.
    The first column, (a), (b), show results for narrow-bandwidth noise, $f_c=0.1\times \Omega_0/2\pi$, while in the second column, (c), (d), we sweep the cut-off frequency $ f_c=10\times\Omega_0/2\pi$ holding $\sigma_{\alpha_I}$=0.05 constant.
    In all cases we choose $\Omega_0= 2\pi\times 1 ~\rm MHz$. 
    }
\end{figure*}

For two-photon gates, the situation is only a little more complicated.
Following the approach leading up to Eq.~(\ref{eq:H2D}), and performing an additional rotation around the $y$ axis, we obtain the approximate two-photon Rabi Hamiltonian,
\begin{equation}
H=\frac{\hbar\tilde\Omega}{2}\sigma_z 
+\frac{\hbar\Delta_+}{2}\sigma_x,
\end{equation}
where we now include intensity noise:
\begin{eqnarray}
\tilde \Omega &=&
\frac{[\Omega_1(1+\alpha_{I1}/2)][\Omega_2(1+\alpha_{I2}/2)]}{\delta} \nonumber \\
&&\approx \tilde\Omega_0\left(1+\frac{\alpha_{I1}}{2}+\frac{\alpha_{I2}}{2}\right) , \label{eq:tildeOmega}
\end{eqnarray}
Here, $\tilde\Omega_0$ is the noise-free version of the two-photon Rabi frequency.
Including intensity noise, the effective detuning is given by
\begin{eqnarray}
\Delta_+ &=& \Delta +
\frac{\Omega_1^2(1+\alpha_{I1}/2)^2-\Omega_2^2(1+\alpha_{I2}/2)^2}{2\delta} \nonumber \\
&&\approx \Delta_{+,0}+
\frac{\Omega_1^2\alpha_{I1}-\Omega_2^2\alpha_{I2}}{2\delta},
\end{eqnarray}
where $\Delta_{+,0}$ is the noise-free detuning, which we set to zero to achieve full-range $x$ rotations, as described in Sec.~\ref{sec:2p}.

\begin{figure*}[t]
    \centering
    \includegraphics[width=5in]{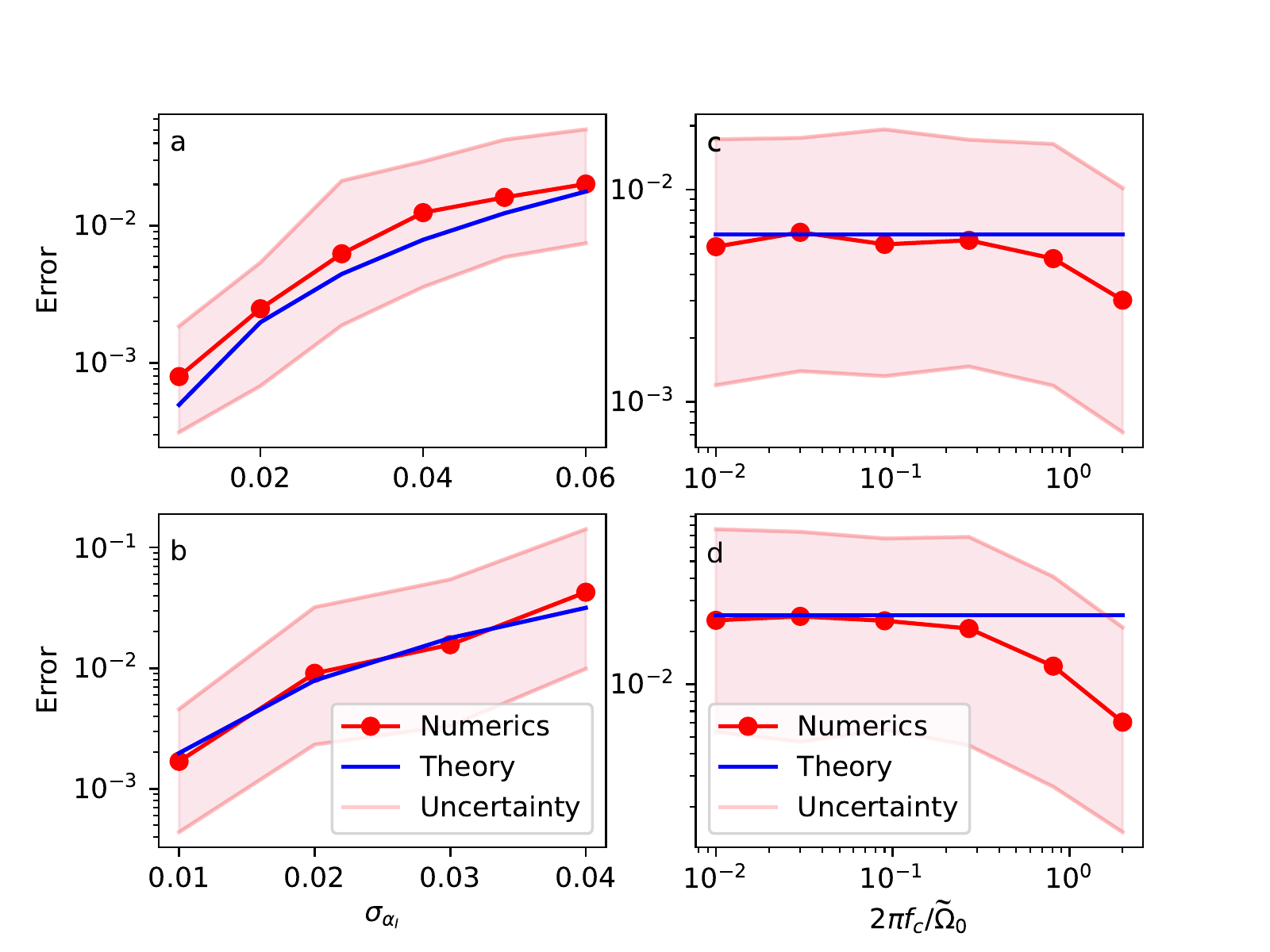}
    \caption{
    \label{fig:int2}
    Rabi errors due to RIN for two-photon gate operations.   
    All panels and simulations parameters are the same as in Fig.~\ref{fig:int1}, but with $\Omega_0$ replaced by $\tilde\Omega_0$. 
    }
\end{figure*}

The complication for two-photon gates is pictured in Fig.~\ref{fig:tilt}.
As consistent with Eq.~(\ref{eq:tildeOmega}), the driving strength along the original Rabi rotation axis is modified by $\delta\Omega_z=\tilde\Omega_0(\alpha_{I1}+\alpha_{I2})/2$.
However, in addition, the rotation axis is tilted by the presence of $\delta\Omega_x=\Delta_+$, where the average detuning $\overline{\Delta_+} =\Delta_{+,0}=0$.
The density matrix can be solved as before, to leading order in $\alpha_{I1}$ and $\alpha_{I2}$, giving
\begin{multline}
\rho(t)=\frac{1}{2}
+\frac{1}{2}\cos\left[\Omega_0\left(1+\frac{\alpha_{I1}}{2}+\frac{\alpha_{I2}}{2}\right)t\right]\sigma_x \\
+\frac{1}{2}\sin\left[\Omega_0\left(1+\frac{\alpha_{I1}}{2}+\frac{\alpha_{I2}}{2}\right)t\right]\sigma_y +\rho_z\sigma_z ,
\end{multline}
where $\rho_z\propto\delta\Omega_x$.
However, since $\rho_\text{ideal}$ does not contain a $\sigma_z$ component, $\rho_z$ does not enter into the final error expression.
The leading-order errors for two-photon gates are therefore found to be additive:
\begin{equation}
\mathcal{E}
=\frac{\pi^2N^2(\sigma_{\alpha_{I1}}^2+\sigma_{\alpha_{I2}}^2)}{4} . \label{eq:2pintensity}
\end{equation}

To test these predictions, we simulate intensity noise by modeling it as finite-bandwidth white noise, using an approach very similar to Sec.~\ref{sec:qausistaticfrequency}.
Specifically, we define the intensity noise power spectral density as
\begin{equation}
S_{\alpha_I}(f) = \left\{ 
\begin{array}{ccc} \vspace{0.03in}
h_I &\quad\text{when}\quad&  |f|\leq f_c, \\ \vspace{0.03in}
0 &\quad\text{when}\quad&  |f|>f_c.
\end{array}
\right. ,
\end{equation} 
where $f_c$ is the bandwidth.
From Eq.~(\ref{eq:sigalph}), we see that $\sigma_{\alpha_I}^2=2h_If_c$,
and since $\alpha_I$ is dimensionless, $h_I$ must have units of time.
In analogy with Eq.~(\ref{eq:randomphifin}), we define
\begin{equation}
\alpha_I(t)= \sum_{j=1}^{M/2}2\sqrt{S_{\alpha_I}(f_j)\Delta f}\cos(2\pi f_jt+\varphi_j) , \label{eq:alphat}
\end{equation}
where $\phi_j\in [0,2\pi]$, as usual.

We now solve Eqs.~(\ref{eq:cg1p}) and (\ref{eq:ce1p}) numerically, using the one-photon Rabi frequency $\Omega(t)\approx\Omega_0[1+\alpha_I(t)/2]$ for a given time series $\alpha_I(t)$, generated from Eq.~(\ref{eq:alphat}). 
The procedure is repeated for many random time series and the results are averaged to compute the fidelity $F$ and the error $\mathcal{E}=1-F$.

The results of these numerical simulations are shown in Fig.~\ref{fig:int1} for $\pi$ and $2\pi$ rotations, and for the cases of narrow-bandwidth noise $f_c\ll \Omega_0/2\pi$ and wide-bandwidth noise $f_c\gg \Omega_0/2\pi$.
Here in panels (a) and (b), $f_c$ is held fixed while $\sigma_{\alpha_I}$ is swept.
Due to the constraints of our noise model, this implies that we sweep the parameter $h_I=\sigma_{\alpha_I}^2/2f_c$.
To determine where the quasistatic approximation breaks down, we also plot our results while holding $\sigma_{\alpha_I}$ fixed and sweeping $f_c$ in panels (c) and (d).
We see that the results are well described by our theoretical predictions in Eq.~(\ref{Eq.intonephoton}) for the low-bandwidth regime, as expected, but diverge quantitatively in the high-bandwidth regime.

For two-photon gates, we follow the same procedure, now using the two-qubit Rabi frequency of Eq.~(\ref{eq:tildeOmega}).
The corresponding results are shown in Fig.~\ref{fig:int2}.
We again obtain good agreement with our theoretical predictions in Eq.~(\ref{eq:2pintensity}) for the low-bandwidth regime, but poor agreement for the high-bandwidth regime.

\section{Summary and Discussion}
\label{sec.discussion}

In this paper, we theoretically investigated errors in quantum gate operations of neutral-atom qubits.
We focused on a noise mechanism that dominates many qubit experiments: phase fluctuations of the driving laser. 
We considered both one and two-photon Rabi oscillations.
We also considered generic noise spectra, such as flat-background (white) noise, and noise peaked at finite frequencies.
We refer to the latter as `servo-bump noise,' due to the common occurrence of noise peaks due to servo-loop feedback circuitry.

We have specifically considered the weak-noise regime, which is typical of modern qubit experiments.
In this limit, we uncover simple relations between the underlying phase-noise spectra and noise spectra measured in self-heterodyne experiments.
These relations are given in Eqs.~(\ref{eq:SEexpand}), (\ref{eq:Siexpand}), and (\ref{eq:Siapprox1}), and they allow us to analyze and fit experimental self-heterodyne data using specific white and servo-bump noise models, as described in Sec.~\ref{sec:laserparameters}. 

The weak-noise limit also allows us to solve a master equation, describing the effects of laser phase noise on Rabi oscillations, which we then use to calculate gate fidelities.
We perform realistic numerical simulations of Rabi gates by generating random time series that include white and servo-bump phase noise.
The results are well explained by our master equation solutions, yielding a deeper understanding of the decoherence process.
Our main results are given in Eqs.~(\ref{eq:error_init}), (\ref{eq:error_white}), (\ref{eq:servo_error}), (\ref{eq:error_white2p}), and (\ref{eq:servo_error2p}).

In the case of servo-bump phase noise, we observe that gate errors are most prominent when the central frequency of the noise peak occurs near the Rabi frequency, as expected for $T_{1\rho}$-type noise mechanisms~\cite{FYan2013}.
For $\pi$ pulses  
the gate-error peak frequency falls in the range $1\leq f_g/(\Omega_0/2\pi)\le1.5$. The one-photon error contributions from white noise and servo bumps are given by Eqs.~(\ref{eq:error_white}) and (\ref{eq:servo_error}) respectively.  For a two-photon drive  the white noise and servo bump errors are given by Eqs.~(\ref{eq:error_white2p}) and (\ref{eq:servo_error2p}), respectively. For a $2\pi$ pulse the gate-error peak frequency is close to $ f_g=\Omega_0/2\pi$. 
Away from these peaks, the gate error  may be suppressed by many orders of magnitude, such that the residual errors are dominated by background noise (e.g., weak white noise).
Generally, the contributions to the gate error from different noise mechanisms are found to be additive in the weak-noise regime.

To demonstrate how these results may be used to guide future experiments, we consider the case of $\pi$-rotations ($N=1/2$), starting in a computational basis state with two-photon driving.  Assuming a Rabi frequency of $\tilde \Omega_0/2\pi=1 ~\rm MHz$,  our results show that a white noise background below $h_0=20~\rm Hz^2/Hz$ would be required on each laser field, to obtain  gate errors below  $10^{-4}$. As shown in Fig.~\ref{fig:servobump}, a locked Ti:Sa laser satisfies this requirement, while other laser types such as semiconductor diode lasers typically have larger frequency noise, even when locked to a reference cavity~\cite{YLi2019}.

If a servo bump is present and its peak frequency occurs near the Rabi frequency of one of the two transitions, then achieving a gate-error level of $10^{-4}$ would require a total (integrated) servo bump noise power of no more than $s_g=0.00016$. 
(Note that $s_g$, as defined here, includes the power in the servo bumps on both sides of the carrier peak.)
In the self-heterodyne noise measurements analyzed in Sec.~\ref{sec:laserparameters}, we observed noise powers of $s_{g1}=0.00013$
 and $s_{g2}=0.00027$.
Moreover, since the peak frequency of the larger servo bump occurs at $f_g=234$~kHz, we can suppress its effect on the fidelity by choosing a Rabi frequency of $\Omega_0/2\pi \sim 117$~kHz, or alternatively, a Rabi frequency larger than 1 - 2~MHz.
In this way, we can realistically expect to achieve gate errors below $10^{-4}$ in this system. For more complex gate operations, such as Rydberg gates for preparing entangled states, longer pulses are involved and the requirements on laser noise are correspondingly more stringent.   A two-qubit entangling Rydberg gate requires about $2\pi$ of ground-Rydberg rotation on each atom. This implies that limiting the laser noise contribution to gate fidelities to below $10^{-4}$ would require a white noise spectrum with a more demanding noise level of $h_0\lesssim 5~\rm Hz^2/Hz$.   

Similar estimates of the RIN level required for a desired gate fidelity can be made.  Assuming the RIN is concentrated at frequencies much lower than the Rabi frequency, Eq.~(\ref{eq:2pintensity}) shows that the gate error is independent of $\Omega_0$. For a two-photon $\pi$ pulse with error at the $10^{-4}$ level, the RIN variance must satisfy $\sigma^2\lesssim 8.\times 10^{-5}$ and for a Rydberg entangling gate with errors below $10^{-4}$ we need $\sigma^2\lesssim 1.\times 10^{-5}$. The data in Fig.~\ref{fig:RINmeasurement} show that this variance level is reached with a well stabilized laser, although the experimentally relevant variance corresponds to the light seen by the atomic qubit, which typically has increased variance due to the instability of optical components, as well as atomic position fluctuations~\cite{Gillen-Christandl2016}.

In the limit of weak noise, the errors due to laser phase noise and RIN are additive. It is apparent that achieving very-high-fidelity, optically driven gate operations puts stringent limits on laser noise parameters. This is true for one-photon and two-photon drives in the ladder configuration considered here. The highest fidelity optically driven gate we are aware of is the demonstration of  Raman gates on trapped $^9$Be$^+$ ions using a two-photon $\Lambda$ configuration, where an average error per gate of $3.8\times 10^{-5}$ was achieved~\cite{Gaebler2016}. Notably the use of a $\Lambda$ configuration with both beams derived from the same laser implies a cancellation of phase noise that is not present in the ladder configuration analyzed here.

\section{Acknowledgement} This material is based on work supported by  NSF Award
2016136 for the QLCI center Hybrid Quantum Architectures and
Networks,
the U.S. Department of Energy Office of Science National Quantum
Information Science Research Centers, and   NSF Award 2210437.

\appendix
\renewcommand\thefigure{\thesection.\arabic{figure}}    
\setcounter{figure}{0}

\section{Fidelity measures}
\label{appendix.fidelity}

In general, the fidelity of a quantum operation depends on the initial state that it is applied to. It is useful to have an expression for the operator fidelity averaged over all possible initial states. It can be shown~\cite{Bowdrey2002} that for a single qubit, the fidelity averaged over all possible initial states can be compactly calculated as the fidelity averaged over just six states:  $\ket{\pm x}, \ket{\pm y},\ket{\pm z}.$ For pure states of a qubit the resulting fidelity can be written as 
$ F=\frac{1}{6}\sum_{j=1}^6\text{Tr}[\langle\rho_j\rangle\rho_{0j}]$ where $\langle\rho_j\rangle$ and $\rho_{0j}$ are the states due to a noisy gate operation, and the ideal states for each of the initial states labeled by $j$.

The fidelity appearing in this way can be expressed in a form that depends on the operator, without specifying the states.  Consider an ideal operator  ${\sf U}_{0}$ and a noisy operator $\sf U$. The fidelity of $\sf U$ with respect to $\sf U_0$ can be expressed as~\cite{Pedersen2007} 
\bea 
F&=&\frac{\textrm{Tr}({\sf U}_0^\dag {\sf U} {\sf U}^\dag 
{\sf U}_0)+\left|\textrm{Tr}({\sf U}_0^\dag {\sf U}) \right|^2}{n(n+1)}\nn\\
&=&\frac{n+\left|\textrm{Tr}({\sf U}_0^\dag {\sf U}) \right|^2 }{n(n+1)},\nn
\eea
where $n$ is the dimension of the Hilbert space and we have used the fact that ${\sf U}_0$ and $\sf U$ are unitary. Clearly when ${\sf U}={\sf U}_0$ we recover $F=1$. 

We can gain some intuition about this expression for the particular case of a rotation operator acting on a single qubit. Let ${\sf U}_0={\sf R}_x(\theta_0)$ and ${\sf U}={\sf R}_x(\theta)$ with $\theta=\theta_0+\delta$ where $\delta$ is a rotation angle error.  After a short calculation we find
\begin{equation}
F=\frac{1}{3}+\frac{2}{3}\cos^2(\delta/2).
\label{Eq.Fstandard}
\end{equation}

An alternative fidelity definition for operations on a single qubit ($n=2$) has been used in Ref.~\cite{Green2013}:
\bea 
F'&=&\frac{\left|\textrm{Tr}({\sf U}_0^\dag  {\sf U}) \right|^2}{4}\nn ,
\eea
 which leads to 
\begin{equation}
F'=\cos^2(\delta/2).
\label{Eq.FBiercuk}
\end{equation}
Although  $F$ and $F'$  agree at $\delta=0$, they provide  different results for finite $\delta$. For this reason our expressions for the average gate fidelity of a one-photon transition in Eq.~(\ref{eq:error_init}), which derive from $F$, the standard definition of the fidelity, differ from the corresponding results that can be derived from  Eqs.~(37a) and (40) in Ref.~\cite{Green2013}, even in the limit of low-bandwidth noise.

\section{Error estimates for arbitrary Rabi gates}
\label{appendix.gates}

In the main text, we computed errors for Rabi gates with gate periods $t=2\pi N/\Omega_0$, for $N=1/2,1,3/2,\dots$.
These special gates were chosen because they can be solved analytically.
However, the master equation formalism developed in this work can be applied to arbitrary Rabi gates.

Let us consider the arbitrary gate period $t=t_g$.
Making the substitutions $\omega_j\rightarrow 2\pi f$, $\Delta f\rightarrow df$, as well as the substitution in Eq.~(\ref{eq:integ}), Eq.~(\ref{eq:rhonon}) can be rewritten as
\begin{widetext}
\begin{multline}
\langle\rho(t_g)\rangle \approx \frac{1}{2}
+\left[\frac{1}{2}\cos(\Omega_0t_g)
-2\pi^2\Omega_0^2\int_0^\infty\!\!\! df\, S_{\delta\nu}(f) \frac{2\cos(\Omega_0t_g)-2\cos(2\pi ft_g)+(\Omega_0^2-4\pi^2f^2)(t_g/\Omega_0)\sin(\Omega_0t_g)}
{(\Omega_0^2-4\pi^2f^2)^2} \right]\sigma_x \\
+\left[\frac{1}{2}\sin(\Omega_0t_g)
-2\pi^2\Omega_0^2\,\,\text{p.v.}\!\!\int_0^\infty \!\!\! df\, S_{\delta\nu}(f)\frac{2\sin(\Omega_0t_g)-2(2\pi f/\Omega_0)\sin(2\pi ft_g)+(\Omega_0^2-4\pi^2f^2)(t_g/\Omega_0)\cos(\Omega_0t_g)}
{(\Omega_0^2-4\pi^2f^2)^2}
\right]\sigma_y , \label{eq:rhononfull}
\end{multline}
where $\text{p.v.}$ stands for principal value (applied at the singularity, $f\rightarrow\Omega_0/2\pi$), and we have made use of the facts that (1) the singularity in the integrand takes the form $(\Omega_0-2\pi f)^{-1}$, and (2) $S_{\delta\nu}(f)$ is generally a smooth function near $f= \Omega_0/2\pi$.
In this form, Eq.~(\ref{eq:rhononfull}) can be solved numerically.

We may also compute the gate error associated with Eq.~(\ref{eq:rhononfull}) by noting that 
\begin{equation}
\rho_\text{ideal}(t_g)= \frac{1}{2}
+\frac{1}{2}\cos(\Omega_0t_g)\sigma_x 
+\frac{1}{2}\sin(\Omega_0t_g)\sigma_y . 
\end{equation}
Defining the gate error as usual by ${\mathcal E}=1-F$, where $F=\text{Tr}[\langle\rho\rangle\rho_\text{ideal}]$, we obtain
\begin{multline}
\mathcal{E} \approx (2\pi\Omega_0)^2\cos(\Omega_0t_g)\int_0^\infty\!\!\! df\, S_{\delta\nu}(f) \frac{2\cos(\Omega_0t_g)-2\cos(2\pi ft_g)+(\Omega_0^2-4\pi^2f^2)(t_g/\Omega_0)\sin(\Omega_0t_g)}
{(\Omega_0^2-4\pi^2f^2)^2}  \\
+(2\pi\Omega_0)^2\sin(\Omega_0t_g)\,\,\text{p.v.}\!\!\int_0^\infty \!\!\! df\, S_{\delta\nu}(f)\frac{2\sin(\Omega_0t_g)-2(2\pi f/\Omega_0)\sin(2\pi ft_g)+(\Omega_0^2-4\pi^2f^2)(t_g/\Omega_0)\cos(\Omega_0t_g)}
{(\Omega_0^2-4\pi^2f^2)^2} , \label{eq:errorfull}
\end{multline}
which can also be solved numerically.
\end{widetext}

\bibliography{optics.bib,qc_refs.bib,rydberg.bib,saffman_refs.bib,atomic.bib} 

\begin{thebibliography}{45}%
\makeatletter
\providecommand \@ifxundefined [1]{%
 \@ifx{#1\undefined}
}%
\providecommand \@ifnum [1]{%
 \ifnum #1\expandafter \@firstoftwo
 \else \expandafter \@secondoftwo
 \fi
}%
\providecommand \@ifx [1]{%
 \ifx #1\expandafter \@firstoftwo
 \else \expandafter \@secondoftwo
 \fi
}%
\providecommand \natexlab [1]{#1}%
\providecommand \enquote  [1]{``#1''}%
\providecommand \bibnamefont  [1]{#1}%
\providecommand \bibfnamefont [1]{#1}%
\providecommand \citenamefont [1]{#1}%
\providecommand \href@noop [0]{\@secondoftwo}%
\providecommand \href [0]{\begingroup \@sanitize@url \@href}%
\providecommand \@href[1]{\@@startlink{#1}\@@href}%
\providecommand \@@href[1]{\endgroup#1\@@endlink}%
\providecommand \@sanitize@url [0]{\catcode `\\12\catcode `\$12\catcode
  `\&12\catcode `\#12\catcode `\^12\catcode `\_12\catcode `\%12\relax}%
\providecommand \@@startlink[1]{}%
\providecommand \@@endlink[0]{}%
\providecommand \url  [0]{\begingroup\@sanitize@url \@url }%
\providecommand \@url [1]{\endgroup\@href {#1}{\urlprefix }}%
\providecommand \urlprefix  [0]{URL }%
\providecommand \Eprint [0]{\href }%
\providecommand \doibase [0]{http://dx.doi.org/}%
\providecommand \selectlanguage [0]{\@gobble}%
\providecommand \bibinfo  [0]{\@secondoftwo}%
\providecommand \bibfield  [0]{\@secondoftwo}%
\providecommand \translation [1]{[#1]}%
\providecommand \BibitemOpen [0]{}%
\providecommand \bibitemStop [0]{}%
\providecommand \bibitemNoStop [0]{.\EOS\space}%
\providecommand \EOS [0]{\spacefactor3000\relax}%
\providecommand \BibitemShut  [1]{\csname bibitem#1\endcsname}%
\let\auto@bib@innerbib\@empty
\bibitem [{\citenamefont {Geva}\ \emph {et~al.}(1995)\citenamefont {Geva},
  \citenamefont {Kosloff},\ and\ \citenamefont {Skinner}}]{Geva1995}%
  \BibitemOpen
  \bibfield  {author} {\bibinfo {author} {\bibfnamefont {E.}~\bibnamefont
  {Geva}}, \bibinfo {author} {\bibfnamefont {R.}~\bibnamefont {Kosloff}}, \
  and\ \bibinfo {author} {\bibfnamefont {J.~L.}\ \bibnamefont {Skinner}},\
  }\bibfield  {title} {\enquote {\bibinfo {title} {On the relaxation of a
  two-level system driven by a strong electromagnetic field},}\ }\href@noop {}
  {\bibfield  {journal} {\bibinfo  {journal} {J. Chem. Phys.}\ }\textbf
  {\bibinfo {volume} {102}},\ \bibinfo {pages} {8541} (\bibinfo {year}
  {1995})}\BibitemShut {NoStop}%
\bibitem [{\citenamefont {Makhlin}\ and\ \citenamefont
  {Shnirman}(2003)}]{Makhlin2003}%
  \BibitemOpen
  \bibfield  {author} {\bibinfo {author} {\bibfnamefont {Yu.}\ \bibnamefont
  {Makhlin}}\ and\ \bibinfo {author} {\bibfnamefont {A.}~\bibnamefont
  {Shnirman}},\ }\bibfield  {title} {\enquote {\bibinfo {title} {Dephasing of
  qubits by transverse low-frequency noise},}\ }\href@noop {} {\bibfield
  {journal} {\bibinfo  {journal} {JETP Lett.}\ }\textbf {\bibinfo {volume}
  {8}},\ \bibinfo {pages} {497} (\bibinfo {year} {2003})}\BibitemShut {NoStop}%
\bibitem [{\citenamefont {Ithier}\ \emph {et~al.}(2005)\citenamefont {Ithier},
  \citenamefont {Collin}, \citenamefont {Joyez}, \citenamefont {Meeson},
  \citenamefont {Vion}, \citenamefont {Esteve}, \citenamefont {Chiarello},
  \citenamefont {Shnirman}, \citenamefont {Makhlin}, \citenamefont {Schriefl},\
  and\ \citenamefont {Sch\"on}}]{Ithier2005}%
  \BibitemOpen
  \bibfield  {author} {\bibinfo {author} {\bibfnamefont {G.}~\bibnamefont
  {Ithier}}, \bibinfo {author} {\bibfnamefont {E.}~\bibnamefont {Collin}},
  \bibinfo {author} {\bibfnamefont {P.}~\bibnamefont {Joyez}}, \bibinfo
  {author} {\bibfnamefont {P.~J.}\ \bibnamefont {Meeson}}, \bibinfo {author}
  {\bibfnamefont {D.}~\bibnamefont {Vion}}, \bibinfo {author} {\bibfnamefont
  {D.}~\bibnamefont {Esteve}}, \bibinfo {author} {\bibfnamefont
  {F.}~\bibnamefont {Chiarello}}, \bibinfo {author} {\bibfnamefont
  {A.}~\bibnamefont {Shnirman}}, \bibinfo {author} {\bibfnamefont
  {Y.}~\bibnamefont {Makhlin}}, \bibinfo {author} {\bibfnamefont
  {J.}~\bibnamefont {Schriefl}}, \ and\ \bibinfo {author} {\bibfnamefont
  {G.}~\bibnamefont {Sch\"on}},\ }\bibfield  {title} {\enquote {\bibinfo
  {title} {Decoherence in a superconducting quantum bit circuit},}\ }\href@noop
  {} {\bibfield  {journal} {\bibinfo  {journal} {Phys. Rev. B}\ }\textbf
  {\bibinfo {volume} {72}},\ \bibinfo {pages} {134519} (\bibinfo {year}
  {2005})}\BibitemShut {NoStop}%
\bibitem [{\citenamefont {Chen}\ \emph {et~al.}(2012)\citenamefont {Chen},
  \citenamefont {Bohnet}, \citenamefont {Weiner},\ and\ \citenamefont
  {Thompson}}]{ZChen2012}%
  \BibitemOpen
  \bibfield  {author} {\bibinfo {author} {\bibfnamefont {Z.}~\bibnamefont
  {Chen}}, \bibinfo {author} {\bibfnamefont {J.~G.}\ \bibnamefont {Bohnet}},
  \bibinfo {author} {\bibfnamefont {J.~M.}\ \bibnamefont {Weiner}}, \ and\
  \bibinfo {author} {\bibfnamefont {J.~K.}\ \bibnamefont {Thompson}},\
  }\bibfield  {title} {\enquote {\bibinfo {title} {General formalism for
  evaluating the impact of phase noise on {B}loch vector rotations},}\ }\href
  {\doibase 10.1103/PhysRevA.86.032313} {\bibfield  {journal} {\bibinfo
  {journal} {Phys. Rev. A}\ }\textbf {\bibinfo {volume} {86}},\ \bibinfo
  {pages} {032313} (\bibinfo {year} {2012})}\BibitemShut {NoStop}%
\bibitem [{\citenamefont {Yan}\ \emph {et~al.}(2013)\citenamefont {Yan},
  \citenamefont {Gustavsson}, \citenamefont {Bylander}, \citenamefont {Jin},
  \citenamefont {Yoshihara}, \citenamefont {Cory}, \citenamefont {Nakamura},
  \citenamefont {Orlando},\ and\ \citenamefont {Oliver}}]{FYan2013}%
  \BibitemOpen
  \bibfield  {author} {\bibinfo {author} {\bibfnamefont {F.}~\bibnamefont
  {Yan}}, \bibinfo {author} {\bibfnamefont {S.}~\bibnamefont {Gustavsson}},
  \bibinfo {author} {\bibfnamefont {J.}~\bibnamefont {Bylander}}, \bibinfo
  {author} {\bibfnamefont {X.}~\bibnamefont {Jin}}, \bibinfo {author}
  {\bibfnamefont {F.}~\bibnamefont {Yoshihara}}, \bibinfo {author}
  {\bibfnamefont {D.~G.}\ \bibnamefont {Cory}}, \bibinfo {author}
  {\bibfnamefont {Y.}~\bibnamefont {Nakamura}}, \bibinfo {author}
  {\bibfnamefont {T.~P.}\ \bibnamefont {Orlando}}, \ and\ \bibinfo {author}
  {\bibfnamefont {W.~D.}\ \bibnamefont {Oliver}},\ }\bibfield  {title}
  {\enquote {\bibinfo {title} {Rotating-frame relaxation as a noise spectrum
  analyser of a superconducting qubit undergoing driven evolution},}\
  }\href@noop {} {\bibfield  {journal} {\bibinfo  {journal} {Nat. Commun.}\
  }\textbf {\bibinfo {volume} {4}},\ \bibinfo {pages} {2337} (\bibinfo {year}
  {2013})}\BibitemShut {NoStop}%
\bibitem [{\citenamefont {Paladino}\ \emph {et~al.}(2014)\citenamefont
  {Paladino}, \citenamefont {Galperin}, \citenamefont {Falci},\ and\
  \citenamefont {Altshuler}}]{Paladino2014}%
  \BibitemOpen
  \bibfield  {author} {\bibinfo {author} {\bibfnamefont {E.}~\bibnamefont
  {Paladino}}, \bibinfo {author} {\bibfnamefont {Y.~M.}\ \bibnamefont
  {Galperin}}, \bibinfo {author} {\bibfnamefont {G.}~\bibnamefont {Falci}}, \
  and\ \bibinfo {author} {\bibfnamefont {B.~L.}\ \bibnamefont {Altshuler}},\
  }\bibfield  {title} {\enquote {\bibinfo {title} {$1/f$ noise: {I}mplications
  for solid-state quantum information},}\ }\href@noop {} {\bibfield  {journal}
  {\bibinfo  {journal} {Rev. Mod. Phys.}\ }\textbf {\bibinfo {volume} {86}},\
  \bibinfo {pages} {361} (\bibinfo {year} {2014})}\BibitemShut {NoStop}%
\bibitem [{\citenamefont {Yoshihara}\ \emph {et~al.}(2014)\citenamefont
  {Yoshihara}, \citenamefont {Nakamura}, \citenamefont {Yan}, \citenamefont
  {Gustavsson}, \citenamefont {Bylander}, \citenamefont {Oliver},\ and\
  \citenamefont {Tsai}}]{Yoshihara2014}%
  \BibitemOpen
  \bibfield  {author} {\bibinfo {author} {\bibfnamefont {F.}~\bibnamefont
  {Yoshihara}}, \bibinfo {author} {\bibfnamefont {Y.}~\bibnamefont {Nakamura}},
  \bibinfo {author} {\bibfnamefont {F.}~\bibnamefont {Yan}}, \bibinfo {author}
  {\bibfnamefont {S.}~\bibnamefont {Gustavsson}}, \bibinfo {author}
  {\bibfnamefont {J.}~\bibnamefont {Bylander}}, \bibinfo {author}
  {\bibfnamefont {W.~D.}\ \bibnamefont {Oliver}}, \ and\ \bibinfo {author}
  {\bibfnamefont {J.-S.}\ \bibnamefont {Tsai}},\ }\bibfield  {title} {\enquote
  {\bibinfo {title} {Flux qubit noise spectroscopy using {R}abi oscillations
  under strong driving conditions},}\ }\href@noop {} {\bibfield  {journal}
  {\bibinfo  {journal} {Phys. Rev. B}\ }\textbf {\bibinfo {volume} {89}},\
  \bibinfo {pages} {020503} (\bibinfo {year} {2014})}\BibitemShut {NoStop}%
\bibitem [{\citenamefont {Jing}\ \emph {et~al.}(2014)\citenamefont {Jing},
  \citenamefont {Huang},\ and\ \citenamefont {Hu}}]{JJing2014}%
  \BibitemOpen
  \bibfield  {author} {\bibinfo {author} {\bibfnamefont {J.}~\bibnamefont
  {Jing}}, \bibinfo {author} {\bibfnamefont {P.}~\bibnamefont {Huang}}, \ and\
  \bibinfo {author} {\bibfnamefont {X.}~\bibnamefont {Hu}},\ }\bibfield
  {title} {\enquote {\bibinfo {title} {Decoherence of an electrically driven
  spin qubit},}\ }\href@noop {} {\bibfield  {journal} {\bibinfo  {journal}
  {Phys. Rev. A}\ }\textbf {\bibinfo {volume} {90}},\ \bibinfo {pages} {022118}
  (\bibinfo {year} {2014})}\BibitemShut {NoStop}%
\bibitem [{\citenamefont {Green}\ \emph {et~al.}(2012)\citenamefont {Green},
  \citenamefont {Uys},\ and\ \citenamefont {Biercuk}}]{Green2012}%
  \BibitemOpen
  \bibfield  {author} {\bibinfo {author} {\bibfnamefont {T.}~\bibnamefont
  {Green}}, \bibinfo {author} {\bibfnamefont {H.}~\bibnamefont {Uys}}, \ and\
  \bibinfo {author} {\bibfnamefont {M.~J.}\ \bibnamefont {Biercuk}},\
  }\bibfield  {title} {\enquote {\bibinfo {title} {High-order noise filtering
  in nontrivial quantum logic gates},}\ }\href@noop {} {\bibfield  {journal}
  {\bibinfo  {journal} {Phys. Rev. Lett.}\ }\textbf {\bibinfo {volume} {109}},\
  \bibinfo {pages} {020501} (\bibinfo {year} {2012})}\BibitemShut {NoStop}%
\bibitem [{\citenamefont {Green}\ \emph {et~al.}(2013)\citenamefont {Green},
  \citenamefont {Sastrawan}, \citenamefont {Uys},\ and\ \citenamefont
  {Biercuk}}]{Green2013}%
  \BibitemOpen
  \bibfield  {author} {\bibinfo {author} {\bibfnamefont {T.~J.}\ \bibnamefont
  {Green}}, \bibinfo {author} {\bibfnamefont {J.}~\bibnamefont {Sastrawan}},
  \bibinfo {author} {\bibfnamefont {H.}~\bibnamefont {Uys}}, \ and\ \bibinfo
  {author} {\bibfnamefont {M.~J.}\ \bibnamefont {Biercuk}},\ }\bibfield
  {title} {\enquote {\bibinfo {title} {Arbitrary quantum control of qubits in
  the presence of universal noise},}\ }\href@noop {} {\bibfield  {journal}
  {\bibinfo  {journal} {New J. Phys.}\ }\textbf {\bibinfo {volume} {15}},\
  \bibinfo {pages} {095004} (\bibinfo {year} {2013})}\BibitemShut {NoStop}%
\bibitem [{\citenamefont {Soare}\ \emph {et~al.}(2014)\citenamefont {Soare},
  \citenamefont {Ball}, \citenamefont {Hayes}, \citenamefont {Sastrawan},
  \citenamefont {Jarratt}, \citenamefont {McLoughlin}, \citenamefont {Zhen},
  \citenamefont {Green},\ and\ \citenamefont {Biercuk}}]{Soare2014}%
  \BibitemOpen
  \bibfield  {author} {\bibinfo {author} {\bibfnamefont {A.}~\bibnamefont
  {Soare}}, \bibinfo {author} {\bibfnamefont {H.}~\bibnamefont {Ball}},
  \bibinfo {author} {\bibfnamefont {D.}~\bibnamefont {Hayes}}, \bibinfo
  {author} {\bibfnamefont {J.}~\bibnamefont {Sastrawan}}, \bibinfo {author}
  {\bibfnamefont {M.~C.}\ \bibnamefont {Jarratt}}, \bibinfo {author}
  {\bibfnamefont {J.~J.}\ \bibnamefont {McLoughlin}}, \bibinfo {author}
  {\bibfnamefont {X.}~\bibnamefont {Zhen}}, \bibinfo {author} {\bibfnamefont
  {T.~J.}\ \bibnamefont {Green}}, \ and\ \bibinfo {author} {\bibfnamefont
  {M.~J.}\ \bibnamefont {Biercuk}},\ }\bibfield  {title} {\enquote {\bibinfo
  {title} {Experimental noise filtering by quantum control},}\ }\href@noop {}
  {\bibfield  {journal} {\bibinfo  {journal} {Nat. Phys.}\ }\textbf {\bibinfo
  {volume} {10}},\ \bibinfo {pages} {825} (\bibinfo {year} {2014})}\BibitemShut
  {NoStop}%
\bibitem [{\citenamefont {Ball}\ \emph {et~al.}(2016)\citenamefont {Ball},
  \citenamefont {Oliver},\ and\ \citenamefont {Biercuk}}]{Ball2016}%
  \BibitemOpen
  \bibfield  {author} {\bibinfo {author} {\bibfnamefont {H.}~\bibnamefont
  {Ball}}, \bibinfo {author} {\bibfnamefont {W.~D.}\ \bibnamefont {Oliver}}, \
  and\ \bibinfo {author} {\bibfnamefont {M.~J.}\ \bibnamefont {Biercuk}},\
  }\bibfield  {title} {\enquote {\bibinfo {title} {The role of master clock
  stability in quantum information processing},}\ }\href@noop {} {\bibfield
  {journal} {\bibinfo  {journal} {npj Qu. Inf.}\ }\textbf {\bibinfo {volume}
  {2}},\ \bibinfo {pages} {16033} (\bibinfo {year} {2016})}\BibitemShut
  {NoStop}%
\bibitem [{\citenamefont {de~L\'es\'eleuc}\ \emph {et~al.}(2018)\citenamefont
  {de~L\'es\'eleuc}, \citenamefont {Barredo}, \citenamefont {Lienhard},
  \citenamefont {Browaeys},\ and\ \citenamefont {Lahaye}}]{deLeseleuc2018}%
  \BibitemOpen
  \bibfield  {author} {\bibinfo {author} {\bibfnamefont {S.}~\bibnamefont
  {de~L\'es\'eleuc}}, \bibinfo {author} {\bibfnamefont {D.}~\bibnamefont
  {Barredo}}, \bibinfo {author} {\bibfnamefont {V.}~\bibnamefont {Lienhard}},
  \bibinfo {author} {\bibfnamefont {A.}~\bibnamefont {Browaeys}}, \ and\
  \bibinfo {author} {\bibfnamefont {T.}~\bibnamefont {Lahaye}},\ }\bibfield
  {title} {\enquote {\bibinfo {title} {Analysis of imperfections in the
  coherent optical excitation of single atoms to {R}ydberg states},}\
  }\href@noop {} {\bibfield  {journal} {\bibinfo  {journal} {Phys. Rev. A}\
  }\textbf {\bibinfo {volume} {97}},\ \bibinfo {pages} {053803} (\bibinfo
  {year} {2018})}\BibitemShut {NoStop}%
\bibitem [{\citenamefont {Zhang}\ \emph {et~al.}(2021)\citenamefont {Zhang},
  \citenamefont {Xie}, \citenamefont {Zhang}, \citenamefont {Wang},
  \citenamefont {Wu}, \citenamefont {Chen}, \citenamefont {Wu},\ and\
  \citenamefont {Chen}}]{MZhang2021}%
  \BibitemOpen
  \bibfield  {author} {\bibinfo {author} {\bibfnamefont {M.}~\bibnamefont
  {Zhang}}, \bibinfo {author} {\bibfnamefont {Y.}~\bibnamefont {Xie}}, \bibinfo
  {author} {\bibfnamefont {J.}~\bibnamefont {Zhang}}, \bibinfo {author}
  {\bibfnamefont {W.}~\bibnamefont {Wang}}, \bibinfo {author} {\bibfnamefont
  {C.}~\bibnamefont {Wu}}, \bibinfo {author} {\bibfnamefont {T.}~\bibnamefont
  {Chen}}, \bibinfo {author} {\bibfnamefont {W.}~\bibnamefont {Wu}}, \ and\
  \bibinfo {author} {\bibfnamefont {P.}~\bibnamefont {Chen}},\ }\bibfield
  {title} {\enquote {\bibinfo {title} {Estimation of the laser frequency noise
  spectrum by continuous dynamical decoupling},}\ }\href@noop {} {\bibfield
  {journal} {\bibinfo  {journal} {Phys. Rev. Appl.}\ }\textbf {\bibinfo
  {volume} {15}},\ \bibinfo {pages} {014033} (\bibinfo {year}
  {2021})}\BibitemShut {NoStop}%
\bibitem [{\citenamefont {Day}\ \emph {et~al.}(2022)\citenamefont {Day},
  \citenamefont {Low}, \citenamefont {White}, \citenamefont {Islam},\ and\
  \citenamefont {Senko}}]{Day2022}%
  \BibitemOpen
  \bibfield  {author} {\bibinfo {author} {\bibfnamefont {M.~L.}\ \bibnamefont
  {Day}}, \bibinfo {author} {\bibfnamefont {P.~J.}\ \bibnamefont {Low}},
  \bibinfo {author} {\bibfnamefont {B.}~\bibnamefont {White}}, \bibinfo
  {author} {\bibfnamefont {R.}~\bibnamefont {Islam}}, \ and\ \bibinfo {author}
  {\bibfnamefont {C.}~\bibnamefont {Senko}},\ }\bibfield  {title} {\enquote
  {\bibinfo {title} {Limits on atomic qubit control from laser noise},}\
  }\href@noop {} {\bibfield  {journal} {\bibinfo  {journal} {npj Qu. Inf.}\
  }\textbf {\bibinfo {volume} {8}},\ \bibinfo {pages} {72} (\bibinfo {year}
  {2022})}\BibitemShut {NoStop}%
\bibitem [{\citenamefont {Levine}\ \emph {et~al.}(2018)\citenamefont {Levine},
  \citenamefont {Keesling}, \citenamefont {Omran}, \citenamefont {Bernien},
  \citenamefont {Schwartz}, \citenamefont {Zibrov}, \citenamefont {Endres},
  \citenamefont {Greiner}, \citenamefont {Vuleti\'c},\ and\ \citenamefont
  {Lukin}}]{Levine2018}%
  \BibitemOpen
  \bibfield  {author} {\bibinfo {author} {\bibfnamefont {H.}~\bibnamefont
  {Levine}}, \bibinfo {author} {\bibfnamefont {A.}~\bibnamefont {Keesling}},
  \bibinfo {author} {\bibfnamefont {A.}~\bibnamefont {Omran}}, \bibinfo
  {author} {\bibfnamefont {H.}~\bibnamefont {Bernien}}, \bibinfo {author}
  {\bibfnamefont {S.}~\bibnamefont {Schwartz}}, \bibinfo {author}
  {\bibfnamefont {A.~S.}\ \bibnamefont {Zibrov}}, \bibinfo {author}
  {\bibfnamefont {M.}~\bibnamefont {Endres}}, \bibinfo {author} {\bibfnamefont
  {M.}~\bibnamefont {Greiner}}, \bibinfo {author} {\bibfnamefont
  {V.}~\bibnamefont {Vuleti\'c}}, \ and\ \bibinfo {author} {\bibfnamefont
  {M.~D.}\ \bibnamefont {Lukin}},\ }\bibfield  {title} {\enquote {\bibinfo
  {title} {High-fidelity control and entanglement of {R}ydberg-atom qubits},}\
  }\href@noop {} {\bibfield  {journal} {\bibinfo  {journal} {Phys. Rev. Lett.}\
  }\textbf {\bibinfo {volume} {121}},\ \bibinfo {pages} {123603} (\bibinfo
  {year} {2018})}\BibitemShut {NoStop}%
\bibitem [{\citenamefont {Okoshi}\ \emph {et~al.}(1980)\citenamefont {Okoshi},
  \citenamefont {Kikuchi},\ and\ \citenamefont {Nakayama}}]{Okoshi1980}%
  \BibitemOpen
  \bibfield  {author} {\bibinfo {author} {\bibfnamefont {T.}~\bibnamefont
  {Okoshi}}, \bibinfo {author} {\bibfnamefont {K.}~\bibnamefont {Kikuchi}}, \
  and\ \bibinfo {author} {\bibfnamefont {A.}~\bibnamefont {Nakayama}},\
  }\bibfield  {title} {\enquote {\bibinfo {title} {Novel method for high
  resolution measurement of laser output spectrum},}\ }\href@noop {} {\bibfield
   {journal} {\bibinfo  {journal} {El. Lett.}\ }\textbf {\bibinfo {volume}
  {16}},\ \bibinfo {pages} {630} (\bibinfo {year} {1980})}\BibitemShut
  {NoStop}%
\bibitem [{\citenamefont {Domenico}\ \emph {et~al.}(2010)\citenamefont
  {Domenico}, \citenamefont {Schilt},\ and\ \citenamefont
  {Thomann}}]{DiDomenico2010}%
  \BibitemOpen
  \bibfield  {author} {\bibinfo {author} {\bibfnamefont {G.~Di}\ \bibnamefont
  {Domenico}}, \bibinfo {author} {\bibfnamefont {S.}~\bibnamefont {Schilt}}, \
  and\ \bibinfo {author} {\bibfnamefont {P.}~\bibnamefont {Thomann}},\
  }\bibfield  {title} {\enquote {\bibinfo {title} {Simple approach to the
  relation between laser frequency noise and laser line shape},}\ }\href@noop
  {} {\bibfield  {journal} {\bibinfo  {journal} {Appl. Opt.}\ }\textbf
  {\bibinfo {volume} {49}},\ \bibinfo {pages} {4801} (\bibinfo {year}
  {2010})}\BibitemShut {NoStop}%
\bibitem [{\citenamefont {Elliott}\ \emph {et~al.}(1982)\citenamefont
  {Elliott}, \citenamefont {Roy},\ and\ \citenamefont {Smith}}]{Elliott1982}%
  \BibitemOpen
  \bibfield  {author} {\bibinfo {author} {\bibfnamefont {D.~S.}\ \bibnamefont
  {Elliott}}, \bibinfo {author} {\bibfnamefont {R.}~\bibnamefont {Roy}}, \ and\
  \bibinfo {author} {\bibfnamefont {S.~J.}\ \bibnamefont {Smith}},\ }\bibfield
  {title} {\enquote {\bibinfo {title} {Extracavity laser band-shape and
  bandwidth modification},}\ }\href@noop {} {\bibfield  {journal} {\bibinfo
  {journal} {Phys. Rev. A}\ }\textbf {\bibinfo {volume} {26}},\ \bibinfo
  {pages} {12} (\bibinfo {year} {1982})}\BibitemShut {NoStop}%
\bibitem [{\citenamefont {Zhu}\ and\ \citenamefont {Hall}(1993)}]{MZhu1993}%
  \BibitemOpen
  \bibfield  {author} {\bibinfo {author} {\bibfnamefont {M.}~\bibnamefont
  {Zhu}}\ and\ \bibinfo {author} {\bibfnamefont {J.~L.}\ \bibnamefont {Hall}},\
  }\bibfield  {title} {\enquote {\bibinfo {title} {Stabilization of optical
  phase/frequency of a laser system: application to a commercial dye laser with
  an external stabilizer},}\ }\href@noop {} {\bibfield  {journal} {\bibinfo
  {journal} {J. Opt. Soc. Am. B}\ }\textbf {\bibinfo {volume} {10}},\ \bibinfo
  {pages} {802} (\bibinfo {year} {1993})}\BibitemShut {NoStop}%
\bibitem [{\citenamefont {Riehle}(2004)}]{Riehle2004}%
  \BibitemOpen
  \bibfield  {author} {\bibinfo {author} {\bibfnamefont {F.}~\bibnamefont
  {Riehle}},\ }\href@noop {} {\emph {\bibinfo {title} {Frequency standards
  basics and applications}}}\ (\bibinfo  {publisher} {Wiley-{VCH}},\ \bibinfo
  {year} {2004})\BibitemShut {NoStop}%
\bibitem [{\citenamefont {Gallion}\ and\ \citenamefont
  {Debarge}(1984)}]{Gallion1984}%
  \BibitemOpen
  \bibfield  {author} {\bibinfo {author} {\bibfnamefont {P.~B.}\ \bibnamefont
  {Gallion}}\ and\ \bibinfo {author} {\bibfnamefont {G.}~\bibnamefont
  {Debarge}},\ }\bibfield  {title} {\enquote {\bibinfo {title} {Quantum phase
  noise and field correlation in single frequency semiconductor laser
  systems},}\ }\href@noop {} {\bibfield  {journal} {\bibinfo  {journal} {IEEE
  J. Qu. Electr.}\ }\textbf {\bibinfo {volume} {20}},\ \bibinfo {pages} {343}
  (\bibinfo {year} {1984})}\BibitemShut {NoStop}%
\bibitem [{\citenamefont {Li}\ \emph {et~al.}(2019)\citenamefont {Li},
  \citenamefont {Fu}, \citenamefont {Zhu}, \citenamefont {Fang}, \citenamefont
  {Zhu}, \citenamefont {Zhong}, \citenamefont {Xu}, \citenamefont {Chen},
  \citenamefont {Wang},\ and\ \citenamefont {Zhan}}]{YLi2019}%
  \BibitemOpen
  \bibfield  {author} {\bibinfo {author} {\bibfnamefont {Y.}~\bibnamefont
  {Li}}, \bibinfo {author} {\bibfnamefont {Z.}~\bibnamefont {Fu}}, \bibinfo
  {author} {\bibfnamefont {L.}~\bibnamefont {Zhu}}, \bibinfo {author}
  {\bibfnamefont {J.}~\bibnamefont {Fang}}, \bibinfo {author} {\bibfnamefont
  {H.}~\bibnamefont {Zhu}}, \bibinfo {author} {\bibfnamefont {J.}~\bibnamefont
  {Zhong}}, \bibinfo {author} {\bibfnamefont {P.}~\bibnamefont {Xu}}, \bibinfo
  {author} {\bibfnamefont {X.}~\bibnamefont {Chen}}, \bibinfo {author}
  {\bibfnamefont {J.}~\bibnamefont {Wang}}, \ and\ \bibinfo {author}
  {\bibfnamefont {M.}~\bibnamefont {Zhan}},\ }\bibfield  {title} {\enquote
  {\bibinfo {title} {Laser frequency noise measurement using an envelope-ratio
  method based on a delayed self-heterodyne interferometer},}\ }\href@noop {}
  {\bibfield  {journal} {\bibinfo  {journal} {Opt. Commun.}\ }\textbf {\bibinfo
  {volume} {435}},\ \bibinfo {pages} {244} (\bibinfo {year}
  {2019})}\BibitemShut {NoStop}%
\bibitem [{\citenamefont {Tsuchida}(2011)}]{Tsuchida2011}%
  \BibitemOpen
  \bibfield  {author} {\bibinfo {author} {\bibfnamefont {H.}~\bibnamefont
  {Tsuchida}},\ }\bibfield  {title} {\enquote {\bibinfo {title} {Laser
  frequency modulation noise measurement by recirculating delayed
  self-heterodyne method},}\ }\href@noop {} {\bibfield  {journal} {\bibinfo
  {journal} {Opt. Lett.}\ }\textbf {\bibinfo {volume} {36}},\ \bibinfo {pages}
  {681} (\bibinfo {year} {2011})}\BibitemShut {NoStop}%
\bibitem [{\citenamefont {Richter}\ \emph {et~al.}(1986)\citenamefont
  {Richter}, \citenamefont {Mandelberg}, \citenamefont {Kruger},\ and\
  \citenamefont {McGrath}}]{Richter1986}%
  \BibitemOpen
  \bibfield  {author} {\bibinfo {author} {\bibfnamefont {L.}~\bibnamefont
  {Richter}}, \bibinfo {author} {\bibfnamefont {H.}~\bibnamefont {Mandelberg}},
  \bibinfo {author} {\bibfnamefont {M.}~\bibnamefont {Kruger}}, \ and\ \bibinfo
  {author} {\bibfnamefont {P.}~\bibnamefont {McGrath}},\ }\bibfield  {title}
  {\enquote {\bibinfo {title} {Linewidth determination from self-heterodyne
  measurements with subcoherence delay times},}\ }\href@noop {} {\bibfield
  {journal} {\bibinfo  {journal} {IEEE J. Qu. Electr.}\ }\textbf {\bibinfo
  {volume} {22}},\ \bibinfo {pages} {2070} (\bibinfo {year}
  {1986})}\BibitemShut {NoStop}%
\bibitem [{\citenamefont {Drever}\ \emph {et~al.}(1983)\citenamefont {Drever},
  \citenamefont {Hall}, \citenamefont {Kowalski}, \citenamefont {Hough},
  \citenamefont {Ford}, \citenamefont {Munley},\ and\ \citenamefont
  {Ward}}]{Drever1983}%
  \BibitemOpen
  \bibfield  {author} {\bibinfo {author} {\bibfnamefont {R.~W.~P.}\
  \bibnamefont {Drever}}, \bibinfo {author} {\bibfnamefont {J.~L.}\
  \bibnamefont {Hall}}, \bibinfo {author} {\bibfnamefont {F.~V.}\ \bibnamefont
  {Kowalski}}, \bibinfo {author} {\bibfnamefont {J.}~\bibnamefont {Hough}},
  \bibinfo {author} {\bibfnamefont {G.~M.}\ \bibnamefont {Ford}}, \bibinfo
  {author} {\bibfnamefont {A.~J.}\ \bibnamefont {Munley}}, \ and\ \bibinfo
  {author} {\bibfnamefont {H.}~\bibnamefont {Ward}},\ }\bibfield  {title}
  {\enquote {\bibinfo {title} {Laser phase and frequency stabilization using an
  optical resonator},}\ }\href@noop {} {\bibfield  {journal} {\bibinfo
  {journal} {Appl. Phys. B}\ }\textbf {\bibinfo {volume} {31}},\ \bibinfo
  {pages} {97} (\bibinfo {year} {1983})}\BibitemShut {NoStop}%
\bibitem [{\citenamefont {Graham}\ \emph {et~al.}(2022)\citenamefont {Graham},
  \citenamefont {Song}, \citenamefont {Scott}, \citenamefont {Poole},
  \citenamefont {Phuttitarn}, \citenamefont {Jooya}, \citenamefont {Eichler},
  \citenamefont {Jiang}, \citenamefont {Marra}, \citenamefont {Grinkemeyer},
  \citenamefont {Kwon}, \citenamefont {Ebert}, \citenamefont {Cherek},
  \citenamefont {Lichtman}, \citenamefont {Gillette}, \citenamefont {Gilbert},
  \citenamefont {Bowman}, \citenamefont {Ballance}, \citenamefont {Campbell},
  \citenamefont {Dahl}, \citenamefont {Crawford}, \citenamefont {Blunt},
  \citenamefont {Rogers}, \citenamefont {Noel},\ and\ \citenamefont
  {Saffman}}]{Graham2022}%
  \BibitemOpen
  \bibfield  {author} {\bibinfo {author} {\bibfnamefont {T.~M.}\ \bibnamefont
  {Graham}}, \bibinfo {author} {\bibfnamefont {Y.}~\bibnamefont {Song}},
  \bibinfo {author} {\bibfnamefont {J.}~\bibnamefont {Scott}}, \bibinfo
  {author} {\bibfnamefont {C.}~\bibnamefont {Poole}}, \bibinfo {author}
  {\bibfnamefont {L.}~\bibnamefont {Phuttitarn}}, \bibinfo {author}
  {\bibfnamefont {K.}~\bibnamefont {Jooya}}, \bibinfo {author} {\bibfnamefont
  {P.}~\bibnamefont {Eichler}}, \bibinfo {author} {\bibfnamefont
  {X.}~\bibnamefont {Jiang}}, \bibinfo {author} {\bibfnamefont
  {A.}~\bibnamefont {Marra}}, \bibinfo {author} {\bibfnamefont
  {B.}~\bibnamefont {Grinkemeyer}}, \bibinfo {author} {\bibfnamefont
  {M.}~\bibnamefont {Kwon}}, \bibinfo {author} {\bibfnamefont {M.}~\bibnamefont
  {Ebert}}, \bibinfo {author} {\bibfnamefont {J.}~\bibnamefont {Cherek}},
  \bibinfo {author} {\bibfnamefont {M.~T.}\ \bibnamefont {Lichtman}}, \bibinfo
  {author} {\bibfnamefont {M.}~\bibnamefont {Gillette}}, \bibinfo {author}
  {\bibfnamefont {J.}~\bibnamefont {Gilbert}}, \bibinfo {author} {\bibfnamefont
  {D.}~\bibnamefont {Bowman}}, \bibinfo {author} {\bibfnamefont
  {T.}~\bibnamefont {Ballance}}, \bibinfo {author} {\bibfnamefont
  {C.}~\bibnamefont {Campbell}}, \bibinfo {author} {\bibfnamefont {E.~D.}\
  \bibnamefont {Dahl}}, \bibinfo {author} {\bibfnamefont {O.}~\bibnamefont
  {Crawford}}, \bibinfo {author} {\bibfnamefont {N.~S.}\ \bibnamefont {Blunt}},
  \bibinfo {author} {\bibfnamefont {B.}~\bibnamefont {Rogers}}, \bibinfo
  {author} {\bibfnamefont {T.}~\bibnamefont {Noel}}, \ and\ \bibinfo {author}
  {\bibfnamefont {M.}~\bibnamefont {Saffman}},\ }\bibfield  {title} {\enquote
  {\bibinfo {title} {Multi-qubit entanglement and algorithms on a neutral-atom
  quantum computer},}\ }\href@noop {} {\bibfield  {journal} {\bibinfo
  {journal} {Nature}\ }\textbf {\bibinfo {volume} {604}},\ \bibinfo {pages}
  {457--462} (\bibinfo {year} {2022})}\BibitemShut {NoStop}%
\bibitem [{\citenamefont {Tucker}\ \emph {et~al.}(1984)\citenamefont {Tucker},
  \citenamefont {Challenor},\ and\ \citenamefont {Carter}}]{Tucker1984}%
  \BibitemOpen
  \bibfield  {author} {\bibinfo {author} {\bibfnamefont {M.~J.}\ \bibnamefont
  {Tucker}}, \bibinfo {author} {\bibfnamefont {P.~G.}\ \bibnamefont
  {Challenor}}, \ and\ \bibinfo {author} {\bibfnamefont {D.~J.~T.}\
  \bibnamefont {Carter}},\ }\bibfield  {title} {\enquote {\bibinfo {title}
  {Numerical simulation of a random sea: a common error and its effect upon
  wave group statistics},}\ }\href@noop {} {\bibfield  {journal} {\bibinfo
  {journal} {Appl. Ocean Res.}\ }\textbf {\bibinfo {volume} {6}},\ \bibinfo
  {pages} {118} (\bibinfo {year} {1984})}\BibitemShut {NoStop}%
\bibitem [{\citenamefont {Saulnier}\ \emph {et~al.}(2009)\citenamefont
  {Saulnier}, \citenamefont {Ricci}, \citenamefont {Falcao},\ and\
  \citenamefont {Clement}}]{Saulnier2009}%
  \BibitemOpen
  \bibfield  {author} {\bibinfo {author} {\bibfnamefont {J.-B.}\ \bibnamefont
  {Saulnier}}, \bibinfo {author} {\bibfnamefont {P.}~\bibnamefont {Ricci}},
  \bibinfo {author} {\bibfnamefont {A.~F.}\ \bibnamefont {Falcao}}, \ and\
  \bibinfo {author} {\bibfnamefont {A.~H.}\ \bibnamefont {Clement}},\
  }\bibfield  {title} {\enquote {\bibinfo {title} {{Mean Power Output
  Estimation of {WEC}s in Simulated Sea-States}},}\ }in\ \href
  {https://hal.archives-ouvertes.fr/hal-01156320} {\emph {\bibinfo {booktitle}
  {{8th European Wave \& Tidal Energy Conference}}}}\ (\bibinfo {address}
  {Uppsala, Sweden},\ \bibinfo {year} {2009})\BibitemShut {NoStop}%
\bibitem [{\citenamefont {Haslwanter}\ \emph {et~al.}(1988)\citenamefont
  {Haslwanter}, \citenamefont {Ritsch}, \citenamefont {Cooper},\ and\
  \citenamefont {Zoller}}]{Haslwanter1988}%
  \BibitemOpen
  \bibfield  {author} {\bibinfo {author} {\bibfnamefont {Th.}\ \bibnamefont
  {Haslwanter}}, \bibinfo {author} {\bibfnamefont {H.}~\bibnamefont {Ritsch}},
  \bibinfo {author} {\bibfnamefont {J.}~\bibnamefont {Cooper}}, \ and\ \bibinfo
  {author} {\bibfnamefont {P.}~\bibnamefont {Zoller}},\ }\bibfield  {title}
  {\enquote {\bibinfo {title} {Laser-noise-induced population fluctuations in
  two- and three-level systems},}\ }\href@noop {} {\bibfield  {journal}
  {\bibinfo  {journal} {Phys. Rev. A}\ }\textbf {\bibinfo {volume} {38}},\
  \bibinfo {pages} {5652--5659} (\bibinfo {year} {1988})}\BibitemShut {NoStop}%
\bibitem [{\citenamefont {Kubo}(1963)}]{Kubo1963}%
  \BibitemOpen
  \bibfield  {author} {\bibinfo {author} {\bibfnamefont {R.}~\bibnamefont
  {Kubo}},\ }\bibfield  {title} {\enquote {\bibinfo {title} {Stochastic
  {L}iouville equations},}\ }\href@noop {} {\bibfield  {journal} {\bibinfo
  {journal} {J. Math. Phys.}\ }\textbf {\bibinfo {volume} {4}},\ \bibinfo
  {pages} {174} (\bibinfo {year} {1963})}\BibitemShut {NoStop}%
\bibitem [{\citenamefont {Bowdrey}\ \emph {et~al.}(2002)\citenamefont
  {Bowdrey}, \citenamefont {Oi}, \citenamefont {Short}, \citenamefont
  {Banaszek},\ and\ \citenamefont {Jones}}]{Bowdrey2002}%
  \BibitemOpen
  \bibfield  {author} {\bibinfo {author} {\bibfnamefont {M.~D.}\ \bibnamefont
  {Bowdrey}}, \bibinfo {author} {\bibfnamefont {D.~K.~L.}\ \bibnamefont {Oi}},
  \bibinfo {author} {\bibfnamefont {A.~J.}\ \bibnamefont {Short}}, \bibinfo
  {author} {\bibfnamefont {K.}~\bibnamefont {Banaszek}}, \ and\ \bibinfo
  {author} {\bibfnamefont {J.~A.}\ \bibnamefont {Jones}},\ }\bibfield  {title}
  {\enquote {\bibinfo {title} {Fidelity of single qubit maps},}\ }\href@noop {}
  {\bibfield  {journal} {\bibinfo  {journal} {Phys. Lett. A}\ }\textbf
  {\bibinfo {volume} {294}},\ \bibinfo {pages} {258} (\bibinfo {year}
  {2002})}\BibitemShut {NoStop}%
\bibitem [{\citenamefont {Nielsen}(2002)}]{Nielsen2002}%
  \BibitemOpen
  \bibfield  {author} {\bibinfo {author} {\bibfnamefont {M.~A.}\ \bibnamefont
  {Nielsen}},\ }\bibfield  {title} {\enquote {\bibinfo {title} {A simple
  formula for the average gate fidelity of a quantum dynamical operation},}\
  }\href@noop {} {\bibfield  {journal} {\bibinfo  {journal} {Phys. Lett. A}\
  }\textbf {\bibinfo {volume} {303}},\ \bibinfo {pages} {249} (\bibinfo {year}
  {2002})}\BibitemShut {NoStop}%
\bibitem [{\citenamefont {Johnson}\ \emph {et~al.}(2008)\citenamefont
  {Johnson}, \citenamefont {Urban}, \citenamefont {Henage}, \citenamefont
  {Isenhower}, \citenamefont {Yavuz}, \citenamefont {Walker},\ and\
  \citenamefont {Saffman}}]{Johnson2008}%
  \BibitemOpen
  \bibfield  {author} {\bibinfo {author} {\bibfnamefont {T.~A.}\ \bibnamefont
  {Johnson}}, \bibinfo {author} {\bibfnamefont {E.}~\bibnamefont {Urban}},
  \bibinfo {author} {\bibfnamefont {T.}~\bibnamefont {Henage}}, \bibinfo
  {author} {\bibfnamefont {L.}~\bibnamefont {Isenhower}}, \bibinfo {author}
  {\bibfnamefont {D.~D.}\ \bibnamefont {Yavuz}}, \bibinfo {author}
  {\bibfnamefont {T.~G.}\ \bibnamefont {Walker}}, \ and\ \bibinfo {author}
  {\bibfnamefont {M.}~\bibnamefont {Saffman}},\ }\bibfield  {title} {\enquote
  {\bibinfo {title} {{R}abi oscillations between ground and {R}ydberg states
  with dipole-dipole atomic interactions},}\ }\href@noop {} {\bibfield
  {journal} {\bibinfo  {journal} {Phys. Rev. Lett.}\ }\textbf {\bibinfo
  {volume} {100}},\ \bibinfo {pages} {113003} (\bibinfo {year}
  {2008})}\BibitemShut {NoStop}%
\bibitem [{\citenamefont {Saffman}\ \emph {et~al.}(2010)\citenamefont
  {Saffman}, \citenamefont {Walker},\ and\ \citenamefont
  {M\o{}lmer}}]{Saffman2010}%
  \BibitemOpen
  \bibfield  {author} {\bibinfo {author} {\bibfnamefont {M.}~\bibnamefont
  {Saffman}}, \bibinfo {author} {\bibfnamefont {T.~G.}\ \bibnamefont {Walker}},
  \ and\ \bibinfo {author} {\bibfnamefont {K.}~\bibnamefont {M\o{}lmer}},\
  }\bibfield  {title} {\enquote {\bibinfo {title} {Quantum information with
  {R}ydberg atoms},}\ }\href@noop {} {\bibfield  {journal} {\bibinfo  {journal}
  {Rev. Mod. Phys.}\ }\textbf {\bibinfo {volume} {82}},\ \bibinfo {pages}
  {2313--2363} (\bibinfo {year} {2010})}\BibitemShut {NoStop}%
\bibitem [{\citenamefont {Graham}\ \emph {et~al.}(2019)\citenamefont {Graham},
  \citenamefont {Kwon}, \citenamefont {Grinkemeyer}, \citenamefont {Marra},
  \citenamefont {Jiang}, \citenamefont {Lichtman}, \citenamefont {Sun},
  \citenamefont {Ebert},\ and\ \citenamefont {Saffman}}]{Graham2019}%
  \BibitemOpen
  \bibfield  {author} {\bibinfo {author} {\bibfnamefont {T.}~\bibnamefont
  {Graham}}, \bibinfo {author} {\bibfnamefont {M.}~\bibnamefont {Kwon}},
  \bibinfo {author} {\bibfnamefont {B.}~\bibnamefont {Grinkemeyer}}, \bibinfo
  {author} {\bibfnamefont {A.}~\bibnamefont {Marra}}, \bibinfo {author}
  {\bibfnamefont {X.}~\bibnamefont {Jiang}}, \bibinfo {author} {\bibfnamefont
  {M.}~\bibnamefont {Lichtman}}, \bibinfo {author} {\bibfnamefont
  {Y.}~\bibnamefont {Sun}}, \bibinfo {author} {\bibfnamefont {M.}~\bibnamefont
  {Ebert}}, \ and\ \bibinfo {author} {\bibfnamefont {M.}~\bibnamefont
  {Saffman}},\ }\bibfield  {title} {\enquote {\bibinfo {title} {Rydberg
  mediated entanglement in a two-dimensional neutral atom qubit array},}\
  }\href@noop {} {\bibfield  {journal} {\bibinfo  {journal} {Phys. Rev. Lett.}\
  }\textbf {\bibinfo {volume} {123}},\ \bibinfo {pages} {230501} (\bibinfo
  {year} {2019})}\BibitemShut {NoStop}%
\bibitem [{\citenamefont {Knoernschild}\ \emph {et~al.}(2010)\citenamefont
  {Knoernschild}, \citenamefont {Zhang}, \citenamefont {Isenhower},
  \citenamefont {Gill}, \citenamefont {Lu}, \citenamefont {Saffman},\ and\
  \citenamefont {Kim}}]{Knoernschild2010}%
  \BibitemOpen
  \bibfield  {author} {\bibinfo {author} {\bibfnamefont {C.}~\bibnamefont
  {Knoernschild}}, \bibinfo {author} {\bibfnamefont {X.~L.}\ \bibnamefont
  {Zhang}}, \bibinfo {author} {\bibfnamefont {L.}~\bibnamefont {Isenhower}},
  \bibinfo {author} {\bibfnamefont {A.~T.}\ \bibnamefont {Gill}}, \bibinfo
  {author} {\bibfnamefont {F.~P.}\ \bibnamefont {Lu}}, \bibinfo {author}
  {\bibfnamefont {M.}~\bibnamefont {Saffman}}, \ and\ \bibinfo {author}
  {\bibfnamefont {J.}~\bibnamefont {Kim}},\ }\bibfield  {title} {\enquote
  {\bibinfo {title} {Independent individual addressing of multiple neutral atom
  qubits with a {MEMS} beam steering system},}\ }\href@noop {} {\bibfield
  {journal} {\bibinfo  {journal} {Appl. Phys. Lett.}\ }\textbf {\bibinfo
  {volume} {97}},\ \bibinfo {pages} {134101} (\bibinfo {year}
  {2010})}\BibitemShut {NoStop}%
\bibitem [{\citenamefont {Akerman}\ \emph {et~al.}(2015)\citenamefont
  {Akerman}, \citenamefont {Navon}, \citenamefont {Kotler}, \citenamefont
  {Glickman},\ and\ \citenamefont {Ozeri}}]{Akerman2015}%
  \BibitemOpen
  \bibfield  {author} {\bibinfo {author} {\bibfnamefont {N.}~\bibnamefont
  {Akerman}}, \bibinfo {author} {\bibfnamefont {N.}~\bibnamefont {Navon}},
  \bibinfo {author} {\bibfnamefont {S.}~\bibnamefont {Kotler}}, \bibinfo
  {author} {\bibfnamefont {Y.}~\bibnamefont {Glickman}}, \ and\ \bibinfo
  {author} {\bibfnamefont {R.}~\bibnamefont {Ozeri}},\ }\bibfield  {title}
  {\enquote {\bibinfo {title} {Universal gate-set for trapped-ion qubits using
  a narrow linewidth diode laser},}\ }\href@noop {} {\bibfield  {journal}
  {\bibinfo  {journal} {New. J. Phys.}\ }\textbf {\bibinfo {volume} {17}},\
  \bibinfo {pages} {113060} (\bibinfo {year} {2015})}\BibitemShut {NoStop}%
\bibitem [{\citenamefont {Kasevich}\ and\ \citenamefont
  {Chu}(1992)}]{Kasevich1992}%
  \BibitemOpen
  \bibfield  {author} {\bibinfo {author} {\bibfnamefont {M.}~\bibnamefont
  {Kasevich}}\ and\ \bibinfo {author} {\bibfnamefont {S.}~\bibnamefont {Chu}},\
  }\bibfield  {title} {\enquote {\bibinfo {title} {Laser cooling below a photon
  recoil with three-level atoms},}\ }\href@noop {} {\bibfield  {journal}
  {\bibinfo  {journal} {Phys. Rev. Lett.}\ }\textbf {\bibinfo {volume} {69}},\
  \bibinfo {pages} {1741} (\bibinfo {year} {1992})}\BibitemShut {NoStop}%
\bibitem [{Note1()}]{Note1}%
  \BibitemOpen
  \bibinfo {note} {The sine and cosine integral functions are defined as
  $\protect \text {Si}(z)=\DOTSI \intop \ilimits@ _0^z(\sin t)/t\protect \,dt$
  and $\protect \text {Ci}(z)=-\DOTSI \intop \ilimits@ _z^\infty (\cos
  t)/t\protect \,dt$.}\BibitemShut {Stop}%
\bibitem [{\citenamefont {Vahala}\ \emph {et~al.}(1983)\citenamefont {Vahala},
  \citenamefont {harder},\ and\ \citenamefont {Yariv}}]{Vahala1983}%
  \BibitemOpen
  \bibfield  {author} {\bibinfo {author} {\bibfnamefont {K.}~\bibnamefont
  {Vahala}}, \bibinfo {author} {\bibfnamefont {Ch.}\ \bibnamefont {harder}}, \
  and\ \bibinfo {author} {\bibfnamefont {A.}~\bibnamefont {Yariv}},\ }\bibfield
   {title} {\enquote {\bibinfo {title} {Observation of relaxation resonance
  effects in the field spectrum of semiconductor lasers},}\ }\href@noop {}
  {\bibfield  {journal} {\bibinfo  {journal} {Appl. Phys. lett.}\ }\textbf
  {\bibinfo {volume} {42}},\ \bibinfo {pages} {211} (\bibinfo {year}
  {1983})}\BibitemShut {NoStop}%
\bibitem [{\citenamefont {Koechner}(1972)}]{Koechner1972}%
  \BibitemOpen
  \bibfield  {author} {\bibinfo {author} {\bibfnamefont {W.}~\bibnamefont
  {Koechner}},\ }\bibfield  {title} {\enquote {\bibinfo {title} {Output
  fluctuations of {CW}-pumped {N}d:{YAG} lasers},}\ }\href@noop {} {\bibfield
  {journal} {\bibinfo  {journal} {IEEE J. Qu. Electr.}\ }\textbf {\bibinfo
  {volume} {8}},\ \bibinfo {pages} {656} (\bibinfo {year} {1972})}\BibitemShut
  {NoStop}%
\bibitem [{\citenamefont {Gillen-Christandl}\ \emph {et~al.}(2016)\citenamefont
  {Gillen-Christandl}, \citenamefont {Gillen}, \citenamefont {Piotrowicz},\
  and\ \citenamefont {Saffman}}]{Gillen-Christandl2016}%
  \BibitemOpen
  \bibfield  {author} {\bibinfo {author} {\bibfnamefont {K.}~\bibnamefont
  {Gillen-Christandl}}, \bibinfo {author} {\bibfnamefont {G.}~\bibnamefont
  {Gillen}}, \bibinfo {author} {\bibfnamefont {M.~J.}\ \bibnamefont
  {Piotrowicz}}, \ and\ \bibinfo {author} {\bibfnamefont {M.}~\bibnamefont
  {Saffman}},\ }\bibfield  {title} {\enquote {\bibinfo {title} {Comparison of
  {G}aussian and super {G}aussian laser beams for addressing atomic qubits},}\
  }\href@noop {} {\bibfield  {journal} {\bibinfo  {journal} {Appl. Phys. B}\
  }\textbf {\bibinfo {volume} {122}},\ \bibinfo {pages} {131} (\bibinfo {year}
  {2016})}\BibitemShut {NoStop}%
\bibitem [{\citenamefont {Gaebler}\ \emph {et~al.}(2016)\citenamefont
  {Gaebler}, \citenamefont {Tan}, \citenamefont {Lin}, \citenamefont {Wan},
  \citenamefont {Bowler}, \citenamefont {Keith}, \citenamefont {Glancy},
  \citenamefont {Coakley}, \citenamefont {Knill}, \citenamefont {Leibfried},\
  and\ \citenamefont {Wineland}}]{Gaebler2016}%
  \BibitemOpen
  \bibfield  {author} {\bibinfo {author} {\bibfnamefont {J.~P.}\ \bibnamefont
  {Gaebler}}, \bibinfo {author} {\bibfnamefont {T.~R.}\ \bibnamefont {Tan}},
  \bibinfo {author} {\bibfnamefont {Y.}~\bibnamefont {Lin}}, \bibinfo {author}
  {\bibfnamefont {Y.}~\bibnamefont {Wan}}, \bibinfo {author} {\bibfnamefont
  {R.}~\bibnamefont {Bowler}}, \bibinfo {author} {\bibfnamefont {A.~C.}\
  \bibnamefont {Keith}}, \bibinfo {author} {\bibfnamefont {S.}~\bibnamefont
  {Glancy}}, \bibinfo {author} {\bibfnamefont {K.}~\bibnamefont {Coakley}},
  \bibinfo {author} {\bibfnamefont {E.}~\bibnamefont {Knill}}, \bibinfo
  {author} {\bibfnamefont {D.}~\bibnamefont {Leibfried}}, \ and\ \bibinfo
  {author} {\bibfnamefont {D.~J.}\ \bibnamefont {Wineland}},\ }\bibfield
  {title} {\enquote {\bibinfo {title} {High-fidelity universal gate set for
  ${^{9}\mathrm{Be}}^{+}$ ion qubits},}\ }\href@noop {} {\bibfield  {journal}
  {\bibinfo  {journal} {Phys. Rev. Lett.}\ }\textbf {\bibinfo {volume} {117}},\
  \bibinfo {pages} {060505} (\bibinfo {year} {2016})}\BibitemShut {NoStop}%
\bibitem [{\citenamefont {Pedersen}\ \emph {et~al.}(2007)\citenamefont
  {Pedersen}, \citenamefont {M\o{}ller},\ and\ \citenamefont
  {M\o{}lmer}}]{Pedersen2007}%
  \BibitemOpen
  \bibfield  {author} {\bibinfo {author} {\bibfnamefont {L.~H.}\ \bibnamefont
  {Pedersen}}, \bibinfo {author} {\bibfnamefont {N.~M.}\ \bibnamefont
  {M\o{}ller}}, \ and\ \bibinfo {author} {\bibfnamefont {K.}~\bibnamefont
  {M\o{}lmer}},\ }\bibfield  {title} {\enquote {\bibinfo {title} {Fidelity of
  quantum operations},}\ }\href@noop {} {\bibfield  {journal} {\bibinfo
  {journal} {Phys. Lett. A}\ }\textbf {\bibinfo {volume} {367}},\ \bibinfo
  {pages} {47} (\bibinfo {year} {2007})}\BibitemShut {NoStop}%
\end{thebibliography}%

\end{document}